\documentclass[11pt]{article}

\usepackage[final]{acl}
\usepackage{times}
\usepackage{amsmath}
\usepackage{latexsym}
\usepackage{adjustbox}
\usepackage{booktabs}
\usepackage{multirow} 
\usepackage{booktabs}
\usepackage{makecell}
\usepackage{tabularx}
\usepackage{siunitx,xcolor}
\usepackage{adjustbox}
\usepackage{amssymb}
\usepackage{pifont}
\usepackage[table,dvipsnames]{xcolor} 
\usepackage{tcolorbox}
\usepackage{subcaption}

\usepackage[T1]{fontenc}

\usepackage[utf8]{inputenc}

\usepackage{microtype}

\usepackage{inconsolata}

\usepackage{graphicx}

\title{
MM-BRIGHT: A Multi-Task Multimodal Benchmark for Reasoning-Intensive Retrieval
}

\author{
  \textbf{Abdelrahman Abdallah$^{1,*}$, Mohamed Darwish Mounis$^{2,*}$, Mahmoud Abdalla$^{3,*}$,} \\
  \textbf{Mahmoud SalahEldin Kasem$^{3,*}$, Mostafa Farouk Senussi$^{3,*}$, Mohamed Mahmoud$^{3,*}$,} \\
  \textbf{Mohammed Ali$^{1,*}$, Adam Jatowt$^{1}$, Hyun-Soo Kang$^{3}$} \\[0.5em]
  $^{1}$University of Innsbruck \quad $^{2}$High Institute for Computer \& Information Systems \\
  $^{3}$Chungbuk National University \\[0.3em]
  \texttt{\{abdelrahman.abdallah,mohammed.ali,adam.jatowt\}@uibk.ac.at} \\[0.3em]
  $^{*}$Equal contribution
}

\begin{document}
\maketitle

\begin{abstract}
Existing retrieval benchmarks primarily consist of text-based queries where keyword or semantic matching is usually sufficient. Many real-world queries contain multimodal elements, particularly, images such as diagrams, charts, and screenshots that require intensive reasoning to identify relevant documents. To address this gap, we introduce \textsc{MM-BRIGHT}, the first multimodal benchmark for reasoning-intensive retrieval. Our dataset consists of 2,803 real-world queries spanning 29 diverse technical domains, with four tasks of increasing complexity: text-to-text, multimodal-to-text, multimodal-to-image, and multimodal-to-multimodal retrieval. Extensive evaluation reveals that state-of-the-art models struggle across all tasks: BM25 achieves only 8.5 nDCG@10 on text-only retrieval, while the best multimodal model Nomic-Vision reaches just 27.6 nDCG@10 on multimodal-to-text retrieval actually underperforming the best text-only model (DiVeR: 32.2). These results highlight substantial headroom and position \textsc{MM-BRIGHT} as a testbed for next-generation retrieval models that better integrate visual reasoning\footnote{Our code and data are available at \url{https://github.com/mm-bright/MM-BRIGHT}. See also our official website: \url{https://mm-bright.github.io/}.}
\end{abstract}

\section{Introduction}
\begin{figure}[t]
    \centering
    \includegraphics[width=0.5\textwidth]{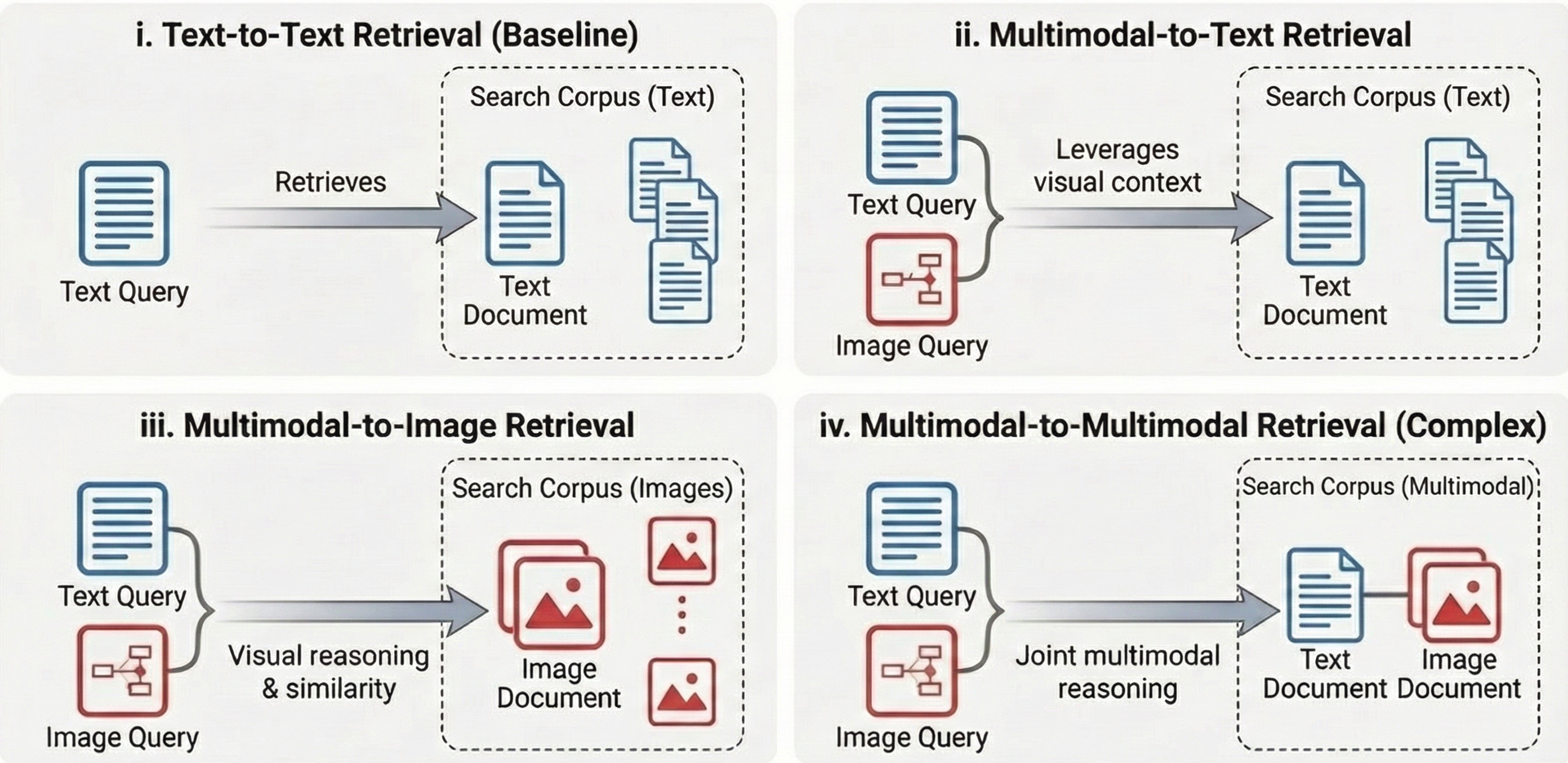}
    \caption{To comprehensively evaluate multimodal retrieval capabilities, we systematically define four retrieval tasks of increasing multimodal complexity. These range from a text-only baseline (i) to complex multimodal-to-multimodal retrieval (iv), requiring different levels of visual reasoning and context integration.}
    \label{fig:main}
\end{figure}
Information retrieval is a fundamental technology that assists users in locating relevant information from extensive corpora, containing documents, web pages, and multimedia content~\citep{abdallah2025tempretriever,nguyen2016ms,thakur2021beir}.
In real-world applications, queries and documents increasingly contain multimodal elements, particularly images such as diagrams, charts, screenshots, and scientific figures that are integral to understanding the information need~\citep{chang2022webqa,meng2025vlm2vec}.
For instance, a software developer troubleshooting a bug might include an error screenshot in their query, or a biologist might need to find research papers containing specific types of microscopy images.
In these scenarios, the visual elements are not merely supplementary; they carry essential information that cannot be adequately captured by text alone.

Despite the prevalence of multimodal queries in real-world applications, existing retrieval benchmarks remain predominantly text-centric.
While recent work has made progress in multimodal retrieval~\citep{wei2024uniir,meng2025vlm2vec,liu2021image}, these benchmarks primarily evaluate surface-level semantic correspondence between queries and documents, where simple visual similarity or object-text matching suffices.
In parallel, the text retrieval community has recognized the importance of reasoning-intensive retrieval, introducing benchmarks like BRIGHT~\citep{su2024bright} and RAR-b~\citep{xiao2024rar} that require deeper logical inference beyond keyword or semantic matching.
However, these benchmarks are limited to text-only queries and documents, leaving a critical gap: \textit{how do retrieval systems perform when both multimodal understanding and intensive reasoning are required simultaneously?}

\newcommand{\cmark}{\ding{51}}%
\newcommand{\xmark}{\ding{55}}%
\begin{table*}[t!]
\centering
\large

\begin{adjustbox}{width=0.8\textwidth,center}
\begin{tabular}{lccccccc}
\toprule
\textbf{Benchmark} & 
\textbf{\#Queries} & 
\textbf{\#Domains} &
\textbf{Modality} & 
\makecell{\textbf{Reasoning-}\\\textbf{Intensive}} & 
\makecell{\textbf{Technical/}\\\textbf{Expert}} & 
\makecell{\textbf{Multi-Task}\\\textbf{Evaluation}} &
\makecell{\textbf{Complex}\\\textbf{Queries}} \\
\midrule
\multicolumn{8}{c}{\textit{Text-Only Reasoning-Intensive Benchmarks}} \\
\midrule
BRIGHT & 1,384 & 12 & Text & \cmark & \cmark & \cmark & \cmark \\ 
RAR-b & 45,745 & 17 & Text & \cmark & \cmark & \xmark & \xmark \\ 
\midrule
\multicolumn{8}{c}{\textit{Multimodal Retrieval Benchmarks}} \\
\midrule
WebQA & 7,540 & Open & IT $\to$ IT & \xmark & \xmark & \xmark & \xmark \\ 
CIRR & 4,148 & Open & IT $\to$ I & \xmark & \xmark & \xmark & \xmark \\ 
UNIIR& 190K & 10 & Mixed & \xmark & \xmark & \cmark & \xmark \\ 
ViDoRe& 3,810 & 10 & T $\to$ IT & \xmark & \cmark & \xmark & \xmark \\
MMEB & 36K & 36 & Mixed & \xmark & \xmark & \cmark & \xmark \\
\midrule
\multicolumn{8}{c}{\textit{Multimodal Reasoning-Intensive Benchmarks}} \\
\midrule
MRMR & 1,502 & 23 & IT $\to$ IT & \cmark & \cmark & \xmark & \xmark \\ 
\textbf{\textsc{MM-BRIGHT} (Ours)} & \textbf{2,803} & \textbf{29} & \textbf{Mixed} & \cmark & \cmark & \cmark & \cmark \\
\bottomrule
\end{tabular}
\end{adjustbox}

\caption{Comparison of \textsc{MM-BRIGHT} with existing multimodal and reasoning-intensive retrieval benchmarks. \textsc{MM-BRIGHT} is the first benchmark combining multimodal queries, reasoning-intensive technical domains, and multiple retrieval task variants. \textbf{Modality legend:} T = text, I = image; IT = image+text; $X \!\to\! Y$ denotes query modality $\to$ retrieved item modality.}
\label{tab:benchmark-comparison}
\end{table*}

In this work, we address this gap by introducing \textbf{MM-BRIGHT}, a new benchmark for multimodal reasoning-intensive retrieval across different domains.
Unlike existing benchmarks that focus on either reasoning \textit{or} multimodality, \textbf{MM-BRIGHT} requires both capabilities together. It consists of 2,803 real-world queries spanning 29 diverse technical domains, sourced from StackExchange, where domain experts ask and answer complex technical questions. These queries are multimodal, 
spanning software engineering, STEM, social sciences, and applied domains (see Table~\ref{tab:dataset1_stats}). The dataset is carefully curated by expert annotators who verify that relevant documents require reasoning rather than simple keyword matching.

To comprehensively evaluate multimodal retrieval capabilities, we systematically define four retrieval tasks of increasing multimodal complexity (\autoref{fig:main}):
\textbf{(1) Query $\to$ Documents}: traditional text-only retrieval, serving as a baseline to understand reasoning intensity without multimodal complexity;
\textbf{(2) Query+Image $\to$ Documents}: multimodal queries retrieving text documents, testing whether models can leverage visual context to improve text retrieval;
\textbf{(3) Query+Image $\to$ Images}: multimodal queries retrieving relevant images, requiring visual reasoning and similarity assessment beyond simple object matching;
\textbf{(4) Query+Image $\to$ Documents+Images}: the most challenging task, retrieving multimodal documents where both text and images must be jointly evaluated for relevance. 

We conduct extensive evaluation with 18 representative retrieval models, including sparse methods, dense retrievers, reasoning-enhanced retrievers, and state-of-the-art multimodal models across diverse architectures. Our experiments reveal that \textsc{MM-BRIGHT} is challenging for all current retrievers. In Task 1, BM25 reaches only 8.5 nDCG@10 and the best model, DiVeR, achieves 32.2. Adding images does not help: in Task 2, the best multimodal model (Nomic-Vision) scores 27.6, below the text-only baseline. Performance is higher for image retrieval (Task 3: GME-2B 45.6) but drops again for multimodal document retrieval (Task 4: CLIP 28.1). Results also vary widely across models and domains, highlighting substantial headroom for reasoning-intensive multimodal retrieval.

\begin{figure*}[h]
  \centering
  \includegraphics[width=0.8\textwidth]{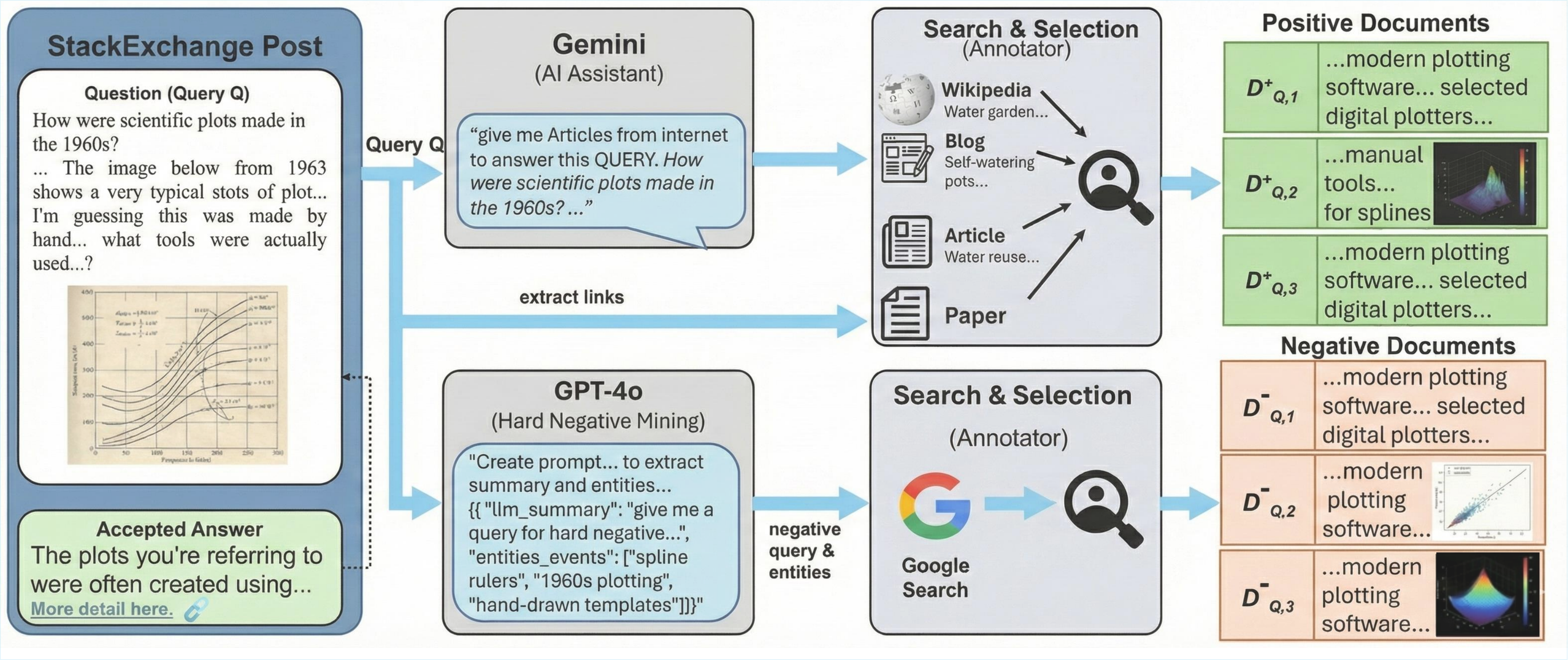}
  \caption{
  \textbf{Overview of the \textsc{MM-BRIGHT} annotation process for Stack Exchange data.} 
  Queries are multimodal Stack Exchange posts containing text and images. 
  Positive documents (text and/or images) are discovered by annotators using Gemini AI assistance or from links in accepted answers, then manually verified for relevance. 
  Negative documents are mined using GPT-4o-generated search queries and entities designed to find similar but actually 
  irrelevant content. Documents can include Wikipedia pages, blogs, articles, research papers, and technical documentation.
  }
  \label{fig:annotation}
\end{figure*}

\section{Related Work}

Traditional retrieval systems rely on lexical or semantic matching~\citep{nguyen2016ms,thakur2021beir}, but many real-world queries require multi-step reasoning. Recent benchmarks such as BRIGHT~\citep{su2024bright} and RAR-b~\citep{xiao2024rar} address this gap for text-only queries, motivating reasoning-aware retrievers~\citep{shao2025reasonir,abdallah2025asrank,abdallah2025dear,das2025rader}. However, these benchmarks remain limited to text-only settings.

As online content becomes increasingly multimodal, retrieval must handle queries and documents combining text and images. Early benchmarks emphasize cross-modal semantic alignment~\citep{abdallah2024arabicaqa,radford2021learning,abdalla2025think,kasem2025attention,liu2021image}, while recent efforts evaluate broader modality combinations~\citep{chang2022webqa,wei2024uniir,jiang2024vlm2vec,zhang2025mrmr}. However, most focus on surface correspondence rather than reasoning-intensive relevance. Some datasets incorporate expert content~\citep{mace2025vidore}, yet they target document QA rather than technical reasoning. \textsc{MM-BRIGHT} addresses this gap by combining multimodal queries with reasoning-intensive relevance across 29 technical domains (see Table~\ref{tab:benchmark-comparison} for detailed comparison).

\section{MM-BRIGHT Dataset}
\label{sec:construct}

We introduce \textsc{MM-BRIGHT}, a multimodal benchmark for reasoning-intensive retrieval across technical domains. In this section, we first formulate the task (\S\ref{sec:formulation}), then detail the data collection process for Stack Exchange (\S\ref{sec:stackexchange}). Data statistics are presented in Tables~\ref{tab:dataset1_stats} and~\ref{tab:dataset2_stats} in appendix~\ref{app:dataset2_stats} .

\subsection{Task Formulation}
\label{sec:formulation}

Given a multimodal query $Q = (Q_{\text{text}}, \{I_1, \ldots, I_k\})$ containing text and images, and a retrieval corpus $\mathcal{D} = \{D_1, \ldots, D_n\}$, retrievers are tasked to find relevant documents $\mathcal{D}^+_Q = \{D_{Q,1}^+, \ldots, D_{Q,m}^+\} \subset \mathcal{D}$ where $m \ll n$. Negative documents are defined as $\mathcal{D}_Q^- = \mathcal{D} \setminus \mathcal{D}_Q^+$. In reasoning-intensive multimodal retrieval, the relevant document set $\mathcal{D}^+_Q$ is connected to query $Q$ through reasoning traces involving visual understanding and logical inference about underlying technical principles, rather than simple visual similarity or keyword matching.

\textsc{MM-BRIGHT} evaluates four retrieval tasks: (1) \textbf{Query $\to$ Documents} (text-only baseline), (2) \textbf{Query+Image $\to$ Documents} (multimodal-to-text), (3) \textbf{Query+Image $\to$ Images} (image retrieval), and (4) \textbf{Query+Image $\to$ Documents+Images} (multimodal retrieval).

\subsection{StackExchange Multimodal Queries}
\label{sec:stackexchange}


StackExchange is a community-driven platform where domain experts ask and answer complex technical questions. Among its 170+ sites, we select 29 diverse technical domains spanning STEM fields (Biology, Chemistry, Physics, Mathematics, Earth Science, Bioacoustics, Bioinformatics, Medical Sciences), computing (Ubuntu, Bitcoin, Cryptography, Quantum Computing, Robotics, Salesforce, GIS, Apple), social sciences (Economics, Psychology, Philosophy, Law, Christianity, Islam), and applied domains (Aviation, Gaming, Project Management, Quantitative Finance, Sustainability, Travel, Academia). StackExchange posts often contain detailed technical descriptions with integral visual elements such as diagrams, code screenshots, and scientific figures. We construct query-document pairs based on user posts and documents referenced in answers (Figure~\ref{fig:annotation}). 

\textbf{Human annotators}\footnote{Five PhD students and one Master's student} browse posts from newest to oldest and select posts with: (1) at least one answer that is either accepted by the user or receives $>10$ votes, and (2) contains one or more images integral to understanding the question. 

\noindent\textbf{Constructing query and positive documents.} For each selected post, annotators combine the title, body text, and images to form the multimodal query $Q$. Annotators visit web pages linked in answers and use Gemini (Google's AI assistant) to discover additional relevant documents. For each web page, they extract passages and images that provide useful information for answering the query. Posts without relevant documents are discarded.  Sources include Wikipedia, technical blogs, research articles, documentation, and news sites.


\noindent\textbf{Constructing hard negative documents.} To prevent models from relying on simple semantic matching, we ensure negative documents are topically related but do not satisfy query requirements. We use GPT-4o to analyze each post and generate a search query designed to find hard negatives, along with entities and events mentioned in the post (prompt details in Appendix~\ref{app:negative_mining_prompt}). Annotators use the generated query to search Google and collect 20 hard negative URLs per query, and extract topically related passages and images.

\begin{figure}[t]
  \centering
  \includegraphics[width=0.95\columnwidth]{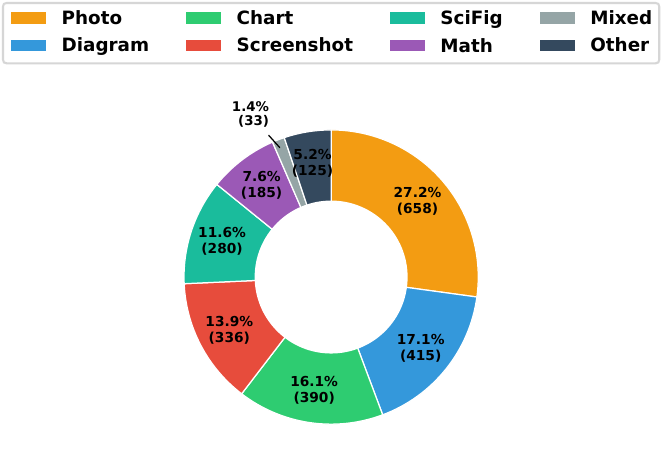}
  \caption{
  \textbf{Distribution of image types in \textsc{MM-BRIGHT}.} 
  }
  \label{fig:image_type_distribution}
\end{figure}

\noindent\textbf{Annotating images for multimodal retrieval.} For Tasks 3 and 4, which involve retrieving images or multimodal documents, we need to determine which images from the corpus are relevant to each query. We scrape all images from positive and negative web pages. For each image scraped from positive documents, we use GPT-4o to classify the image as positive (relevant) or negative (irrelevant) by providing the query, positive passages, ground truth answer, and the image itself (prompt details in Appendix~\ref{app:image_annotation_prompt}). GPT-4o evaluates whether the image directly illustrates concepts, provides visual evidence, or depicts technical content discussed in the query and positive passages. Each classification includes a detailed rationale explaining the decision. This process yields 7,621 annotated images across 1,218 queries (Table~\ref{tab:dataset2_stats}).


To ensure the gold-standard quality of the benchmark, this AI-driven process is followed by rigorous human verification. Domain-knowledgeable students review the GPT-4o classifications and rationales, which are then further verified by expert reviewers. Only annotations that receive unanimous approval from the human experts are retained, ensuring high-quality relevance judgments for both documents and images. This process yields 7,621 verified images across 1,218 queries (Table~\ref{tab:dataset2_stats}). Additional annotation guidelines are provided in Appendix~\ref{app:annotation_guidelines}.

\subsection{Image Type Diversity}
\label{sec:image_types_Diversity}

To understand the visual reasoning challenges in \textsc{MM-BRIGHT}, we analyze the distribution of image types across our 1,585 query images using GPT-4o classification (Prompt in Appendix~\ref{fig:image_classification_prompt}). As shown in Figure~\ref{fig:image_type_distribution}, \textsc{MM-BRIGHT} exhibits substantial diversity across eight categories: photos (27.2\%), diagrams (17.1\%), charts/graphs (16.1\%), screenshots (13.9\%), scientific figures (11.6\%), mathematical notation (7.6\%), mixed, and others. This diversity ensures evaluation across varied visual reasoning challenges from interpreting technical schematics and data visualizations to understanding scientific imagery and UI elements, rather than focusing on a single image type. The distribution varies significantly by domain (Appendix~\ref{app:image_types}): Biology and Aviation are dominated by photos (68.1\% and 75.2\%), Quantum Computing primarily contains diagrams (61.1\%), Economics consists mostly of charts/graphs (83.0\%), while Ask Ubuntu contains predominantly screenshots (95.1\%). This domain-specific variation reflects authentic technical communication patterns and prevents models from succeeding through image type-specific heuristics.
\begin{figure}[t]
  \centering
  \includegraphics[width=0.50\textwidth]{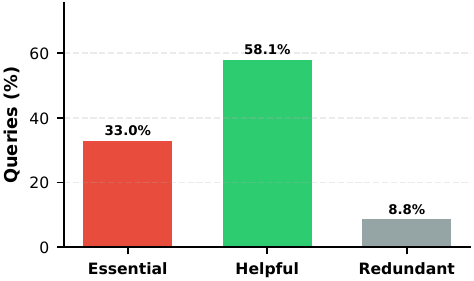}
  \caption{\textbf{Image essentiality distribution in \textsc{MM-BRIGHT}.} 
  }
  \label{fig:essentiality_distribution}
\end{figure}

\subsection{Image Essentiality Analysis}
\label{sec:image_essentiality}

To understand the role of visual information in reasoning-intensive retrieval, we use GPT-4o to classify each query's images into three categories: \textbf{Essential} (images critical for understanding the query; without them, the query would be incomplete or ambiguous), \textbf{Helpful} (images providing useful context but not strictly necessary), and \textbf{Redundant} (images duplicating text information or providing no retrieval value). Figure~\ref{fig:essentiality_distribution} shows that of 1,585 queries, 33.0\% contain essential images, 58.1\% contain helpful images, and only 8.8\% contain redundant images. The high proportion of essential and helpful images ($\sim$90\% combined) demonstrates that visual information in \textsc{MM-BRIGHT} is genuinely important for understanding queries rather than being merely decorative.

\begin{table}[t!]

\centering
\small
\resizebox{0.40\textwidth}{!}{
\begin{tabular}{l|rrr r|rr}
\toprule
& \multicolumn{4}{c|}{\textbf{Total Number}} & \multicolumn{2}{c}{\textbf{Avg. Length}} \\
\cmidrule{2-7}
\textbf{Dataset} &
\multicolumn{1}{c}{$\mathbf{Q}$} &
\multicolumn{1}{c}{$\boldsymbol{\mathcal{D}}$} &
\multicolumn{1}{c}{$\boldsymbol{\mathcal{D}^+}$} &
\multicolumn{1}{c|}{$\boldsymbol{\mathcal{I}^+}$} &
\multicolumn{1}{c}{$\mathbf{Q}$} &
\multicolumn{1}{c}{$\boldsymbol{\mathcal{D}}$} \\
\midrule

\rowcolor{gray!12}\multicolumn{7}{c}{\textit{\textbf{STEM \& Life Sciences}}} \\
\midrule
Academia & 26 & 60{,}050  & 3.5 & 1.77 & 437.5 & 534.8 \\
Bioacoustics & 41 & 29{,}812 & 2.8 & 2.17 & 454.1 & 994.5 \\
Bioinformatics & 90 & 45{,}545 & 1.5 & 1.62 & 489.4 & 481.2 \\
Biology & 99 & 89{,}435 & 1.2 & 2.96 & 347.3 & 472.3 \\
Chemistry & 65 & 36{,}043 & 2.5 & 2.54 & 499.7 & 545.2 \\
Earthscience & 85 & 73{,}451 & 2.8 & 2.15 & 364.3 & 465.9 \\
Math & 45 & 151{,}867 & 1.9 & 2.64 & 648.0 & 472.9 \\
Medicalsciences & 55 & 240{,}844 & 2.3 & 1.85 & 384.4 & 509.8 \\
Physics & 100 & 338{,}291 & 2.3 & 2.45 & 447.5 & 565.2 \\
\midrule

\rowcolor{gray!12}\multicolumn{7}{c}{\textit{\textbf{Software \& Technical Systems}}} \\
\midrule
Apple & 14 & 29{,}285 & 2.3 & 2.14 & 400.4 & 562.1 \\
Askubuntu & 35 & 90{,}198 & 1.5 & 2.09 & 309.9 & 519.8 \\
Bitcoin & 64 & 29{,}595 & 2.8 & 1.48 & 316.0 & 523.1 \\
Crypto & 74 & 24{,}054 & 1.2 & 1.50 & 578.2 & 695.2 \\
Gis & 44 & 20{,}705 & 1.3 & 2.98 & 332.5 & 556.3 \\
Quantumcomputing & 88 & 127{,}009 & 2.3 & 1.84 & 460.7 & 532.0 \\
Robotics & 30 & 11{,}185 & 3.0 & 2.33 & 685.2 & 497.7 \\
Salesforce & 10 & 8{,}890 & 1.8 & 2.50 & 549.1 & 358.3 \\
\midrule

\rowcolor{gray!12}\multicolumn{7}{c}{\textit{\textbf{Social Sciences \& Humanities}}} \\
\midrule
Christianity & 30 & 37{,}875 & 2.2 & 1.47 & 282.7 & 463.0 \\
Economics & 31 & 18{,}431 & 2.0 & 1.84 & 347.6 & 317.1 \\
Islam & 27 & 14{,}079 & 4.3 & 1.33 & 372.9 & 566.9 \\
Law & 30 & 26{,}142 & 2.7 & 1.23 & 544.5 & 891.5 \\
Philosophy & 50 & 137{,}860 & 2.6 & 1.58 & 503.4 & 513.7 \\
Psychology & 87 & 328{,}520 & 2.9 & 1.67 & 507.3 & 499.5 \\
\midrule

\rowcolor{gray!12}\multicolumn{7}{c}{\textit{\textbf{Applied Domains}}} \\
\midrule
Aviation & 125 & 203{,}938 & 2.2 & 2.41 & 287.4 & 739.7 \\
Gaming & 26 & 68{,}321 & 1.3 & 1.85 & 244.8 & 415.6 \\
Pm & 50 & 93{,}376 & 2.1 & 1.56 & 360.4 & 445.8 \\
Quant & 34 & 64{,}044 & 2.2 & 1.38 & 443.2 & 454.6 \\
Sustainability & 62 & 32{,}365 & 3.4 & 1.61 & 397.5 & 345.1 \\
Travel & 68 & 68{,}063 & 1.5 & 1.84 & 358.7 & 415.6 \\
\midrule

\textbf{Total} & \textbf{1{,}585} & \textbf{2{,}499{,}273} & -- & \textbf{2.01} & -- & -- \\
\bottomrule
\end{tabular}
}

\caption{
\textbf{Data statistics for Dataset 1 (Text Retrieval Task).}
This dataset supports Task 1: Query $\to$ Documents. 
For each domain, we report the number of queries ($\mathbf{Q}$), corpus size ($\boldsymbol{\mathcal{D}}$),
average positive documents per query ($\boldsymbol{\mathcal{D}^+}$),
average images per query ($\boldsymbol{\mathcal{I}^+}$),
and average token length for queries and documents (GPT-2 tokenizer).
Statistics for Dataset 2 (Tasks 2--4) are provided in Appendix~\ref{app:dataset2_stats}.
}
\label{tab:dataset1_stats}
\end{table}

\subsection{Dataset Quality Assessment}
\label{sec:dataset_quality_assessment}

To check the quality of human annotation for the queries and the usefulness of positive evidence in \textsc{MM-BRIGHT}, we conduct an automatic quality audit using GPT-4o as an LLM judge. We evaluate query and positive documents pairs across all domains using four 1--5 Likert criteria: \textit{Readability} (is the query well-formed), \textit{Clarity} (is the information need unambiguous), \textit{Evidence usefulness} (do the passages help reasoning toward an answer), and \textit{Evidence sufficiency} (is the evidence sufficient to answer). Overall, the dataset receives high scores for Readability (4.47) and Clarity (4.18), and the evidence is typically sufficient (4.10), while usefulness is (3.80), reflecting that many queries require multi-step reasoning rather than direct lexical overlap. We provide the full prompt and complete per-domain results in Appendix~\ref{app:llm_quality_full}.

\begin{table*}[t]
\centering

\begin{minipage}[t]{0.60\textwidth}
\centering


\resizebox{\linewidth}{0.18\textheight}{%
\begin{tabular}{lccccccccccc}
\toprule
\textbf{Domain} & \textbf{BM25} & \textbf{Contriever} & \textbf{DiVeR} & \textbf{E5} & \textbf{GritLM} & \textbf{OpenAI} & \textbf{Qwen} & \textbf{Qwen2} & \textbf{Rader} & \textbf{ReasonIR} & \textbf{SFR} \\
\midrule
\rowcolor{gray!12} \multicolumn{12}{c}{\textit{Software \& Technical Systems
}} \\
\midrule
\textbf{Acad} & 9.1 & 23.3 & \textbf{31.6} & 24.6 & 21.1 & 24.3 & 14.7 & 22.1 & 19.8 & \underline{27.2} & 23.7 \\
\textbf{Bio} & 4.8 & 18.5 & \underline{27.7} & 18.2 & 23.8 & \textbf{27.8} & 14.6 & 21.0 & 20.3 & 22.0 & 25.2 \\
\textbf{Chem} & 7.8 & 24.9 & \underline{33.8} & 25.4 & 24.9 & \textbf{33.9} & 17.8 & 25.7 & 25.0 & 28.4 & 28.3 \\
\textbf{Phys} & 4.0 & 10.7 & \textbf{19.9} & 14.6 & 12.0 & 15.0 & 10.4 & 16.0 & 16.1 & \underline{16.1} & 16.0 \\
\textbf{Math} & 2.0 & 20.7 & \textbf{34.2} & 18.1 & 24.8 & 28.8 & 15.7 & 16.1 & \underline{29.8} & 16.7 & 26.3 \\
\textbf{Earth} & 4.7 & 21.8 & \textbf{34.6} & 29.1 & 22.9 & 29.5 & 19.6 & 31.1 & 25.0 & \underline{31.5} & 30.5 \\
\textbf{BioAc} & 8.5 & 20.7 & \textbf{27.4} & 20.5 & 16.6 & 22.2 & 16.6 & 23.2 & 18.6 & 14.7 & \underline{24.2} \\
\textbf{BioInf} & 5.3 & 18.1 & 32.7 & 18.9 & 30.2 & \underline{33.7} & 15.8 & 20.3 & \textbf{37.2} & 23.9 & 28.1 \\
\textbf{Med} & 12.6 & 21.8 & 36.1 & 34.7 & 26.1 & \underline{37.0} & 30.4 & 31.2 & 29.2 & \textbf{38.4} & 31.6 \\
\midrule
\rowcolor{gray!12}\multicolumn{12}{c}{\textit{Software \& Technical Systems
}} \\
\midrule
\textbf{Apple} & 1.0 & 17.7 & 22.8 & 24.8 & 24.4 & \underline{26.5} & 18.2 & \textbf{28.6} & 16.5 & 19.4 & 25.1 \\
\textbf{Ubuntu} & 17.4 & 25.8 & \textbf{44.6} & 29.0 & \underline{39.8} & 34.6 & 35.0 & 39.2 & 30.8 & 34.1 & 27.3 \\
\textbf{BTC} & 3.3 & 13.3 & 30.4 & \underline{32.4} & 22.8 & 25.0 & 22.1 & \textbf{33.0} & 19.0 & 31.9 & 26.6 \\
\textbf{Crypto} & 0.8 & 12.4 & \textbf{24.3} & 10.2 & 17.7 & 17.9 & 10.5 & 7.8 & \underline{23.4} & 15.3 & 17.5 \\
\textbf{GIS} & 1.3 & 15.9 & \textbf{31.7} & 25.0 & 24.5 & \underline{30.9} & 21.1 & 25.2 & 28.0 & 27.5 & 29.0 \\
\textbf{QC} & 3.8 & 7.9 & 13.0 & 11.4 & 10.8 & \textbf{14.0} & 5.9 & 9.6 & 9.0 & 9.2 & \underline{13.6} \\
\textbf{Robot} & 5.0 & 17.7 & \textbf{33.4} & 20.2 & 20.3 & 26.2 & 20.7 & 23.6 & \underline{30.5} & 25.7 & 27.9 \\
\textbf{Sales} & 3.9 & 17.0 & 38.6 & 25.7 & 29.0 & \underline{42.1} & 29.1 & 37.4 & 40.0 & \textbf{52.3} & 26.9 \\
\midrule
\rowcolor{gray!12}\multicolumn{12}{c}{\textit{Social Sciences \& Humanities
}} \\
\midrule
\textbf{Econ} & 4.0 & 15.0 & \textbf{31.9} & 27.7 & 17.7 & 26.9 & 18.3 & 21.8 & \underline{28.5} & 22.7 & 26.8 \\
\textbf{Psych} & 5.3 & 20.0 & \underline{29.3} & 22.8 & 21.2 & \textbf{32.2} & 21.0 & 24.5 & 20.5 & 27.7 & 27.2 \\
\textbf{Phil} & 4.1 & 16.1 & 17.4 & 21.3 & 19.1 & \underline{22.1} & 15.6 & 17.2 & 17.5 & 21.4 & \textbf{23.9} \\
\textbf{Law} & 6.2 & 49.7 & 52.3 & 50.2 & 42.5 & 51.4 & 47.1 & \textbf{59.3} & 44.2 & \underline{57.8} & 45.4 \\
\textbf{Christ} & 30.3 & 21.3 & \underline{35.2} & 29.9 & 28.9 & 21.5 & 24.3 & \textbf{37.4} & 14.1 & 26.6 & 24.7 \\
\textbf{Islam} & 9.8 & 21.0 & \textbf{34.9} & 27.5 & 28.8 & 27.3 & 17.7 & 34.3 & 20.6 & \underline{34.3} & 27.5 \\
\midrule
\rowcolor{gray!12}\multicolumn{12}{c}{\textit{Applied Domains
}} \\
\midrule
\textbf{Aviat} & 1.1 & 21.2 & 29.5 & 21.9 & 29.2 & \underline{30.4} & 12.6 & \textbf{31.3} & 21.7 & 25.3 & 28.0 \\
\textbf{Game} & 36.0 & 23.0 & \underline{50.4} & 42.1 & 50.1 & 45.1 & 48.2 & \textbf{56.8} & 35.0 & 44.8 & 38.4 \\
\textbf{PM} & 18.0 & 24.9 & \textbf{40.4} & 31.0 & 34.8 & 24.5 & \underline{39.9} & 36.5 & 27.7 & 34.7 & 27.8 \\
\textbf{Sustain} & 11.2 & 25.5 & \underline{34.8} & 30.6 & 23.3 & 26.5 & 19.5 & 34.0 & 20.2 & \textbf{36.4} & 29.4 \\
\textbf{Travel} & 22.2 & 25.7 & \underline{38.3} & 24.8 & 30.7 & 37.0 & 24.0 & 32.4 & 28.2 & \textbf{38.6} & 28.5 \\
\textbf{Quant} & 2.3 & 12.3 & 23.2 & 22.5 & 16.4 & 22.1 & 16.9 & 18.6 & \underline{25.5} & \textbf{25.7} & 24.7 \\
\midrule
\textbf{Avg.} & 8.5 & 20.1 & \textbf{32.2} & 25.3 & 25.3 & \underline{28.8} & 21.5 & 28.1 & 24.9 & 28.6 & 26.9 \\
\bottomrule
\end{tabular}}
\captionof{table}{
\textbf{Task 1: Text-only retrieval (Query $\to$ Documents).}
nDCG@10 scores for 11 text retrieval models across 29 domains.
Best in \textbf{bold}, second best \underline{underlined}.
}
\label{tab:task1_results}
\end{minipage}
\hfill
\begin{minipage}[t]{0.37\textwidth}
\centering

\resizebox{\linewidth}{0.18\textheight}{%
\begin{tabular}{lccccccc}
\toprule
\textbf{Domain} & \textbf{BGE-VL} & \textbf{CLIP} & \textbf{GME-2B} & \textbf{GME-7B} & \textbf{\makecell{Jina\\CLIP}} & \textbf{Nomic} & \textbf{SigLIP} \\
\midrule
\rowcolor{gray!12}\multicolumn{8}{c}{\textit{STEM \& Life Sciences
}} \\
\midrule
\textbf{Acad} & 4.2 & 4.8 & 16.2 & \textbf{27.6} & 22.3 & \underline{22.6} & 3.6 \\
\textbf{Bio} & 5.7 & 14.8 & \underline{22.9} & 15.2 & 20.5 & \textbf{26.9} & 11.9 \\
\textbf{Chem} & 10.8 & 9.6 & 27.2 & 21.9 & \textbf{30.6} & \underline{30.6} & 11.6 \\
\textbf{Phys} & 6.8 & 6.1 & 13.3 & 14.0 & \underline{14.4} & \textbf{17.2} & 7.3 \\
\textbf{Math} & 13.1 & 17.9 & 16.4 & 9.3 & \underline{27.0} & \textbf{34.0} & 15.3 \\
\textbf{Earth} & 10.1 & 10.9 & 20.5 & \underline{26.2} & 24.6 & \textbf{30.1} & 11.8 \\
\textbf{BioAc} & 13.3 & 11.4 & 10.5 & 13.4 & \underline{19.4} & \textbf{23.4} & 14.8 \\
\textbf{BioInf} & 11.6 & 9.4 & 21.1 & 19.2 & \underline{23.7} & \textbf{33.8} & 16.8 \\
\textbf{Med} & 12.6 & 9.8 & 22.7 & 19.0 & \underline{26.8} & \textbf{33.9} & 9.1 \\
\midrule
\rowcolor{gray!12}\multicolumn{8}{c}{\textit{Software \& Technical Systems
}} \\
\midrule
\textbf{Apple} & 7.2 & 12.3 & 23.9 & 17.0 & \underline{24.3} & \textbf{28.7} & 4.4 \\

\textbf{Ubuntu} & 11.6 & 5.5 & 25.9 & \underline{34.2} & 26.1 & \textbf{34.3} & 12.6 \\
\textbf{BTC} & 8.9 & 8.3 & 18.2 & 19.6 & \underline{22.6} & \textbf{22.7} & 10.0 \\
\textbf{Crypto} & 11.3 & 14.8 & 9.8 & 7.1 & \underline{15.5} & \textbf{22.4} & 10.2 \\
\textbf{QC} & 4.5 & 2.6 & 5.9 & 5.6 & \underline{10.8} & \textbf{12.1} & 2.6 \\
\textbf{Robot} & 16.1 & 10.6 & 15.8 & 18.7 & \underline{19.0} & \textbf{30.3} & 14.3 \\
\textbf{Sales} & 14.2 & 2.3 & 31.1 & \textbf{47.3} & \underline{32.3} & 26.2 & 6.5 \\
\midrule
\rowcolor{gray!12}\multicolumn{8}{c}{\textit{Social Sciences \& Humanities
}} \\
\midrule
\textbf{Econ} & 9.5 & 6.0 & 10.0 & 12.6 & \underline{13.5} & \textbf{21.1} & 9.8 \\
\textbf{Psych} & 6.4 & 8.7 & 15.6 & 18.6 & \underline{20.8} & \textbf{23.9} & 7.9 \\
\textbf{Phil} & 2.4 & 5.4 & 15.2 & 18.0 & \underline{19.4} & \textbf{21.7} & 7.0 \\
\textbf{Law} & 10.2 & 19.7 & 30.7 & 35.0 & \underline{35.3} & \textbf{47.6} & 16.4 \\
\textbf{Christ} & 8.9 & 15.0 & 20.0 & \underline{26.5} & 21.0 & \textbf{30.9} & 13.0 \\
\textbf{Islam} & 12.0 & 10.7 & 25.8 & \textbf{32.0} & 24.3 & \underline{28.9} & 6.5 \\
\midrule
\rowcolor{gray!12}\multicolumn{8}{c}{\textit{Applied Domains
}} \\
\midrule
\textbf{Aviat} & 9.6 & 15.4 & 16.2 & 17.0 & \textbf{24.3} & \underline{24.1} & 9.2 \\
\textbf{Game} & 17.5 & 19.1 & 41.6 & \underline{43.9} & \textbf{45.6} & 43.1 & 21.4 \\
\textbf{GIS} & 13.8 & 13.1 & 15.5 & 15.6 & \underline{20.3} & \textbf{25.8} & 16.5 \\
\textbf{PM} & 8.6 & 8.9 & 21.9 & \textbf{33.2} & 20.5 & \underline{27.6} & 12.4 \\
\textbf{Sustain} & 10.1 & 9.0 & 16.7 & \textbf{25.6} & 24.3 & \underline{24.7} & 11.5 \\
\textbf{Travel} & 10.1 & 16.1 & 23.9 & \underline{30.8} & 26.6 & \textbf{36.7} & 13.1 \\
\textbf{Quant} & 8.1 & 2.1 & 12.4 & \underline{15.3} & 11.6 & \textbf{16.2} & 5.8 \\
\midrule
\textbf{Avg.} & 10.0 & 10.4 & 19.5 & 22.0 & \underline{23.0} & \textbf{27.6} & 10.8 \\
\bottomrule
\end{tabular}}
\captionof{table}{
\textbf{Task 2: Multimodal query to text retrieval (Query+Image $\to$ Documents).}
nDCG@10 scores for 7 multimodal retrieval models across 29 domains.
}
\label{tab:task2_results}
\end{minipage}

\end{table*}

\begin{table}[t]
\centering
\large
\resizebox{0.4\textwidth}{0.18\textheight}{%
\begin{tabular}{lccccccc}
\toprule
\textbf{Domain} & \textbf{BGE-VL} & \textbf{CLIP} & \textbf{GME-2B} & \textbf{GME-7B} & \textbf{\makecell{Jina\\CLIP}} & \textbf{Nomic} & \textbf{SigLIP} \\
\midrule
\rowcolor{gray!12}\multicolumn{8}{c}{\textit{STEM \& Life Sciences
}} \\
\midrule
\textbf{Acad} & 41.7 & 38.1 & \textbf{59.2} & 38.3 & 42.7 & 43.0 & \underline{45.6} \\
\textbf{Bio} & 42.1 & 48.6 & \textbf{58.6} & 53.3 & 38.5 & 38.7 & \underline{56.4} \\
\textbf{Chem} & 18.2 & 15.6 & \textbf{40.1} & 30.7 & 11.9 & 14.8 & \underline{34.6} \\
\textbf{Phys} & 29.6 & 27.8 & \underline{35.6} & 29.6 & 24.3 & 24.6 & \textbf{38.6} \\
\textbf{Math} & 28.2 & 33.1 & \textbf{48.8} & 35.5 & 29.1 & 32.5 & \underline{48.0} \\
\textbf{Earth} & 33.2 & 40.0 & \underline{44.5} & 38.4 & 32.5 & 27.5 & \textbf{45.9} \\
\textbf{BioAc} & 22.5 & \underline{46.1} & 37.3 & 28.2 & 41.9 & 40.6 & \textbf{49.0} \\
\textbf{BioInf} & 28.0 & 14.2 & \textbf{51.1} & \underline{36.9} & 13.6 & 10.9 & 32.7 \\
\textbf{Med} & 55.0 & 50.2 & \textbf{66.7} & \underline{63.8} & 41.9 & 39.7 & 63.6 \\
\midrule
\rowcolor{gray!12}\multicolumn{8}{c}{\textit{Software \& Technical Systems
}} \\
\midrule
\textbf{Apple} & 37.1 & 28.2 & \textbf{67.4} & 42.3 & 28.6 & 25.6 & \underline{60.1} \\

\textbf{Ubuntu} & 29.1 & 37.3 & \underline{58.8} & \textbf{60.8} & 34.5 & 26.6 & 52.3 \\
\textbf{BTC} & 15.0 & 19.1 & \textbf{32.3} & 15.1 & 20.0 & 15.1 & \underline{30.6} \\
\textbf{Crypto} & 17.1 & 17.5 & \textbf{28.9} & 15.1 & 13.0 & 8.6 & \underline{26.8} \\
\textbf{QC} & 6.8 & 5.0 & \underline{9.7} & 5.1 & 6.7 & 7.9 & \textbf{13.9} \\
\textbf{Robot} & 21.3 & 15.6 & \textbf{29.6} & 19.5 & 14.0 & 14.0 & \underline{28.5} \\
\textbf{Sales} & \underline{59.6} & 47.7 & 58.6 & 55.9 & 39.2 & 30.9 & \textbf{72.5} \\
\midrule
\rowcolor{gray!12}\multicolumn{8}{c}{\textit{Social Sciences \& Humanities
}} \\
\midrule
\textbf{Econ} & 39.0 & 39.3 & \underline{44.7} & 36.6 & 32.5 & 30.1 & \textbf{52.1} \\
\textbf{Psych} & 30.0 & 37.7 & \textbf{47.0} & 30.3 & 35.4 & 28.9 & \underline{44.9} \\
\textbf{Phil} & 21.0 & 14.6 & \textbf{27.3} & \underline{24.1} & 13.8 & 19.4 & 24.1 \\
\textbf{Law} & 61.7 & 67.7 & \underline{70.2} & 54.1 & 45.6 & 49.2 & \textbf{76.1} \\
\textbf{Christ} & 32.5 & \underline{39.8} & 34.3 & 38.6 & 34.8 & 29.1 & \textbf{40.5} \\
\textbf{Islam} & 22.7 & 29.5 & \textbf{41.0} & 30.4 & 31.4 & 21.8 & \underline{37.1} \\
\midrule
\rowcolor{gray!12}\multicolumn{8}{c}{\textit{Applied Domains
}} \\
\midrule
\textbf{Aviat} & 29.7 & \underline{35.0} & 33.8 & 29.3 & 23.7 & 26.6 & \textbf{41.9} \\
\textbf{Game} & 35.1 & 50.9 & \underline{58.3} & \textbf{73.8} & 48.9 & 40.9 & 53.2 \\
\textbf{GIS} & 28.0 & 39.4 & \underline{43.0} & 32.0 & 33.4 & 30.3 & \textbf{44.9} \\
\textbf{PM} & 21.7 & 26.1 & \textbf{46.8} & 33.7 & 30.6 & 24.8 & \underline{45.0} \\
\textbf{Sustain} & 35.7 & 35.2 & \underline{48.5} & 39.6 & 40.1 & 31.0 & \textbf{55.1} \\
\textbf{Travel} & 51.0 & 50.5 & \underline{66.1} & 59.3 & 52.9 & 34.4 & \textbf{68.6} \\
\textbf{Quant} & 24.2 & 23.8 & \textbf{33.4} & 21.9 & 26.7 & 18.3 & \underline{30.3} \\
\midrule
\textbf{Avg.} & 31.6 & 33.6 & \textbf{45.6} & 37.0 & 30.4 & 27.1 & \underline{45.3} \\
\bottomrule
\end{tabular}}
\caption{
\textbf{Task 3: Multimodal query to image retrieval (Query+Image $\to$ Images).} 
nDCG@10 scores for 7 multimodal retrieval models across 29 domains. 
}
\label{tab:task3_results}
\end{table}

\begin{table}[t]
\centering
\large
\resizebox{0.4\textwidth}{0.18\textheight}{%
\begin{tabular}{lccccc}
\toprule
\textbf{Domain} & \textbf{BGE-VL} & \textbf{CLIP} & \textbf{GME-2B} & \textbf{GME-7B} & \textbf{SigLIP} \\
\midrule
\rowcolor{gray!12}\multicolumn{6}{c}{\textit{STEM \& Life Sciences
}} \\
\midrule
\textbf{Acad} & 4.0 & 18.3 & \textbf{22.7} & \underline{21.4} & 12.0 \\
\textbf{Bio} & 3.2 & 9.2 & \underline{10.9} & 5.7 & \textbf{16.0} \\
\textbf{Chem} & 9.7 & 17.6 & \textbf{31.9} & 20.7 & \underline{25.2} \\
\textbf{Phys} & 9.0 & \textbf{24.7} & 17.6 & 13.3 & \underline{23.6} \\
\textbf{Math} & 17.6 & \underline{38.9} & 19.0 & 12.0 & \textbf{43.4} \\
\textbf{Earth} & 8.2 & \textbf{39.5} & 22.3 & 19.0 & \underline{32.0} \\
\textbf{BioAc} & 17.2 & \textbf{46.1} & 23.0 & 15.4 & \underline{34.7} \\
\textbf{BioInf} & 17.4 & 22.2 & \underline{23.8} & 14.6 & \textbf{29.4} \\
\textbf{Med} & 11.3 & \textbf{38.1} & 25.6 & 19.5 & \underline{31.8} \\
\midrule
\rowcolor{gray!12}\multicolumn{6}{c}{\textit{Software \& Technical Systems
}} \\
\midrule
\textbf{Apple} & 9.7 & \textbf{38.2} & \underline{33.6} & 21.8 & 15.7 \\

\textbf{Ubuntu} & 13.4 & \underline{32.3} & 26.3 & 28.2 & \textbf{33.8} \\
\textbf{BTC} & 10.0 & 15.4 & \textbf{23.1} & 18.4 & \underline{19.9} \\
\textbf{Crypto} & \underline{12.6} & \textbf{19.5} & 10.6 & 6.2 & 11.8 \\
\textbf{QC} & 4.0 & \underline{7.9} & 5.7 & 3.8 & \textbf{13.1} \\
\textbf{Robot} & 14.4 & 15.1 & \underline{25.5} & 17.5 & \textbf{31.6} \\
\textbf{Sales} & 12.4 & 25.7 & \textbf{45.6} & \underline{42.3} & 23.0 \\
\midrule
\rowcolor{gray!12}\multicolumn{6}{c}{\textit{Social Sciences \& Humanities
}} \\
\midrule
\textbf{Econ} & 5.4 & \underline{31.3} & 11.0 & 6.6 & \textbf{31.7} \\
\textbf{Psych} & 7.3 & \textbf{33.7} & \underline{21.8} & 13.9 & 18.7 \\
\textbf{Phil} & 3.5 & 12.4 & \textbf{19.5} & 13.9 & \underline{14.3} \\
\textbf{Law} & 5.6 & \textbf{29.4} & 24.1 & \underline{27.5} & 16.0 \\
\textbf{Christ} & 8.5 & \textbf{29.8} & 21.4 & \underline{22.5} & 17.1 \\
\textbf{Islam} & 13.5 & 20.9 & \underline{31.1} & \textbf{32.1} & 17.5 \\
\midrule
\rowcolor{gray!12}\multicolumn{6}{c}{\textit{Applied Domains
}} \\
\midrule
\textbf{Aviat} & 10.8 & \textbf{38.9} & 26.3 & 16.8 & \underline{36.9} \\
\textbf{Game} & 10.0 & \textbf{54.0} & 44.5 & 29.6 & \underline{51.1} \\
\textbf{GIS} & 10.5 & \underline{38.4} & 19.8 & 15.8 & \textbf{42.0} \\
\textbf{PM} & 9.8 & \underline{21.5} & 18.6 & 20.8 & \textbf{27.6} \\
\textbf{Sustain} & 11.2 & \textbf{38.1} & \underline{31.2} & 27.8 & 30.8 \\
\textbf{Travel} & 11.7 & \textbf{36.7} & 32.2 & 29.8 & \underline{34.0} \\
\textbf{Quant} & 7.4 & \textbf{20.6} & \underline{17.1} & 9.7 & 12.1 \\
\midrule
\textbf{Avg.} & 10.0 & \textbf{28.1} & 23.6 & 18.8 & \underline{25.8} \\
\bottomrule
\end{tabular}}
\caption{
\textbf{Task 4: Multimodal document retrieval (Query+Image $\to$ Documents+Images).} 
nDCG@10 scores for 5 multimodal retrieval models.
}
\label{tab:task4_results}
\end{table}

\section{Experiments}
\label{sec:experiments}

\subsection{Experimental Setup}

We evaluate 18 representative retrieval models across diverse architectures, including top performers from recent benchmarks~\citep{abdallah2025rerankarena,abdallah2025rankify,muennighoff2023mteb,thakur2021beir,su2024bright}.

\noindent\textbf{Text-only retrieval models (Task 1).} We evaluate BM25~\citep{robertson2009probabilistic} as our sparse baseline, dense retrievers trained on large-scale corpora: Contriever~\citep[110M;][]{izacard2021unsupervised}, E5-Mistral~\citep[7.1B;][]{wang2022text}, GritLM~\citep[7.1B;][]{muennighoff2024generative}, SFR-Embedding-Mistral~\citep[7.1B;][]{meng2024sfrembedding}, and gte-Qwen2.5~\citep[7.6B;][]{li2023towards}, as well as reasoning-enhanced retrievers: ReasonIR~\citep{shao2025reasonir}, DiVeR~\citep{diver}, and Rader~\citep{das2025rader}. We also include OpenAI's proprietary model~\citep{achiam2023gpt}.
\noindent\textbf{Multimodal retrieval models (Tasks 2--4).} We evaluate 7 multimodal retrievers: contrastive vision-language models CLIP~\citep{radford2021learning} and SigLIP~\citep{zhai2023sigmoid}, and multimodal embedding models BGE-VL~\citep{zhou2024megapairs}, Jina-CLIP~\citep{koukounas2024jina}, Nomic-Vision~\citep{nussbaum2024nomic}, and GME-Qwen2-VL~\citep[2B and 7B;][]{zhang2024gme}. All evaluated multimodal retrievers, including the GME models, encode a query using only its first image; multiple query images are not combined. Since multi-image queries carry 2.01 images on average and 47\% of image-bearing queries have more than one image, the reported scores understate what a retriever able to consume every query image could achieve.

\noindent\textbf{Evaluation metrics.} Following prior work~\citep{thakur2021beir,nguyen2016ms,su2024bright}, we use nDCG@10 as the primary metric. Tasks 1--3 use binary relevance labels following BEIR~\citep{thakur2021beir}. Task 4 uses graded relevance: rel=2 for a gold passage paired with its corresponding positive image, rel=1 for a gold passage paired with no image, and rel=0 otherwise, including a gold passage paired with an image that is present but not annotated positive. Pairs formed from an image annotated negative are excluded from the ranking rather than scored rel=0, and where an image is annotated both positive and negative for a query the positive label takes precedence. 

\subsection{Main Results}

\paragraph{Reasoning-intensive multimodal retrieval poses substantial challenges for all current models.}
Tables~\ref{tab:task1_results}--\ref{tab:task4_results} show that \textsc{MM-BRIGHT} is difficult for all retrieval models. In Task 1 (Query $\to$ Documents), we use text-only queries without images to measure reasoning difficulty before adding visual elements. BM25 achieves only 8.5 nDCG@10, showing that keyword matching fails on reasoning-intensive queries. Dense retrievers perform better (E5: 25.3, SFR: 26.9), but reasoning-enhanced models achieve the best results: DiVeR reaches 32.2 nDCG@10 and ReasonIR reaches 28.6. However, these scores are much lower than the 50+ nDCG@10 typically seen on BEIR~\citep{thakur2021beir}, indicating that \textsc{MM-BRIGHT} requires different capabilities than standard retrieval benchmarks. The proprietary OpenAI model (28.8) performs similarly to open-source reasoning models, suggesting that model size alone does not solve this task.

\paragraph{Adding images hurts retrieval performance instead of helping.}
Comparing Tasks 1 and 2 reveals an unexpected finding: multimodal models perform worse when given images. The best multimodal model on Task 2 (Query+Image $\to$ Documents) is Nomic-Vision with 27.6 nDCG@10, which is lower than the best text-only model on Task 1 (DiVeR: 32.2). This happens even though multimodal models have access to additional visual information from query images. BGE-VL performs particularly poorly (10.0), matching BM25 despite being a vision-language model. Jina-CLIP (23.0) and GME-7B (22.0) achieve moderate scores, but no multimodal model beats the text-only baseline. We believe this is because current multimodal models are trained mainly on simple image-text matching rather than visual reasoning. Understanding technical diagrams, scientific visualizations, or error screenshots requires deeper reasoning than matching objects to text descriptions.

\paragraph{Performance across visual and joint tasks.} 
Results vary significantly based on the retrieval objective. In Task 3 (Query+Image $\to$ Images), 
models perform best (GME-2B: 45.6 nDCG@10), leveraging visual similarity to match query 
images to the corpus. Conversely, the more complex Task 4 (Query+Image $\to$ Documents+Images) 
proves much harder (CLIP: 28.1). Task 4 uses a distinct graded relevance 
scale ($rel=2$ for complete pairs), hence the results are not directly comparable to the binary 
Task 3. Nevertheless, they reveal a specific failure in modality alignment. Even when models 
identify relevant information, they rarely retrieve the complete text-image pair, 
typically finding one modality but failing to align both into a unified evidence block.

\paragraph{Different models have different strengths, but all are inconsistent.}
Each multimodal model excels at different tasks. Nomic-Vision performs best on Task 2 (27.6) but poorly on Task 3 (27.1). GME-2B shows the opposite pattern: excellent on Task 3 (45.6) but mediocre on Task 2 (19.5). CLIP achieves balanced but never top performance across all tasks (Task 2: 10.4, Task 3: 33.6, Task 4: 28.1). These inconsistencies indicate that current models have narrow specializations rather than general multimodal reasoning ability. BGE-VL performs poorly on both Tasks 2 and 4 (10.0 on both), showing that complex architecture does not guarantee good performance.

\section{Additional Analysis}

\begin{figure}[ht!]
  \centering
  \includegraphics[width=.5\textwidth]{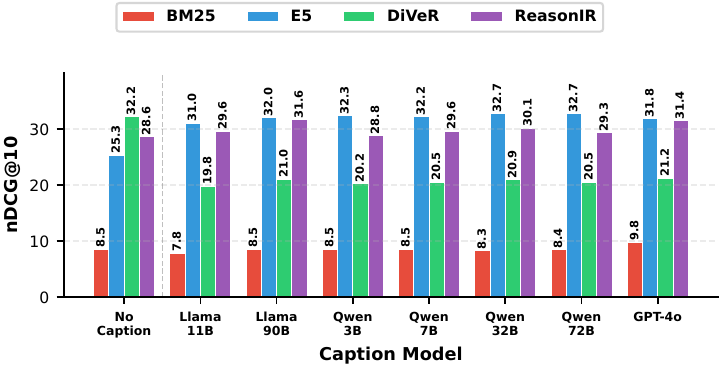}
  \caption{
  \textbf{Impact of image captioning on retrieval performance across different retriever types.} 
  We augment text-only queries with image captions generated by various vision-language models and measure nDCG@10 on Task 1. 
  }
  \label{fig:caption_impact}
\end{figure}

\paragraph{Image captions reveal fundamental differences in retrieval paradigms.}
\label{sec:caption_experiments}
To understand whether visual information helps reasoning-intensive retrieval, we augment text queries with image captions generated by various vision-language models (Llama-3.2-11B/90B, Qwen-2.5-3B/7B/32B/72B, GPT-4o). Figure~\ref{fig:caption_impact} shows a striking divergence: while semantic dense retrievers like E5 improve dramatically with captions (+7.4 nDCG@10, from 25.3 to 32.7), reasoning-enhanced models like DiVeR suffer severe performance degradation (-12.0 points, from 32.2 to 20.2). This suggests that image captions, while providing useful semantic information, introduce noise that disrupts reasoning-based retrieval strategies. BM25 shows modest improvements with better caption quality (8.5 → 9.8 with GPT-4o), benefiting from expanded lexical coverage. ReasonIR maintains stable performance across caption models (29-31 nDCG@10), indicating some robustness to caption variations. These patterns suggest that current caption-based approaches cannot replace true multimodal reasoning, and that reasoning-enhanced retrievers may require different strategies for incorporating visual information. Complete results across all domains are provided in Appendix~\ref{app:caption_results}.

\begin{figure}[t]
  \centering
  \includegraphics[width=0.45\textwidth]{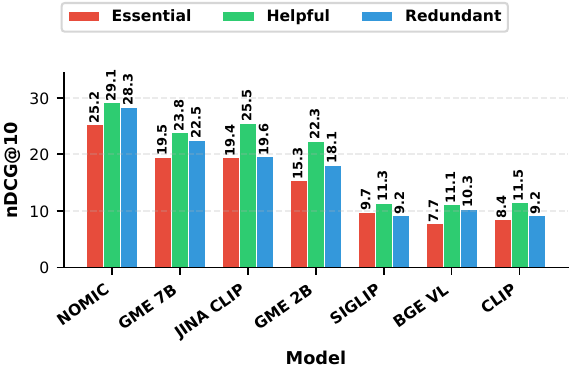}
  \caption{
  \textbf{Multimodal retrieval performance by image essentiality.} 
  All models achieve their best performance on queries with helpful images (green bars) but perform worse when images are essential (red bars) for understanding the query. 
  }
  \label{fig:image_essentiality_performance}
\end{figure}

\begin{figure}[t]
  \centering
  \includegraphics[width=\columnwidth]{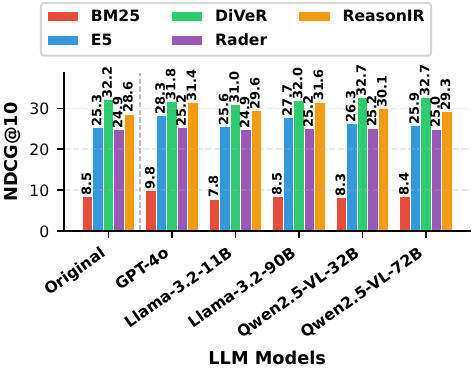}
  \caption{
  \textbf{Impact of query reformulation on retrieval performance across different LLM reformulators.} 
  We compare original queries against reformulated queries generated by six vision-language models. 
  }
  \label{fig:reformulation_bars}
\end{figure}
\paragraph{Multimodal models fail when visual information is most critical.}
To understand why adding images degrades retrieval in Task 2, we analyze performance by image essentiality. Figure~\ref{fig:image_essentiality_performance} shows a consistent trend: all multimodal models perform worst when images are essential for understanding the query. Performance is highest for queries with helpful images (20.4 nDCG@10 on average), but drops sharply for essential images (15.0), even below redundant images (16.7). This inverted pattern indicates that current multimodal retrievers struggle to identify and use critical visual evidence, relying instead on surface-level image-text associations. For example, Nomic-Vision shows the largest gap, scoring 29.1 on helpful images but only 25.2 on essential images. This helps explain why Task 2 underperforms Task 1, especially for the 33.0\% of queries where visual information is indispensable (Figure~\ref{fig:essentiality_distribution}).

\begin{table}[t]
\resizebox{0.8\linewidth}{!}{
\begin{tabular}{l c}
\toprule
\textbf{Retrieval setting} & \textbf{Average score} \\
\midrule
None (no retrieval) & 61.85 \\
Oracle (ground-truth positives) & \textbf{67.17} \\
GME-Qwen2-VL-7B & 60.71 \\
Nomic-Vision & 59.90 \\
BGE-VL-Large & 58.65 \\
\bottomrule
\end{tabular}}
\centering
\caption{End-to-end QA with RAG: Llama-3.2-90B as generator, GPT-4 as judge (0--100).}
\label{tab:rag_qa_results}
\end{table}

\paragraph{Query reformulation provides limited gains for reasoning-intensive retrieval.}
We test whether vision-language models can improve retrieval by reformulating queries with explicit reasoning before retrieval, using GPT-4o (see the Prompt in Figure~\ref{fig:reformulation_prompt}), Llama-3.2 (11B and 90B), and Qwen2.5-VL (3B, 7B, 32B, 72B). As shown in Figure~\ref{fig:reformulation_bars}, reformulation provides modest gains for semantic retrievers such as E5 (25.3 to 28.3 with GPT-4o) and consistent improvements for Rader (+2.8 with GPT-4o), but offers little benefit for the strongest reasoning-focused retriever DiVeR (32.2 to 31.8 with GPT-4o). BM25 improves slightly with high-quality reformulations (8.5 to 9.8 with GPT-4o) but can degrade with smaller models, suggesting sensitivity to reformulation quality. Larger reformulators do not consistently outperform smaller ones, indicating that query reformulation is not a reliable solution for reasoning-intensive multimodal retrieval; full results are in Appendix~\ref{sec:reformulation_appendix}.
\paragraph{Retrieval augmentation improves end-to-end QA, but a large oracle gap remains.}
To assess whether retrieval translates into better downstream question answering, we run an end-to-end RAG pipeline: Llama-3.2-90B-VL generates answers using either no retrieved evidence (\textit{None}), oracle positives (\textit{Oracle}), or top-$5$ documents retrieved by different multimodal retrievers. Following ~\citet{su2024bright}, we evaluate answer correctness with a GPT-4 judge that scores the generated answer against the reference answer on a 0--100 scale (Appendix~\ref{app:rag_judge_prompt}). As shown in Table~\ref{tab:rag_qa_results}, oracle evidence yields the best performance (67.17), while retrieval-based settings are lower, with the best retriever still trailing oracle by 6.46 points. 


\section{Conclusion}
We introduced \textsc{MM-BRIGHT}, a multimodal benchmark for reasoning-intensive retrieval across 29 technical domains, covering four tasks that range from text-only retrieval to multimodal document retrieval. Our evaluation of 18 models shows that current retrievers struggle across all settings, and that adding images often degrades performance when visual information is critical. These results highlight substantial headroom and suggest that future retrieval systems must better integrate visual understanding with multi-step technical reasoning.

\bibliography{custom}

\makeatletter
\newcommand{\apxline}[6]{%
  \@dottedtocline{#1}{#2}{#3}{\numberline{#4}#5}{#6}%
}
\makeatother

\clearpage
\appendix

\section*{Appendix Contents}
\vspace{-0.6em}

\begingroup
\setlength{\parindent}{0pt}
\setlength{\parskip}{0pt}
\noindent
\begin{minipage}[t]{0.90\textwidth}
\apxline{0}{0em}{6.0em}{\textbf{Appendix A}}{Dataset Construction and Annotation Protocol}{\pageref{app:appendixA}}
\apxline{1}{1.6em}{3.2em}{A.1}{Dataset Task 2 Statistics}{\pageref{app:dataset2_stats}}
\apxline{1}{1.6em}{3.2em}{A.2}{Query Selection and Filtering}{\pageref{app:annotation_guidelines}}
\apxline{2}{3.6em}{4.0em}{A.2.1}{Post Selection Criteria}{\pageref{app:post_selection}}
\apxline{2}{3.6em}{4.0em}{A.2.2}{Constructing Queries}{\pageref{app:constructing_queries}}
\apxline{1}{1.6em}{3.2em}{A.3}{Positive Document Construction}{\pageref{app:positive_docs}}
\apxline{1}{1.6em}{3.2em}{A.4}{Hard Negative Mining}{\pageref{app:hard_negative}}
\apxline{1}{1.6em}{3.2em}{A.5}{Quality Control and Review}{\pageref{app:qc_review}}
\apxline{1}{1.6em}{3.2em}{A.6}{Domain-Specific Annotation Notes}{\pageref{app:domain_notes}}
\apxline{1}{1.6em}{3.2em}{A.7}{Prompt Templates}{\pageref{app:prompt_templates}}
\apxline{2}{3.6em}{4.0em}{A.7.1}{Hard Negative Mining Prompt}{\pageref{app:negative_mining_prompt}}
\apxline{2}{3.6em}{4.0em}{A.7.2}{Image Relevance Annotation Prompt}{\pageref{app:image_annotation_prompt}}

\vspace{0.35em}
\apxline{0}{0em}{6.0em}{\textbf{Appendix B}}{Caption-based Query Augmentation Results (Task 1)}{\pageref{app:caption_results}}

\vspace{0.35em}
\apxline{0}{0em}{6.0em}{\textbf{Appendix C}}{Image Type Breakdown by Domain}{\pageref{app:image_types}}
\apxline{1}{1.6em}{3.2em}{C.1}{Image Type Taxonomy and Classifier Setup}{\pageref{app:image_types_taxonomy}}
\apxline{1}{1.6em}{3.2em}{C.2}{Per-domain Distributions (Figures and Prompt)}{\pageref{sec:image_types}}

\vspace{0.35em}
\apxline{0}{0em}{6.0em}{\textbf{Appendix D}}{Image Essentiality Analysis}{\pageref{app:essentiality_full}}
\apxline{1}{1.6em}{3.2em}{D.1}{Essentiality Labeling Method}{\pageref{app:essentiality_method}}
\apxline{1}{1.6em}{3.2em}{D.2}{Prompt Template}{\pageref{app:essentiality_prompt}}
\apxline{1}{1.6em}{3.2em}{D.3}{Domain-Level Analysis}{\pageref{app:essentiality_domain}}
\apxline{1}{1.6em}{3.2em}{D.4}{Implications}{\pageref{app:essentiality_implications}}
\apxline{1}{1.6em}{3.2em}{D.5}{Per-model Results by Essentiality}{\pageref{app:essentiality_details}}

\vspace{0.35em}
\apxline{0}{0em}{6.0em}{\textbf{Appendix E}}{Query Reformulation Detailed Results}{\pageref{sec:reformulation_appendix}}
\apxline{1}{1.6em}{3.2em}{E.1}{Reformulation Prompt}{\pageref{app:reformulation_prompt}}
\apxline{1}{1.6em}{3.2em}{E.2}{Per-domain Results (All Reformulation Models)}{\pageref{app:reformulation_results}}

\vspace{0.35em}
\apxline{0}{0em}{6.0em}{\textbf{Appendix F}}{LLM-based Dataset Quality Assessment}{\pageref{app:llm_quality_full}}

\vspace{0.35em}
\apxline{0}{0em}{6.0em}{\textbf{Appendix G}}{RAG answer evaluation prompt (GPT-4 judge)}{\pageref{app:rag_judge_prompt}}

\vspace{0.35em}
\apxline{0}{0em}{6.0em}{\textbf{Appendix H}}{Dataset Examples}{\pageref{app:dataset_examples}}
\end{minipage}

\endgroup

\clearpage

\section{Dataset Construction and Annotation Protocol}
\label{app:appendixA}

\subsection{Dataset Task  2 Statistics}
\label{app:dataset2_stats}

Table~\ref{tab:dataset2_stats} presents detailed statistics for Dataset 2, which supports the multimodal retrieval tasks: Task 2 (Query+Image $\to$ Documents), Task 3 (Query+Image $\to$ Images), and Task 4 (Query+Image $\to$ Documents+Images). This dataset comprises 1,218 queries with 7,621 annotated positive images across 29 domains. Compared to Dataset 1, this subset contains only queries where image annotations were completed for multimodal evaluation.

\begin{table}[h!]
\caption{
\textbf{Data statistics for Dataset 2 (Multimodal Retrieval Tasks).}
This dataset supports Tasks 2--4: (2) Query+Image $\to$ Documents, (3) Query+Image $\to$ Images, and (4) Query+Image $\to$ Documents+Images.
For each domain, we report the number of queries ($\mathbf{Q}$), corpus size ($\boldsymbol{\mathcal{D}}$),
average positive documents per query ($\boldsymbol{\mathcal{D}^+}$),
average images per query ($\boldsymbol{\mathcal{I}^+}$),
average token length for queries and documents (GPT-2 tokenizer),
and total annotated positive images (\textbf{Total Img.}).
}
\label{tab:dataset2_stats}
\centering
\small
\resizebox{0.4\textwidth}{!}{
\begin{tabular}{l|rrr r|rr|r}
\toprule
& \multicolumn{4}{c|}{\textbf{Total Number}} & \multicolumn{2}{c|}{\textbf{Avg. Length}} & \textbf{Total} \\
\cmidrule{2-8}
\textbf{Dataset} &
\multicolumn{1}{c}{$\mathbf{Q}$} &
\multicolumn{1}{c}{$\boldsymbol{\mathcal{D}}$} &
\multicolumn{1}{c}{$\boldsymbol{\mathcal{D}^+}$} &
\multicolumn{1}{c|}{$\boldsymbol{\mathcal{I}^+}$} &
\multicolumn{1}{c}{$\mathbf{Q}$} &
\multicolumn{1}{c|}{$\boldsymbol{\mathcal{D}}$} &
\multicolumn{1}{c}{\textbf{Img.}} \\
\midrule

\rowcolor{gray!12}\multicolumn{8}{c}{\textit{\textbf{STEM \& Life Sciences}}} \\
\midrule
Academia & 22 & 60{,}050  & 3.7 & 1.27 & 341.6 & 365.1 & 94 \\
Bioacoustics & 39 & 29{,}812 & 2.8 & 2.21 & 464.1 & 294.5 & 260 \\
Bioinformatics & 46 & 45{,}545 & 1.4 & 1.57 & 454.2 & 466.0 & 166 \\
Biology & 64 & 89{,}435 & 1.2 & 2.94 & 374.6 & 637.4 & 425 \\
Chemistry & 51 & 36{,}043 & 2.6 & 2.43 & 421.5 & 953.2 & 565 \\
Earthscience & 72 & 73{,}451 & 2.9 & 2.18 & 367.3 & 665.3 & 675 \\
Math & 39 & 151{,}867 & 1.9 & 2.74 & 699.4 & 950.2 & 315 \\
Medicalsciences & 43 & 240{,}844 & 2.4 & 1.79 & 376.3 & 479.6 & 149 \\
Physics & 81 & 338{,}291 & 2.5 & 2.44 & 414.8 & 1{,}032.7 & 594 \\
\midrule

\rowcolor{gray!12}\multicolumn{8}{c}{\textit{\textbf{Software \& Technical Systems}}} \\
\midrule
Apple & 13 & 29{,}285 & 2.3 & 2.23 & 414.6 & 457.7 & 99 \\
Askubuntu & 28 & 90{,}198 & 1.5 & 2.18 & 326.9 & 546.9 & 122 \\
Bitcoin & 54 & 29{,}595 & 2.9 & 1.46 & 292.0 & 414.9 & 198 \\
Crypto & 51 & 24{,}054 & 1.3 & 1.47 & 585.7 & 1{,}031.5 & 142 \\
Gis & 34 & 20{,}705 & 1.4 & 2.76 & 312.1 & 625.9 & 202 \\
Quantumcomputing & 65 & 127{,}009 & 2.6 & 1.74 & 463.8 & 762.2 & 434 \\
Robotics & 24 & 11{,}185 & 3.3 & 2.62 & 672.1 & 700.2 & 450 \\
Salesforce & 8 & 8{,}890 & 2.0 & 2.62 & 452.2 & 497.1 & 32 \\
\midrule

\rowcolor{gray!12}\multicolumn{8}{c}{\textit{\textbf{Social Sciences \& Humanities}}} \\
\midrule
Christianity & 22 & 37{,}875 & 2.3 & 1.50 & 291.3 & 1{,}001.2 & 124 \\
Economics & 28 & 18{,}431 & 2.1 & 1.89 & 360.1 & 601.9 & 93 \\
Islam & 24 & 14{,}079 & 4.5 & 1.38 & 368.1 & 533.9 & 158 \\
Law & 24 & 26{,}142 & 2.8 & 1.25 & 513.8 & 755.2 & 33 \\
Philosophy & 38 & 137{,}860 & 2.8 & 1.50 & 507.7 & 814.0 & 124 \\
Psychology & 46 & 328{,}520 & 3.2 & 1.50 & 427.1 & 674.2 & 251 \\
\midrule

\rowcolor{gray!12}\multicolumn{8}{c}{\textit{\textbf{Applied Domains}}} \\
\midrule
Aviation & 111 & 203{,}938 & 2.3 & 2.36 & 294.6 & 709.2 & 863 \\
Gaming & 19 & 68{,}321 & 1.5 & 1.95 & 247.6 & 418.8 & 107 \\
Pm & 43 & 93{,}376 & 2.2 & 1.63 & 359.3 & 388.3 & 187 \\
Quant & 28 & 64{,}044 & 2.5 & 1.36 & 456.9 & 431.6 & 119 \\
Sustainability & 55 & 32{,}365 & 3.6 & 1.60 & 399.6 & 493.3 & 405 \\
Travel & 46 & 68{,}063 & 1.7 & 1.78 & 362.3 & 418.9 & 235 \\
\midrule

\textbf{Total} & \textbf{1{,}218} & \textbf{2{,}499{,}273} & -- & \textbf{1.99} & -- & -- & \textbf{7{,}621} \\
\bottomrule
\end{tabular}
}
\end{table}

\subsection{Query selection and filtering}
\label{app:annotation_guidelines}

This section provides detailed guidelines for annotators constructing the \textsc{MM-BRIGHT} dataset. The annotation process involves selecting StackExchange posts, identifying relevant documents, and mining hard negatives.

\subsubsection{Post Selection Criteria}
\label{app:post_selection}
Annotators browse Stack Exchange posts from newest to oldest within their assigned domain and select posts meeting ALL of the following criteria:

\textbf{Required Criteria:}
\begin{enumerate}
    \item \textbf{High-quality answer}: The post must have at least one answer that is either:
    \begin{itemize}
        \item Accepted by the question author (marked with green checkmark), OR
        \item Has received more than 10 upvotes
    \end{itemize}
    
    \item \textbf{Contains images}: The post must include at least one image that is \textit{integral} to understanding the question. Images should be:
    \begin{itemize}
        \item Technical diagrams, charts, screenshots, or scientific figures
        \item Essential for understanding the problem (not merely decorative)
        \item Clearly visible and not corrupted
    \end{itemize}

    \item \textbf{Technical complexity}: The question requires reasoning beyond simple keyword matching to answer. Avoid questions that can be answered by direct fact lookup.
\end{enumerate}

\textbf{Exclusion Criteria:}
\begin{itemize}
    \item Posts with only decorative, meme, or low-quality images
    \item Opinion-based or subjective questions without technical content
    \item Posts where all answers are speculative or lack authoritative sources
    \item Duplicate or near-duplicate questions
    \item Posts with images that violate copyright or contain inappropriate content
\end{itemize}

\subsubsection{Constructing Queries}
\label{app:constructing_queries}
For each selected post, construct the multimodal query as follows:

\textbf{Step 1: Extract text content}
\begin{itemize}
    \item Combine the post title and body text
    \item Preserve technical terminology, code snippets, and formatting where relevant
    \item Remove HTML artifacts and ensure readability
\end{itemize}

\textbf{Step 2: Extract images}
\begin{itemize}
    \item Include all images from the post body that are integral to the question
    \item Maintain image order as they appear in the post
    \item Ensure images are high quality and clearly visible
    \item Record image paths/URLs for reference
\end{itemize}

\subsection{Positive document construction}
\label{app:positive_docs}
Positive documents must provide critical information that helps \textit{reason through} the query, not just mention related keywords. Follow these steps:

\textbf{Step 1: Discover candidate documents}

Use TWO methods to find candidate documents:

\textbf{Method A - Answer links:}
\begin{itemize}
    \item Visit all external URLs linked in accepted or highly-voted answers
    \item Check if the linked page is still accessible (not 404 or paywalled)
\end{itemize}

\textbf{Method B - AI-assisted discovery:}
\begin{itemize}
    \item Use Gemini (Google AI) with the following prompt template:
    \item[] \textit{"Give me articles from the internet to answer this query: [paste full question text and describe images]"}
    \item Gemini will suggest relevant web pages - visit these suggestions
\end{itemize}

\textbf{Step 2: Evaluate relevance}

For each candidate web page, extract passages that meet the relevance criteria:

\textbf{A document/passage is POSITIVE if it:}
\begin{itemize}
    \item \textbf{Provides critical concepts or theories} that explain the phenomenon or problem described in the query
    \item \textbf{Contains technical documentation, code, or formulas} directly applicable to solving the problem
    \item \textbf{Explains underlying principles} that bridge the query to the solution (not just surface-level description)
    \item \textbf{Offers reasoning steps or logical connections} that help derive the answer
\end{itemize}

\textbf{A document/passage is NOT positive if it:}
\begin{itemize}
    \item Only shares keywords or topic with the query without providing reasoning support
    \item Provides tangential or background information that doesn't help answer the specific question
    \item Contains the answer directly stated without explanation (we want documents that help users \textit{reason} to the answer)
    \item Is primarily promotional, opinion-based, or lacks technical rigor
\end{itemize}

\textbf{Step 3: Extract passages}
\begin{itemize}
    \item For each positive web page, identify and extract relevant passages
    \item Each passage should be self-contained and coherent (typically 1-5 paragraphs)
    \item Include sufficient context for the passage to be understandable independently
    \item If an entire article is relevant, you may include the full text
    \item Extract and preserve any images from positive documents that help convey the information
\end{itemize}

\textbf{Step 4: Record metadata}
\begin{itemize}
    \item Source URL of the document
    \item Type of source (Wikipedia, blog, research article, documentation, news, etc.)
    \item Date accessed
    \item Brief justification for why this document is relevant
\end{itemize}

\subsection{Hard negative mining}
\label{app:hard_negative}
Hard negatives are documents that are topically related but do not satisfy the specific requirements of the query. These are crucial for preventing models from relying on simple semantic matching.

\textbf{Step 1: Generate hard negative search query}

\begin{itemize}
    \item Use the GPT-4o prompt (Appendix~\ref{app:negative_mining_prompt}) to generate:
    \begin{enumerate}
        \item A search query designed to find semantically similar but irrelevant content
        \item List of entities and events from the post
    \end{enumerate}
    \item The LLM will output JSON with \texttt{llm\_summary} and \texttt{entities\_events}
\end{itemize}

\textbf{Step 2: Collect hard negative URLs}

\begin{itemize}
    \item Use the generated \texttt{llm\_summary} as your Google search query
    \item Additionally search using combinations of \texttt{entities\_events}
    \item Collect exactly \textbf{20 URLs} per query that are:
    \begin{itemize}
        \item Topically related to the query domain
        \item Semantically similar to the query
        \item BUT do not provide the specific reasoning or concepts needed to answer the query
    \end{itemize}
\end{itemize}

\textbf{Step 3: Extract hard negative passages}

For each hard negative URL:
\begin{itemize}
    \item Extract 20 passages that are topically related but not helpful for answering the query
    \item Hard negatives should be challenging - they might discuss the same general topic but miss the specific technical details needed
    \item Avoid completely unrelated content (e.g., if query is about Python programming, don't use passages about cooking)
\end{itemize}

\subsection{Quality control and review}
\label{app:qc_review}
\textbf{Self-check before submission:}
\begin{enumerate}
    \item Does every selected post have at least one image integral to understanding the question?
    \item Did you verify that answers have $>10$ votes or are accepted?
    \item Can you clearly explain WHY each positive document helps reason through the query?
    \item Are hard negatives actually challenging (not obviously irrelevant)?
    \item Have you collected exactly 20 hard negative URLs per query?
    \item Are all images clearly visible and properly referenced?
\end{enumerate}

\textbf{Annotation review process:}
\begin{itemize}
    \item Initial annotations are reviewed by two PhD students/domain experts
    \item Reviewers check: (1) relevance of positive documents, (2) quality of hard negatives, (3) appropriateness of selected posts
    \item Only annotations with \textbf{unanimous approval} from all reviewers are retained
    \item If disagreement occurs, discuss with team and reach consensus or discard the example
\end{itemize}

\textbf{Common Mistakes to Avoid}

\begin{enumerate}
    \item \textbf{Selecting posts without integral images}: Images must be necessary for understanding the question, not decorative
    
    \item \textbf{Including answers as positive documents}: We want documents that help users reason, not documents that directly state the answer
    
    \item \textbf{Too-easy negatives}: Hard negatives should be semantically similar. Don't include completely off-topic documents
    
    \item \textbf{Insufficient justification}: Always document why a document is relevant - this helps maintain consistency
    
    \item \textbf{Ignoring source quality}: Prefer authoritative sources (official documentation, peer-reviewed articles, reputable blogs) over forums or unverified content
    
    \item \textbf{Extracting too-short passages}: Passages should have enough context to be understandable independently
\end{enumerate}

\subsection{Domain-specific annotation notes}
\label{app:domain_notes}
\textbf{For STEM domains (Biology, Chemistry, Physics, Math):}
\begin{itemize}
    \item Prioritize peer-reviewed sources and textbooks
    \item Include equations, diagrams, and technical figures when relevant
    \item Ensure positive documents explain the underlying scientific principles
\end{itemize}

\textbf{For Computing domains (Ubuntu, Programming, etc.):}
\begin{itemize}
    \item Include official documentation as positive sources
    \item Screenshots of code/errors are often integral images
    \item Hard negatives can be solutions to similar but distinct technical problems
\end{itemize}

\textbf{For Social Sciences (Economics, Psychology, Law, etc.):}
\begin{itemize}
    \item Prefer academic sources and authoritative analyses
    \item Charts and graphs in queries often require interpretation
    \item Hard negatives should discuss related concepts but miss the specific application
\end{itemize}

\textbf{For Applied domains (Aviation, Gaming, etc.):}
\begin{itemize}
    \item Balance between technical documentation and practical guides
    \item Visual elements often show specific scenarios or configurations
    \item Hard negatives can discuss the same domain but different specific cases
\end{itemize}

\subsection{Prompt templates}
\label{app:prompt_templates}
\subsubsection{Hard Negative Mining Prompt}
\label{app:negative_mining_prompt}

We use GPT-4o to analyze Stack Exchange posts and generate search queries designed to find challenging negative documents. The complete prompt is shown below:

\begin{figure*}[!t]
\centering
\begin{tcolorbox}[
  colback=gray!5,
  colframe=gray!75,
  title=GPT-4o Prompt for Hard Negative Mining,
  width=\textwidth,
]
\small\ttfamily
You are a document annotator specializing in hard negative mining for information retrieval.

Your task: Analyze the following Stack Exchange question and extract key information for finding challenging negative passages from web search.

\textbf{Question Title:} \{title\}

\textbf{Question Body:} \{clean\_body\}...

\textbf{Tags:} \{tags\}

Generate a structured analysis with:
\begin{enumerate}
  \item Give me a query for hard negative to use it to search on google
  \item All entities (people, places, organizations, concepts, technologies, etc.)
  \item All events, actions, or processes mentioned
\end{enumerate}

Output ONLY a valid JSON object with this exact structure:

\{
  "llm\_summary": "give me a query for hard negative to use it to search on google...",
  "entities\_events": ["entity1", "entity2", "event1", "event2", ...]
\}

Rules:
\begin{itemize}
  \item Summary must be EXACTLY 32 words (count carefully)
  \item List ALL relevant entities and events
  \item Include technical terms, concepts, and domain-specific vocabulary
  \item Extract named entities (people, places, companies, technologies)
  \item Include temporal events and processes
  \item Output ONLY valid JSON, no explanations
\end{itemize}
\end{tcolorbox}
\end{figure*}

The LLM generates a search query designed to retrieve topically similar but technically irrelevant content, along with entities and events that help construct effective negative search queries. Annotators use the generated query to collect 20 hard negative URLs per query from Google search.

\subsubsection{Image relevance annotation prompt}
\label{app:image_annotation_prompt}

For Tasks 3 and 4, we use GPT-4o to annotate whether scraped images from web pages are relevant to each query. The complete prompt is shown below:

\begin{figure*}[!t]
\centering
\begin{tcolorbox}[
  colback=gray!5,
  colframe=gray!75,
  title=GPT-4o Prompt for Image Annotation,
  width=\textwidth,
]
\small\ttfamily
You are an expert annotator for a multimodal information retrieval dataset. Your task is to determine whether an image is RELEVANT to a given query and its positive passages.

\textbf{Task Context:}
\begin{itemize}
  \item Query: The user's information need/question
  \item Positive Passages: Text passages that correctly answer the query
  \item Answer: The ground truth answer to the query
  \item Image: An image that may or may not be relevant
\end{itemize}

\textbf{Your Task:}

Analyze whether the image has a meaningful relationship to the query, positive passages, or answer. The image comes from some passage in the corpus (not necessarily a positive passage), and you need to determine if it's relevant.

\textbf{Classification Criteria:}

\textbf{POSITIVE (relevant)} if the image:
\begin{itemize}
  \item Directly illustrates concepts, examples, or scenarios mentioned in the query or positive passages
  \item Shows diagrams, charts, plots, or visualizations that support the answer
  \item Provides visual evidence or examples that help answer the query
  \item Contains figures, tables, or screenshots referenced or described in the positive passages
  \item Depicts technical concepts, code snippets, or UI elements discussed in the context
\end{itemize}

\textbf{NEGATIVE (irrelevant)} if the image:
\begin{itemize}
  \item Is completely unrelated to the query, answer, or positive passages
  \item Shows generic/decorative content not meaningful to the information need
  \item Contains advertisements, logos, or irrelevant UI elements
  \item Is corrupted, blank, or illegible
  \item Only tangentially relates but doesn't help answer the query
  \item Depicts different topics, concepts, or domains
\end{itemize}

\textbf{Query:} \{query\}

\textbf{Positive Passages (Ground Truth):} \{positive\_passages\}

\textbf{Answer:} \{answer\}

\textbf{Instructions:}
\begin{enumerate}
  \item Carefully examine the provided image
  \item Compare it against the query, positive passages, and answer
  \item Determine if it's POSITIVE (relevant) or NEGATIVE (irrelevant)
  \item Provide a clear, specific rationale explaining your decision
\end{enumerate}

\textbf{Output Format (JSON only, no additional text):}

\{
  "label": "positive" or "negative",
  "rationale": "Detailed explanation of why this image is relevant or irrelevant to the query and positive passages"
\}
\end{tcolorbox}
\end{figure*}

GPT-4o processes each query-image pair and outputs a JSON classification with rationale. This process yields 7,621 annotated images across 1,218 queries in Dataset 2.

\clearpage

\section{Caption-based Query Augmentation Results (Task 1)}
\label{app:caption_results}
In this appendix, we provide comprehensive results for all retrieval models on Task 1 (Query $\to$ Documents) when augmenting text queries with image captions generated by different vision-language models. We evaluate seven caption generation models of varying sizes and capabilities: Llama-3.2-11B, Llama-3.2-90B, Qwen-2.5-3B, Qwen-2.5-7B, Qwen-2.5-32B, Qwen-2.5-72B, and GPT-4o. For each query, we use the vision-language model to generate a detailed description of the image(s), then concatenate this caption with the original text query before retrieval. These experiments reveal how different retriever architectures respond to vision-augmented queries: semantic retrievers benefit from rich captions while reasoning-enhanced models show performance degradation, suggesting fundamental differences in how these systems process multimodal information.

The comprehensive results in Tables~\ref{tab:results_Llama-3.2-11B}--\ref{tab:results_gpt4o} reveal several important patterns:

\paragraph{Semantic retrievers benefit consistently from captions.} Models like E5 and Contriever show substantial and consistent improvements across all caption models (+5.7 to +7.4 points for E5), indicating that textual descriptions of visual content align well with semantic embedding spaces trained on text corpora.

\paragraph{Reasoning-enhanced retrievers are disrupted by captions.} DiVeR experiences severe performance degradation with all caption models (-12.0 to -12.4 points), suggesting that its reasoning mechanisms are sensitive to input format and may be optimized for concise, human-written queries rather than verbose generated descriptions.

\paragraph{Caption quality shows diminishing returns.} E5 performance plateaus at ~32-33 nDCG@10 across Qwen-3B through Qwen-72B, with minimal differences (±0.5 points) despite 24× parameter scaling. This suggests that retrieval bottlenecks lie in the retriever architecture rather than caption quality.

\paragraph{Domain-specific patterns persist.} The relative difficulty across domains remains consistent regardless of caption model: Quantum Computing remains challenging (best: 14.1 with GPT-4o) while Law remains easier (best: 63.7 with Qwen-7B). This indicates that visual information alone cannot overcome inherent domain complexity.

These findings have important implications for multimodal retrieval system design. While caption-based approaches offer a simple way to incorporate visual information into text-only retrievers, they fundamentally cannot replace true multimodal reasoning. Future work should explore retrieval architectures that can natively process both visual and textual signals without relying on intermediate text generation.

\begin{table*}[t]
\centering
\small
\caption{\textbf{Retrieval performance on Task 1 with Llama-3.2-11B image captions.} Results show nDCG@10 across all 29 domains when text queries are augmented with captions generated by Llama-3.2-11B. Compared to the no-caption baseline (Table~\ref{tab:task1_results}), semantic dense retrievers like E5 show substantial improvements (+5.7 points), while reasoning-enhanced models like DiVeR experience significant performance drops (-12.4 points). This pattern suggests that automatically generated captions, while semantically informative, may introduce noise that disrupts reasoning-based retrieval strategies.}
\label{tab:results_Llama-3.2-11B}
\resizebox{\textwidth}{!}{%
\begin{tabular}{lcccccccccc}
\toprule
\textbf{Domain} & \textbf{BM25} & \textbf{Contriever} & \textbf{DiVeR} & \textbf{E5} & \textbf{GritLM} & \textbf{Qwen} & \textbf{Qwen2} & \textbf{Rader} & \textbf{ReasonIR} & \textbf{SFR} \\
\midrule
\rowcolor{gray!12}\multicolumn{11}{c}{\textit{STEM}} \\
\midrule
\textbf{Bio} & 4.4 & 18.7 & 16.8 & \textbf{28.1} & 18.3 & 17.7 & 21.5 & 21.0 & 20.7 & \underline{25.5} \\
\textbf{Chem} & 8.7 & 24.5 & 26.6 & \textbf{33.4} & 28.2 & 16.9 & 26.1 & 24.6 & \underline{28.8} & 28.0 \\
\textbf{Phys} & 3.0 & 10.4 & 9.4 & \textbf{17.9} & 14.7 & 11.1 & 15.7 & 15.7 & \underline{15.9} & 15.5 \\
\textbf{Math} & 1.0 & 22.9 & 18.0 & \textbf{29.1} & 15.5 & 14.7 & 14.1 & \underline{25.1} & 16.6 & 24.8 \\
\textbf{Earth} & 6.1 & 22.2 & 21.1 & \underline{33.0} & 30.5 & 20.5 & \textbf{35.9} & 27.7 & 30.4 & 31.9 \\
\textbf{BioAc} & 7.5 & 18.0 & 19.0 & \textbf{24.2} & 20.1 & 18.9 & 23.0 & 16.8 & 15.8 & \underline{23.3} \\
\textbf{BioInf} & 4.9 & 23.8 & 19.4 & \underline{29.1} & 15.6 & 15.4 & 22.1 & \textbf{31.7} & 22.4 & 27.4 \\
\textbf{Med} & 12.5 & 28.0 & 25.0 & 33.3 & \underline{34.8} & 31.8 & 33.5 & 29.6 & \textbf{40.3} & 34.6 \\
\midrule
\rowcolor{gray!12}\multicolumn{11}{c}{\textit{Computing}} \\
\midrule
\textbf{Ubuntu} & 17.0 & 28.3 & 26.3 & \textbf{47.3} & 32.6 & 32.1 & \underline{43.8} & 31.7 & 38.0 & 31.2 \\
\textbf{BTC} & 3.6 & 21.2 & 14.6 & 28.2 & 32.8 & 23.6 & \textbf{37.2} & 19.9 & \underline{34.1} & 26.8 \\
\textbf{Crypto} & 0.0 & 17.4 & 8.2 & \underline{19.3} & 8.8 & 10.5 & 7.2 & \textbf{20.6} & 13.2 & 15.4 \\
\textbf{QC} & 4.0 & 9.7 & 7.3 & 11.0 & \underline{11.1} & 6.0 & 9.6 & 8.3 & 9.3 & \textbf{12.6} \\
\textbf{Robot} & 3.4 & 23.0 & 17.5 & \underline{31.9} & 21.2 & 22.0 & 28.8 & 28.3 & \textbf{33.4} & 29.1 \\
\textbf{Sales} & 3.3 & 21.8 & 18.8 & 42.8 & 27.7 & 33.5 & 37.7 & \underline{45.3} & \textbf{51.2} & 26.4 \\
\midrule
\rowcolor{gray!12}\multicolumn{11}{c}{\textit{Social Sci.}} \\
\midrule
\textbf{Econ} & 1.2 & 12.7 & 17.0 & \textbf{29.6} & 26.7 & 19.2 & \underline{29.0} & 26.5 & 25.7 & 25.8 \\
\textbf{Psych} & 4.0 & 23.4 & 19.2 & \underline{27.6} & 22.9 & 21.5 & 25.7 & 20.7 & 23.8 & \textbf{28.1} \\
\textbf{Phil} & 4.0 & 18.4 & 16.0 & 18.2 & 21.5 & 16.9 & 19.6 & 19.5 & \underline{21.7} & \textbf{23.9} \\
\textbf{Law} & 4.7 & 38.8 & 50.5 & 52.9 & 49.3 & 50.4 & \textbf{62.6} & 45.5 & \underline{62.0} & 44.7 \\
\textbf{Christ} & 31.1 & 15.9 & 24.6 & \textbf{38.3} & 33.8 & 24.2 & \underline{37.1} & 16.4 & 29.5 & 26.6 \\
\textbf{Islam} & 9.3 & 26.5 & 18.1 & 33.5 & 27.7 & 17.3 & \textbf{38.7} & 19.6 & \underline{34.4} & 28.3 \\
\midrule
\rowcolor{gray!12}\multicolumn{11}{c}{\textit{Applied}} \\
\midrule
\textbf{Aviat} & 1.4 & 21.1 & 20.2 & 24.4 & 21.6 & 14.1 & \textbf{33.9} & 19.7 & 25.5 & \underline{29.2} \\
\textbf{Game} & 34.3 & 26.7 & 30.3 & 50.5 & 44.8 & 45.4 & \underline{53.9} & 41.6 & \textbf{54.3} & 39.4 \\
\textbf{GIS} & 0.7 & 21.8 & 14.4 & \underline{28.6} & 24.4 & 18.3 & 24.3 & 28.6 & 24.3 & \textbf{28.7} \\
\textbf{PM} & 16.8 & 28.0 & 23.4 & \textbf{42.1} & 33.4 & \underline{42.1} & 39.3 & 28.5 & 39.4 & 30.4 \\
\textbf{Sustain} & 9.0 & 17.1 & 22.5 & 32.7 & 29.7 & 20.1 & \textbf{37.1} & 19.3 & \underline{36.5} & 29.7 \\
\textbf{Travel} & 23.1 & 19.6 & 24.7 & \underline{36.8} & 25.3 & 27.6 & 35.3 & 27.7 & \textbf{40.5} & 29.1 \\
\textbf{Apple} & 0.0 & 25.9 & 11.8 & \underline{26.2} & 21.0 & 21.6 & \textbf{28.7} & 15.5 & 19.0 & 23.5 \\
\textbf{Acad} & 4.7 & 24.2 & 22.7 & \textbf{28.9} & 26.3 & 16.3 & 23.6 & 22.3 & \underline{27.1} & 24.7 \\
\textbf{Quant} & 2.6 & 14.8 & 10.5 & 20.5 & 22.0 & 17.2 & 23.1 & \textbf{24.6} & 23.5 & \underline{24.5} \\
\midrule
\midrule
\textbf{Avg.} & 7.8 & 21.6 & 19.8 & \textbf{31.0} & 25.6 & 22.3 & \underline{29.9} & 24.9 & 29.6 & 27.2 \\
\bottomrule
\end{tabular}}
\end{table*}

\begin{table*}[t]
\centering
\small
\caption{\textbf{Retrieval performance on Task 1 with Llama-3.2-90B image captions.} Results show nDCG@10 across all domains using captions from the larger Llama-3.2-90B model. The 90B model generates more detailed and accurate captions compared to the 11B variant, leading to further improvements for semantic retrievers (E5: 32.0 vs 31.0 with 11B). However, reasoning-enhanced retrievers remain significantly below their no-caption baseline performance, indicating that caption quality alone cannot resolve the fundamental mismatch between semantic descriptions and reasoning-based relevance assessment.}
\label{tab:results_Llama-3.2-90B}
\resizebox{\textwidth}{!}{%
\begin{tabular}{lcccccccccc}
\toprule
\textbf{Domain} & \textbf{BM25} & \textbf{Contriever} & \textbf{DiVeR} & \textbf{E5} & \textbf{GritLM} & \textbf{Qwen} & \textbf{Qwen2} & \textbf{Rader} & \textbf{ReasonIR} & \textbf{SFR} \\
\midrule
\rowcolor{gray!12}\multicolumn{11}{c}{\textit{STEM}} \\
\midrule
\textbf{Bio} & 5.0 & 19.8 & 18.2 & \textbf{29.3} & 18.0 & 17.5 & 22.5 & 21.0 & 21.4 & \underline{25.8} \\
\textbf{Chem} & 8.4 & 25.4 & 24.0 & \textbf{32.7} & 28.1 & 15.7 & 25.4 & 25.7 & 29.2 & \underline{29.7} \\
\textbf{Phys} & 3.5 & 11.5 & 11.8 & \textbf{19.5} & 15.4 & 12.5 & 17.6 & 16.9 & \underline{19.2} & 16.9 \\
\textbf{Math} & 1.9 & 25.7 & 24.0 & \textbf{30.2} & 18.4 & 16.7 & 15.8 & 25.5 & 19.5 & \underline{28.0} \\
\textbf{Earth} & 4.5 & 21.9 & 22.1 & 31.7 & 29.9 & 21.7 & \textbf{37.2} & 25.7 & \underline{33.6} & 32.1 \\
\textbf{BioAc} & 7.3 & 16.8 & 19.1 & \underline{25.4} & 24.3 & 19.1 & 25.2 & 17.9 & 17.8 & \textbf{26.5} \\
\textbf{BioInf} & 5.3 & 23.3 & 18.4 & 29.3 & 21.0 & 16.2 & 21.4 & \textbf{34.9} & 24.5 & \underline{29.7} \\
\textbf{Med} & 12.9 & 27.6 & 26.2 & 34.2 & \underline{36.8} & 32.3 & 34.4 & 29.3 & \textbf{43.8} & 33.5 \\
\midrule
\rowcolor{gray!12}\multicolumn{11}{c}{\textit{Computing}} \\
\midrule
\textbf{Ubuntu} & 18.5 & 28.6 & 26.9 & \textbf{47.4} & 33.7 & 37.7 & 44.3 & 33.3 & \underline{44.6} & 33.5 \\
\textbf{BTC} & 3.7 & 20.9 & 15.0 & 29.0 & 33.9 & 22.9 & \textbf{36.3} & 21.7 & \underline{34.0} & 28.3 \\
\textbf{Crypto} & 0.2 & 17.2 & 11.4 & \underline{20.0} & 10.9 & 10.1 & 7.5 & \textbf{20.1} & 15.0 & 17.7 \\
\textbf{QC} & 4.0 & 9.9 & 6.7 & 11.9 & \underline{12.2} & 6.3 & 9.6 & 8.5 & 9.1 & \textbf{13.5} \\
\textbf{Robot} & 3.4 & 23.3 & 18.1 & \textbf{32.6} & 22.6 & 21.6 & 27.4 & 29.5 & \underline{30.8} & 29.1 \\
\textbf{Sales} & 3.2 & 37.5 & 23.4 & \underline{46.2} & 36.8 & 40.8 & 37.5 & 44.2 & \textbf{51.7} & 34.1 \\
\midrule
\rowcolor{gray!12}\multicolumn{11}{c}{\textit{Social Sci.}} \\
\midrule
\textbf{Econ} & 2.4 & 11.3 & 17.6 & \textbf{29.8} & 27.7 & 19.7 & 26.4 & \underline{29.2} & 24.2 & 27.5 \\
\textbf{Psych} & 5.0 & 23.4 & 19.8 & \textbf{28.8} & 24.8 & 22.9 & 27.9 & 21.4 & 27.4 & \underline{28.5} \\
\textbf{Phil} & 4.0 & 17.3 & 16.9 & 17.5 & \underline{21.7} & 13.9 & 19.8 & 18.4 & 21.6 & \textbf{25.5} \\
\textbf{Law} & 5.8 & 38.2 & 47.9 & 54.2 & 51.0 & 47.8 & \textbf{62.9} & 41.3 & \underline{62.5} & 45.8 \\
\textbf{Christ} & 32.3 & 17.7 & 22.9 & \textbf{41.0} & 33.5 & 21.9 & \underline{37.0} & 16.7 & 30.2 & 27.4 \\
\textbf{Islam} & 9.2 & 25.9 & 19.8 & 33.9 & 32.6 & 17.9 & \textbf{38.9} & 19.2 & \underline{34.7} & 29.5 \\
\midrule
\rowcolor{gray!12}\multicolumn{11}{c}{\textit{Applied}} \\
\midrule
\textbf{Aviat} & 0.6 & 19.8 & 21.8 & 25.0 & 24.7 & 14.0 & \textbf{35.1} & 19.9 & 28.4 & \underline{30.9} \\
\textbf{Game} & 38.9 & 41.8 & 40.4 & 54.1 & 47.5 & 45.2 & \underline{56.1} & 38.5 & \textbf{56.5} & 43.1 \\
\textbf{GIS} & 1.2 & 24.6 & 14.1 & \textbf{30.4} & 26.7 & 21.3 & 26.6 & 26.9 & 29.2 & \underline{29.2} \\
\textbf{PM} & 20.7 & 28.4 & 25.3 & \underline{43.6} & 33.5 & \textbf{45.8} & 40.7 & 29.5 & 40.0 & 30.2 \\
\textbf{Sustain} & 11.1 & 18.8 & 23.4 & 35.0 & 30.6 & 20.5 & \textbf{38.5} & 20.4 & \underline{38.5} & 30.5 \\
\textbf{Travel} & 23.8 & 22.1 & 26.2 & 39.5 & 29.5 & 27.0 & \underline{40.0} & 29.3 & \textbf{43.9} & 31.4 \\
\textbf{Apple} & 0.0 & \underline{28.9} & 15.4 & 23.9 & 26.2 & 23.7 & \textbf{29.1} & 15.5 & 27.5 & 27.6 \\
\textbf{Acad} & 7.4 & 23.3 & 23.0 & \underline{29.5} & 29.1 & 19.2 & 23.8 & 23.5 & \textbf{31.8} & 25.9 \\
\textbf{Quant} & 1.1 & 16.7 & 10.4 & 22.2 & 21.9 & 17.8 & 24.6 & \textbf{26.0} & 25.0 & \underline{25.1} \\
\midrule
\midrule
\textbf{Avg.} & 8.5 & 23.0 & 21.0 & \textbf{32.0} & 27.7 & 23.1 & 30.7 & 25.2 & \underline{31.6} & 28.8 \\
\bottomrule
\end{tabular}}
\end{table*}

\begin{table*}[t]
\centering
\small
\caption{\textbf{Retrieval performance on Task 1 with Qwen-2.5-3B image captions.} Results show nDCG@10 using captions generated by Qwen-2.5-3B, the smallest model in the Qwen series. Despite its compact size, Qwen-2.5-3B produces captions that substantially boost semantic retriever performance (E5: 32.3), achieving comparable results to much larger caption models. The consistent pattern of reasoning-enhanced model degradation (DiVeR: 20.2) persists across all Qwen model sizes, suggesting this phenomenon is independent of caption model architecture or scale.}
\label{tab:results_Qwen-2.5-3B}
\resizebox{\textwidth}{!}{%
\begin{tabular}{lcccccccccc}
\toprule
\textbf{Domain} & \textbf{BM25} & \textbf{Contriever} & \textbf{DiVeR} & \textbf{E5} & \textbf{GritLM} & \textbf{Qwen} & \textbf{Qwen2} & \textbf{Rader} & \textbf{ReasonIR} & \textbf{SFR} \\
\midrule
\rowcolor{gray!12}\multicolumn{11}{c}{\textit{STEM}} \\
\midrule
\textbf{Bio} & 4.8 & 19.3 & 18.5 & \textbf{27.8} & 18.2 & 14.6 & 21.0 & 20.3 & 22.0 & \underline{25.2} \\
\textbf{Chem} & 7.8 & 24.3 & 25.0 & \textbf{33.8} & 25.6 & 17.7 & 25.8 & 24.4 & \underline{28.6} & 28.1 \\
\textbf{Phys} & 4.0 & 11.3 & 10.7 & \textbf{19.9} & 14.6 & 10.4 & 16.0 & 16.1 & \underline{16.1} & 16.0 \\
\textbf{Math} & 2.0 & 22.7 & 20.7 & \textbf{34.2} & 18.1 & 15.7 & 16.1 & \underline{29.8} & 16.7 & 26.3 \\
\textbf{Earth} & 4.7 & 21.9 & 21.8 & \textbf{34.5} & 29.1 & 19.6 & 31.1 & 25.0 & \underline{31.5} & 30.5 \\
\textbf{BioAc} & 8.5 & 18.7 & 20.8 & \textbf{25.8} & 20.9 & 17.3 & 22.5 & 18.6 & 15.6 & \underline{24.1} \\
\textbf{BioInf} & 5.3 & 23.1 & 18.1 & \underline{32.6} & 18.9 & 15.8 & 20.3 & \textbf{37.2} & 23.9 & 28.1 \\
\textbf{Med} & 12.6 & 27.1 & 21.8 & \underline{36.1} & 34.7 & 30.4 & 31.2 & 29.2 & \textbf{38.4} & 31.6 \\
\midrule
\rowcolor{gray!12}\multicolumn{11}{c}{\textit{Computing}} \\
\midrule
\textbf{Ubuntu} & 17.3 & 27.3 & 26.0 & \textbf{43.8} & 28.9 & 35.1 & \underline{40.7} & 32.7 & 35.6 & 26.9 \\
\textbf{BTC} & 3.4 & 21.1 & 14.2 & 29.8 & \underline{32.6} & 22.1 & \textbf{32.7} & 19.9 & 32.2 & 27.0 \\
\textbf{Crypto} & 0.8 & 18.3 & 12.4 & \textbf{24.3} & 10.2 & 10.5 & 7.8 & \underline{23.4} & 15.3 & 17.5 \\
\textbf{QC} & 3.8 & 9.6 & 7.6 & \underline{13.2} & 11.7 & 5.7 & 9.4 & 9.3 & 9.3 & \textbf{13.9} \\
\textbf{Robot} & 5.0 & 20.9 & 17.7 & \textbf{34.9} & 20.2 & 19.7 & 23.6 & \underline{30.5} & 26.7 & 27.9 \\
\textbf{Sales} & 3.9 & 27.9 & 17.0 & \underline{40.9} & 25.6 & 29.1 & 37.4 & 40.0 & \textbf{52.3} & 26.8 \\
\midrule
\rowcolor{gray!12}\multicolumn{11}{c}{\textit{Social Sci.}} \\
\midrule
\textbf{Econ} & 4.0 & 11.6 & 15.0 & \textbf{31.8} & 27.7 & 18.3 & 21.6 & \underline{28.5} & 22.7 & 26.7 \\
\textbf{Psych} & 5.6 & 23.1 & 19.9 & \textbf{29.4} & 23.0 & 20.9 & 24.8 & 20.4 & 27.3 & \underline{27.7} \\
\textbf{Phil} & 4.1 & 19.1 & 16.1 & 17.4 & 21.3 & 15.6 & 17.2 & 17.5 & \underline{21.4} & \textbf{23.9} \\
\textbf{Law} & 6.2 & 38.3 & 49.2 & 52.5 & 50.2 & 47.8 & \textbf{59.5} & 43.9 & \underline{57.3} & 45.6 \\
\textbf{Christ} & 30.0 & 16.7 & 21.2 & \textbf{38.1} & 29.8 & 25.5 & \underline{36.4} & 14.1 & 26.7 & 24.0 \\
\textbf{Islam} & 9.8 & 26.4 & 21.1 & \textbf{35.8} & 29.9 & 17.4 & 34.9 & 20.6 & \underline{35.0} & 28.2 \\
\midrule
\rowcolor{gray!12}\multicolumn{11}{c}{\textit{Applied}} \\
\midrule
\textbf{Aviat} & 1.1 & 19.2 & 21.2 & \underline{29.3} & 21.9 & 12.6 & \textbf{31.3} & 21.7 & 25.3 & 28.0 \\
\textbf{Game} & 35.6 & 30.9 & 24.1 & \underline{50.2} & 41.4 & 47.5 & \textbf{58.0} & 35.0 & 45.2 & 38.4 \\
\textbf{GIS} & 1.3 & 20.2 & 15.6 & \textbf{31.3} & 24.7 & 20.5 & 24.4 & 28.0 & 26.6 & \underline{28.8} \\
\textbf{PM} & 18.4 & 27.1 & 25.4 & \textbf{40.7} & 31.6 & \underline{40.0} & 36.5 & 29.7 & 37.2 & 28.4 \\
\textbf{Sustain} & 11.4 & 17.4 & 24.1 & \underline{34.3} & 30.9 & 20.3 & 33.6 & 20.5 & \textbf{36.6} & 29.4 \\
\textbf{Travel} & 22.2 & 21.3 & 25.7 & \underline{38.3} & 24.8 & 24.0 & 32.4 & 28.2 & \textbf{38.6} & 28.5 \\
\textbf{Apple} & 1.0 & \textbf{30.3} & 20.4 & 22.8 & 24.3 & 20.6 & \underline{28.5} & 16.5 & 19.4 & 24.5 \\
\textbf{Acad} & 9.8 & 22.9 & 23.9 & \textbf{30.3} & 25.7 & 14.2 & 22.3 & 19.2 & \underline{27.2} & 23.7 \\
\textbf{Quant} & 2.3 & 16.7 & 10.6 & 22.3 & 22.5 & 16.9 & 18.4 & \underline{25.5} & \textbf{25.7} & 24.7 \\
\midrule
\midrule
\textbf{Avg.} & 8.5 & 21.9 & 20.2 & \textbf{32.3} & 25.5 & 21.6 & 28.1 & 25.0 & \underline{28.8} & 26.9 \\
\bottomrule
\end{tabular}}
\end{table*}

\begin{table*}[t]
\centering
\small
\caption{\textbf{Retrieval performance on Task 1 with Qwen-2.5-7B image captions.} Results show nDCG@10 using captions from Qwen-2.5-7B. The 7B model maintains strong performance for semantic retrievers (E5: 32.2), with minimal differences compared to the 3B variant. This suggests that for caption-augmented retrieval, caption quality may plateau beyond a certain model size, and further scaling does not necessarily improve retrieval performance. BM25 shows slight variations across Qwen model sizes, indicating sensitivity to caption phrasing and lexical choices.}
\label{tab:results_Qwen-2.5-7B}
\resizebox{\textwidth}{!}{%
\begin{tabular}{lcccccccccc}
\toprule
\textbf{Domain} & \textbf{BM25} & \textbf{Contriever} & \textbf{DiVeR} & \textbf{E5} & \textbf{GritLM} & \textbf{Qwen} & \textbf{Qwen2} & \textbf{Rader} & \textbf{ReasonIR} & \textbf{SFR} \\
\midrule
\rowcolor{gray!12}\multicolumn{11}{c}{\textit{STEM}} \\
\midrule
\textbf{Bio} & 4.8 & 19.3 & 18.5 & \textbf{27.7} & 18.2 & 14.6 & 21.0 & 20.3 & 22.0 & \underline{25.2} \\
\textbf{Chem} & 8.4 & 24.6 & 24.2 & \textbf{33.1} & 28.5 & 15.9 & 25.6 & 25.4 & 28.8 & \underline{28.8} \\
\textbf{Phys} & 4.0 & 11.3 & 10.7 & \textbf{19.9} & 14.6 & 10.4 & 16.0 & 16.1 & \underline{16.1} & 16.0 \\
\textbf{Math} & 1.9 & 24.2 & 20.7 & \textbf{30.1} & 16.0 & 15.4 & 17.6 & 24.8 & 16.7 & \underline{27.0} \\
\textbf{Earth} & 4.7 & 21.9 & 21.8 & \textbf{35.0} & 29.1 & 19.6 & 31.1 & 25.0 & \underline{31.5} & 30.5 \\
\textbf{BioAc} & 8.0 & 18.0 & 21.5 & \textbf{23.8} & 20.8 & 16.6 & 22.9 & 18.2 & 16.3 & \underline{23.6} \\
\textbf{BioInf} & 5.3 & 23.1 & 18.1 & \underline{32.6} & 18.9 & 15.8 & 20.3 & \textbf{37.2} & 23.9 & 28.1 \\
\textbf{Med} & 12.6 & 27.1 & 21.8 & \underline{36.1} & 34.7 & 30.4 & 31.2 & 29.2 & \textbf{38.4} & 31.6 \\
\midrule
\rowcolor{gray!12}\multicolumn{11}{c}{\textit{Computing}} \\
\midrule
\textbf{Ubuntu} & 17.7 & 27.8 & 28.6 & \textbf{45.2} & 32.0 & 33.6 & \underline{41.9} & 33.1 & 39.6 & 30.3 \\
\textbf{BTC} & 3.9 & 20.7 & 14.1 & 29.4 & 33.0 & 22.2 & \textbf{35.6} & 19.4 & \underline{33.8} & 29.1 \\
\textbf{Crypto} & 0.8 & 18.3 & 12.4 & \textbf{24.2} & 10.2 & 10.5 & 7.8 & \underline{23.4} & 15.3 & 17.5 \\
\textbf{QC} & 4.0 & 9.6 & 9.0 & \underline{11.2} & 10.7 & 6.4 & 8.9 & 8.0 & 8.6 & \textbf{12.2} \\
\textbf{Robot} & 4.4 & 21.0 & 18.0 & \textbf{32.3} & 21.0 & 19.8 & 25.9 & 27.9 & \underline{30.8} & 29.3 \\
\textbf{Sales} & 3.9 & 25.3 & 18.5 & \underline{44.8} & 29.3 & 31.8 & 36.2 & 44.5 & \textbf{51.1} & 31.8 \\
\midrule
\rowcolor{gray!12}\multicolumn{11}{c}{\textit{Social Sci.}} \\
\midrule
\textbf{Econ} & 3.7 & 12.2 & 15.6 & \textbf{30.8} & \underline{29.5} & 16.8 & 24.6 & 27.9 & 25.6 & 26.3 \\
\textbf{Psych} & 5.7 & 23.3 & 19.8 & 28.1 & 23.0 & 21.0 & \underline{28.2} & 20.2 & 26.9 & \textbf{28.8} \\
\textbf{Phil} & 4.1 & 19.1 & 16.1 & 17.4 & 21.3 & 15.6 & 17.2 & 17.5 & \underline{21.4} & \textbf{23.9} \\
\textbf{Law} & 5.6 & 38.4 & 46.9 & 52.5 & 50.3 & 47.0 & \textbf{63.7} & 44.1 & \underline{59.6} & 45.3 \\
\textbf{Christ} & 30.4 & 17.1 & 22.4 & \textbf{40.2} & 33.8 & 24.5 & \underline{37.2} & 17.1 & 29.4 & 27.3 \\
\textbf{Islam} & 10.3 & 26.1 & 20.9 & 33.1 & 28.9 & 18.7 & \textbf{39.6} & 19.6 & \underline{34.7} & 28.4 \\
\midrule
\rowcolor{gray!12}\multicolumn{11}{c}{\textit{Applied}} \\
\midrule
\textbf{Aviat} & 1.1 & 19.2 & 21.2 & \underline{29.2} & 21.9 & 12.6 & \textbf{31.3} & 21.7 & 25.3 & 28.0 \\
\textbf{Game} & 37.6 & 33.9 & 30.7 & \underline{53.2} & 48.0 & 43.7 & \textbf{56.3} & 37.3 & 51.8 & 42.1 \\
\textbf{GIS} & 1.3 & 21.9 & 15.2 & \underline{27.8} & 22.7 & 20.3 & 23.9 & 27.5 & 23.8 & \textbf{28.8} \\
\textbf{PM} & 17.2 & 27.0 & 25.3 & \underline{41.0} & 30.8 & \textbf{43.7} & 40.2 & 30.2 & 38.0 & 29.2 \\
\textbf{Sustain} & 11.2 & 18.0 & 25.5 & \underline{34.7} & 30.6 & 19.5 & 34.0 & 20.2 & \textbf{36.4} & 29.4 \\
\textbf{Travel} & 22.2 & 21.3 & 25.7 & \underline{38.3} & 24.8 & 24.0 & 32.4 & 28.2 & \textbf{38.6} & 28.5 \\
\textbf{Apple} & 0.0 & \textbf{29.1} & 18.4 & 26.3 & 26.5 & 20.4 & \underline{28.0} & 16.2 & 22.5 & 25.9 \\
\textbf{Acad} & 9.1 & 22.9 & 23.3 & \textbf{32.4} & 24.6 & 14.7 & 22.1 & 19.8 & \underline{27.2} & 23.7 \\
\textbf{Quant} & 2.3 & 16.9 & 10.4 & 22.0 & 22.9 & 16.9 & 23.7 & 24.1 & \underline{24.6} & \textbf{24.8} \\
\midrule
\midrule
\textbf{Avg.} & 8.5 & 22.0 & 20.5 & \textbf{32.2} & 26.1 & 21.5 & 29.1 & 25.0 & \underline{29.6} & 27.6 \\
\bottomrule
\end{tabular}}
\end{table*}

\begin{table*}[t]
\centering
\small
\caption{\textbf{Retrieval performance on Task 1 with Qwen-2.5-32B image captions.} Results show nDCG@10 using captions from the larger Qwen-2.5-32B model. The 32B model achieves the highest E5 performance (32.7) among all Qwen variants, suggesting that larger models may produce slightly more informative captions for semantic matching. However, the marginal gains (+0.4 over 7B) are small relative to the computational cost increase. Reasoning-enhanced retrievers show no improvement with larger caption models, remaining at ~20-21 nDCG@10 across all Qwen sizes.}
\label{tab:results_Qwen-2.5-32B}
\resizebox{\textwidth}{!}{%
\begin{tabular}{lcccccccccc}
\toprule
\textbf{Domain} & \textbf{BM25} & \textbf{Contriever} & \textbf{DiVeR} & \textbf{E5} & \textbf{GritLM} & \textbf{Qwen} & \textbf{Qwen2} & \textbf{Rader} & \textbf{ReasonIR} & \textbf{SFR} \\
\midrule
\rowcolor{gray!12}\multicolumn{11}{c}{\textit{STEM}} \\
\midrule
\textbf{Bio} & 4.8 & 19.3 & 18.5 & \textbf{27.8} & 18.2 & 14.6 & 21.0 & 20.3 & 22.0 & \underline{25.2} \\
\textbf{Chem} & 9.0 & 24.2 & 24.7 & \textbf{33.2} & 28.7 & 17.0 & 26.7 & 26.3 & 28.7 & \underline{29.9} \\
\textbf{Phys} & 4.0 & 11.3 & 10.7 & \textbf{19.9} & 14.6 & 10.4 & 16.0 & 16.1 & \underline{16.1} & 16.0 \\
\textbf{Math} & 2.0 & 22.7 & 20.7 & \textbf{34.2} & 18.1 & 15.7 & 16.1 & \underline{29.8} & 16.7 & 26.3 \\
\textbf{Earth} & 4.7 & 21.9 & 21.8 & \textbf{34.5} & 29.1 & 19.6 & 31.1 & 25.0 & \underline{31.5} & 30.5 \\
\textbf{BioAc} & 8.5 & 17.9 & 20.7 & \textbf{27.4} & 20.5 & 16.6 & 23.2 & 18.6 & 14.7 & \underline{24.2} \\
\textbf{BioInf} & 5.3 & 23.1 & 18.1 & \underline{32.9} & 18.9 & 15.8 & 20.3 & \textbf{37.2} & 23.9 & 28.1 \\
\textbf{Med} & 12.6 & 27.1 & 21.8 & \underline{36.1} & 34.7 & 30.4 & 31.2 & 29.2 & \textbf{38.4} & 31.6 \\
\midrule
\rowcolor{gray!12}\multicolumn{11}{c}{\textit{Computing}} \\
\midrule
\textbf{Ubuntu} & 16.2 & 28.8 & 28.5 & \textbf{45.6} & 33.5 & 36.7 & \underline{43.5} & 32.7 & 42.6 & 31.2 \\
\textbf{BTC} & 3.3 & 21.2 & 13.3 & 30.4 & \underline{32.4} & 22.1 & \textbf{33.0} & 19.0 & 31.9 & 26.6 \\
\textbf{Crypto} & 0.8 & 18.3 & 12.4 & \textbf{24.3} & 10.2 & 10.5 & 7.8 & \underline{23.4} & 15.3 & 17.5 \\
\textbf{QC} & 3.6 & 10.4 & 8.0 & 11.1 & \underline{11.7} & 6.4 & 10.0 & 7.7 & 8.4 & \textbf{12.6} \\
\textbf{Robot} & 3.5 & 23.5 & 20.2 & \underline{32.7} & 23.7 & 22.6 & 27.6 & 29.7 & \textbf{33.1} & 28.6 \\
\textbf{Sales} & 2.9 & 37.9 & 21.8 & \underline{47.2} & 28.0 & 35.8 & 37.5 & 42.1 & \textbf{51.7} & 30.3 \\
\midrule
\rowcolor{gray!12}\multicolumn{11}{c}{\textit{Social Sci.}} \\
\midrule
\textbf{Econ} & 1.7 & 11.4 & 15.9 & \textbf{30.7} & 27.9 & 19.4 & 25.7 & \underline{28.0} & 25.9 & 27.1 \\
\textbf{Psych} & 5.3 & 23.3 & 20.0 & \textbf{29.3} & 22.8 & 21.0 & 24.5 & 20.5 & \underline{27.7} & 27.2 \\
\textbf{Phil} & 4.1 & 19.1 & 16.1 & 17.3 & 21.3 & 15.6 & 17.2 & 17.5 & \underline{21.4} & \textbf{23.9} \\
\textbf{Law} & 6.2 & 38.8 & 49.7 & 52.3 & 50.2 & 47.1 & \textbf{59.3} & 44.2 & \underline{57.8} & 45.4 \\
\textbf{Christ} & 30.9 & 18.1 & 22.5 & \textbf{41.1} & 35.0 & 22.1 & \underline{35.9} & 15.7 & 33.5 & 28.2 \\
\textbf{Islam} & 8.9 & 24.0 & 21.0 & 33.5 & 31.5 & 19.5 & \textbf{40.1} & 19.0 & \underline{35.1} & 29.3 \\
\midrule
\rowcolor{gray!12}\multicolumn{11}{c}{\textit{Applied}} \\
\midrule
\textbf{Aviat} & 1.1 & 19.2 & 21.2 & \underline{29.3} & 21.9 & 12.6 & \textbf{31.3} & 21.7 & 25.3 & 28.0 \\
\textbf{Game} & 36.8 & 43.6 & 37.8 & \underline{55.4} & 47.1 & 44.7 & \textbf{57.7} & 41.1 & 54.1 & 42.2 \\
\textbf{GIS} & 1.3 & 20.2 & 15.9 & \textbf{31.7} & 25.0 & 21.1 & 25.2 & 28.0 & 27.5 & \underline{29.0} \\
\textbf{PM} & 18.0 & 26.9 & 24.9 & \textbf{40.4} & 31.0 & \underline{39.9} & 36.5 & 27.7 & 34.7 & 27.8 \\
\textbf{Sustain} & 11.2 & 18.0 & 25.5 & \underline{34.4} & 30.6 & 19.5 & 34.0 & 20.2 & \textbf{36.4} & 29.4 \\
\textbf{Travel} & 22.2 & 21.3 & 25.7 & \underline{38.3} & 24.8 & 24.0 & 32.4 & 28.2 & \textbf{38.6} & 28.5 \\
\textbf{Apple} & 0.0 & 24.5 & 15.7 & 25.8 & 24.1 & 21.4 & \textbf{29.3} & 18.7 & \underline{28.2} & 27.5 \\
\textbf{Acad} & 9.1 & 22.9 & 23.3 & \textbf{31.8} & 24.6 & 14.7 & 22.1 & 19.8 & \underline{27.2} & 23.7 \\
\textbf{Quant} & 2.2 & 14.2 & 9.2 & 20.1 & 22.2 & 17.5 & \underline{26.0} & 23.4 & 24.4 & \textbf{29.3} \\
\midrule
\midrule
\textbf{Avg.} & 8.3 & 22.5 & 20.9 & \textbf{32.7} & 26.3 & 21.9 & 29.0 & 25.2 & \underline{30.1} & 27.8 \\
\bottomrule
\end{tabular}}
\end{table*}

\begin{table*}[t]
\centering
\small
\caption{\textbf{Retrieval performance on Task 1 with Qwen-2.5-72B image captions.} Results show nDCG@10 using captions from Qwen-2.5-72B, the largest open-source caption model we evaluate. The 72B model achieves the highest overall average for E5 (32.7, tied with 32B), confirming that caption quality plateaus and larger models offer diminishing returns for retrieval. Interestingly, Qwen2 shows its strongest performance with this caption model across multiple domains (Law: 59.3, Gaming: 56.0), suggesting some retrievers may benefit from the most detailed captions available.}
\label{tab:results_Qwen-2.5-72B}
\resizebox{\textwidth}{!}{%
\begin{tabular}{lcccccccccc}
\toprule
\textbf{Domain} & \textbf{BM25} & \textbf{Contriever} & \textbf{DiVeR} & \textbf{E5} & \textbf{GritLM} & \textbf{Qwen} & \textbf{Qwen2} & \textbf{Rader} & \textbf{ReasonIR} & \textbf{SFR} \\
\midrule
\multicolumn{11}{c}{\textit{STEM}} \\
\midrule
\textbf{Bio} & 4.8 & 19.3 & 18.5 & \textbf{27.8} & 18.2 & 14.6 & 21.0 & 20.3 & 22.0 & \underline{25.2} \\
\textbf{Chem} & 7.8 & 24.3 & 24.9 & \textbf{33.4} & 25.4 & 17.8 & 25.7 & 25.0 & \underline{28.4} & 28.3 \\
\textbf{Phys} & 4.0 & 11.3 & 10.7 & \textbf{19.9} & 14.6 & 10.4 & 16.0 & 16.1 & \underline{16.1} & 16.0 \\
\textbf{Math} & 2.0 & 22.7 & 20.7 & \textbf{34.2} & 18.1 & 15.7 & 16.1 & \underline{29.8} & 16.7 & 26.3 \\
\textbf{Earth} & 4.7 & 21.9 & 21.8 & \textbf{34.8} & 29.1 & 19.6 & 31.1 & 25.0 & \underline{31.5} & 30.5 \\
\textbf{BioAc} & 8.5 & 17.9 & 20.7 & \textbf{27.5} & 20.5 & 16.6 & 23.2 & 18.6 & 14.7 & \underline{24.2} \\
\textbf{BioInf} & 5.3 & 23.1 & 18.1 & \underline{32.6} & 18.9 & 15.8 & 20.3 & \textbf{37.2} & 23.9 & 28.1 \\
\textbf{Med} & 12.6 & 27.1 & 21.8 & \underline{36.1} & 34.7 & 30.4 & 31.2 & 29.2 & \textbf{38.4} & 31.6 \\
\midrule
\rowcolor{gray!12}\multicolumn{11}{c}{\textit{Computing}} \\
\midrule
\textbf{Ubuntu} & 17.4 & 27.5 & 25.8 & \textbf{44.5} & 29.0 & 35.0 & \underline{39.2} & 30.8 & 34.1 & 27.3 \\
\textbf{BTC} & 3.3 & 21.2 & 13.3 & 30.4 & \underline{32.4} & 22.1 & \textbf{33.0} & 19.0 & 31.9 & 26.6 \\
\textbf{Crypto} & 0.8 & 18.3 & 12.4 & \textbf{24.3} & 10.2 & 10.5 & 7.8 & \underline{23.4} & 15.3 & 17.5 \\
\textbf{QC} & 4.0 & 9.9 & 6.8 & 11.0 & \underline{11.2} & 7.0 & 9.3 & 7.7 & 8.9 & \textbf{13.0} \\
\textbf{Robot} & 5.0 & 20.9 & 17.7 & \textbf{34.6} & 20.2 & 20.7 & 23.6 & \underline{30.5} & 25.7 & 27.9 \\
\textbf{Sales} & 3.2 & 24.6 & 18.3 & \underline{46.8} & 30.4 & 28.9 & 36.2 & 41.1 & \textbf{51.7} & 31.3 \\
\midrule
\rowcolor{gray!12}\multicolumn{11}{c}{\textit{Social Sci.}} \\
\midrule
\textbf{Econ} & 4.0 & 11.6 & 15.0 & \textbf{31.9} & 27.7 & 18.3 & 21.8 & \underline{28.5} & 22.7 & 26.8 \\
\textbf{Psych} & 5.3 & 23.3 & 20.0 & \textbf{29.5} & 22.8 & 21.0 & 24.5 & 20.5 & \underline{27.7} & 27.2 \\
\textbf{Phil} & 4.1 & 19.1 & 16.1 & 17.4 & 21.3 & 15.6 & 17.2 & 17.5 & \underline{21.4} & \textbf{23.9} \\
\textbf{Law} & 6.2 & 38.8 & 49.7 & 52.4 & 50.2 & 47.1 & \textbf{59.3} & 44.2 & \underline{57.8} & 45.4 \\
\textbf{Christ} & 32.7 & 17.9 & 21.6 & \textbf{38.5} & 33.3 & 22.0 & \underline{35.9} & 17.8 & 30.5 & 27.4 \\
\textbf{Islam} & 10.3 & 26.0 & 22.7 & 34.3 & 30.9 & 19.2 & \textbf{39.9} & 19.6 & \underline{36.4} & 29.5 \\
\midrule
\rowcolor{gray!12}\multicolumn{11}{c}{\textit{Applied}} \\
\midrule
\textbf{Aviat} & 1.1 & 19.2 & 21.2 & \underline{29.2} & 21.9 & 12.6 & \textbf{31.3} & 21.7 & 25.3 & 28.0 \\
\textbf{Game} & 32.9 & 33.8 & 29.7 & 51.8 & 47.3 & 43.8 & \textbf{56.0} & 37.0 & \underline{54.8} & 41.6 \\
\textbf{GIS} & 1.3 & 20.2 & 15.9 & \textbf{31.7} & 25.0 & 21.1 & 25.2 & 28.0 & 27.5 & \underline{29.0} \\
\textbf{PM} & 18.0 & 26.9 & 24.9 & \textbf{40.3} & 31.0 & \underline{39.9} & 36.5 & 27.7 & 34.7 & 27.8 \\
\textbf{Sustain} & 11.2 & 18.0 & 25.5 & \underline{34.7} & 30.6 & 19.5 & 34.0 & 20.2 & \textbf{36.4} & 29.4 \\
\textbf{Travel} & 22.2 & 21.3 & 25.7 & \underline{38.3} & 24.8 & 24.0 & 32.4 & 28.2 & \textbf{38.6} & 28.5 \\
\textbf{Apple} & 1.1 & \textbf{28.7} & 19.1 & 26.2 & 23.9 & 20.5 & \underline{26.7} & 14.3 & 24.6 & 26.2 \\
\textbf{Acad} & 9.1 & 22.9 & 23.3 & \textbf{31.6} & 24.6 & 14.7 & 22.1 & 19.8 & \underline{27.2} & 23.7 \\
\textbf{Quant} & 2.3 & 15.2 & 12.3 & 23.2 & 22.5 & 16.9 & 18.6 & \underline{25.5} & \textbf{25.7} & 24.7 \\
\midrule
\midrule
\textbf{Avg.} & 8.4 & 21.8 & 20.5 & \textbf{32.7} & 25.9 & 21.4 & 28.1 & 25.0 & \underline{29.3} & 27.3 \\
\bottomrule
\end{tabular}}
\end{table*}

\begin{table*}[t]
\centering
\small
\caption{\textbf{Retrieval performance on Task 1 with GPT-4o image captions.} Results show nDCG@10 using captions from the proprietary GPT-4o model, which generally produces the most detailed and accurate image descriptions. GPT-4o captions yield the highest BM25 performance (9.8) across all caption models, likely due to superior lexical diversity and coverage. Surprisingly, E5 performs slightly worse with GPT-4o (31.8) compared to Qwen models (32.7), suggesting that extremely detailed captions may introduce irrelevant information. Reasoning-enhanced models show marginal improvements with GPT-4o (DiVeR: 21.2, ReasonIR: 31.4), but remain far below their no-caption performance.}
\label{tab:results_gpt4o}
\resizebox{\textwidth}{!}{%
\begin{tabular}{lcccccccccc}
\toprule
\textbf{Domain} & \textbf{BM25} & \textbf{Contriever} & \textbf{DiVeR} & \textbf{E5} & \textbf{GritLM} & \textbf{Qwen} & \textbf{Qwen2} & \textbf{Rader} & \textbf{ReasonIR} & \textbf{SFR} \\
\midrule
\rowcolor{gray!12}\multicolumn{11}{c}{\textit{STEM}} \\
\midrule
\textbf{Bio} & 5.5 & 20.3 & 18.9 & \textbf{31.7} & 19.5 & 18.5 & 25.3 & 21.7 & 23.2 & \underline{26.5} \\
\textbf{Chem} & 8.1 & 26.0 & 25.9 & \textbf{34.2} & 29.8 & 17.6 & 26.5 & 25.7 & 29.4 & \underline{30.8} \\
\textbf{Phys} & 3.9 & 11.6 & 11.3 & \underline{19.9} & 15.3 & 11.4 & 17.0 & 17.3 & \textbf{20.2} & 16.9 \\
\textbf{Math} & 2.0 & 22.8 & 19.3 & \textbf{30.6} & 19.0 & 16.8 & 16.1 & 24.6 & 18.0 & \underline{27.3} \\
\textbf{Earth} & 5.6 & 21.7 & 23.8 & 33.3 & 32.1 & 21.8 & \textbf{36.3} & 26.9 & \underline{33.9} & 33.6 \\
\textbf{BioAc} & 8.5 & 17.5 & 20.2 & \textbf{25.2} & 22.7 & 16.6 & 23.1 & 18.7 & 17.0 & \underline{25.0} \\
\textbf{BioInf} & 5.3 & 23.8 & 18.2 & \underline{29.7} & 20.9 & 16.8 & 20.5 & \textbf{35.6} & 24.7 & 29.5 \\
\textbf{Med} & 14.7 & 31.1 & 25.8 & 33.6 & \underline{36.5} & 30.9 & 35.9 & 30.6 & \textbf{45.4} & 32.8 \\
\midrule
\rowcolor{gray!12}\multicolumn{11}{c}{\textit{Computing}} \\
\midrule
\textbf{Ubuntu} & 22.8 & 29.2 & 25.9 & \textbf{47.4} & 33.5 & 37.4 & 42.8 & 31.6 & \underline{43.6} & 31.8 \\
\textbf{BTC} & 5.1 & 21.0 & 15.9 & 29.3 & 33.9 & 22.6 & \textbf{35.9} & 20.3 & \underline{34.8} & 27.8 \\
\textbf{Crypto} & 0.7 & 17.2 & 11.0 & \underline{20.8} & 11.7 & 10.4 & 7.9 & \textbf{22.2} & 15.3 & 16.4 \\
\textbf{QC} & 4.1 & 10.6 & 7.9 & 11.4 & \underline{12.3} & 6.7 & 10.4 & 8.8 & 9.0 & \textbf{14.1} \\
\textbf{Robot} & 3.8 & 23.4 & 18.0 & \textbf{32.7} & 23.2 & 21.7 & 27.0 & 26.9 & \underline{28.8} & 27.4 \\
\textbf{Sales} & 3.9 & 34.6 & 16.8 & \underline{46.6} & 40.7 & 33.8 & 36.2 & 40.6 & \textbf{52.5} & 35.0 \\
\midrule
\rowcolor{gray!12}\multicolumn{11}{c}{\textit{Social Sci.}} \\
\midrule
\textbf{Econ} & 2.3 & 11.6 & 16.0 & \underline{29.5} & \textbf{31.6} & 19.6 & 25.9 & 28.4 & 26.7 & 28.2 \\
\textbf{Psych} & 5.6 & 23.9 & 20.0 & \underline{28.6} & 25.2 & 22.7 & 28.4 & 21.1 & 28.6 & \textbf{30.0} \\
\textbf{Phil} & 4.7 & 20.1 & 18.3 & 19.1 & \underline{22.5} & 14.4 & 20.6 & 17.8 & 22.0 & \textbf{26.4} \\
\textbf{Law} & 7.6 & 38.0 & 48.3 & 52.1 & 52.7 & 47.1 & \textbf{63.6} & 44.5 & \underline{61.3} & 46.4 \\
\textbf{Christ} & 35.2 & 20.0 & 24.7 & \underline{36.7} & 35.4 & 23.3 & \textbf{37.0} & 16.9 & 31.0 & 28.3 \\
\textbf{Islam} & 11.9 & 25.9 & 20.0 & 35.0 & 31.7 & 20.0 & \textbf{39.6} & 21.2 & \underline{36.5} & 30.4 \\
\midrule
\rowcolor{gray!12}\multicolumn{11}{c}{\textit{Applied}} \\
\midrule
\textbf{Aviat} & 1.4 & 19.5 & 21.2 & 25.5 & 24.4 & 14.9 & \textbf{34.7} & 20.9 & 28.6 & \underline{30.5} \\
\textbf{Game} & 44.5 & 35.3 & 35.8 & 48.3 & 47.8 & 48.3 & \textbf{55.9} & 38.2 & \underline{50.6} & 44.5 \\
\textbf{GIS} & 1.2 & 21.1 & 16.9 & \textbf{30.2} & 25.9 & 19.4 & 26.2 & 28.1 & 24.9 & \underline{29.7} \\
\textbf{PM} & 20.7 & 27.8 & 25.1 & \underline{41.7} & 35.1 & \textbf{42.3} & 41.0 & 31.1 & 40.9 & 29.3 \\
\textbf{Sustain} & 12.8 & 18.5 & 26.4 & 34.8 & 31.2 & 21.9 & \underline{36.9} & 20.5 & \textbf{37.5} & 30.8 \\
\textbf{Travel} & 27.3 & 23.0 & 27.6 & 39.4 & 29.2 & 26.5 & \underline{39.5} & 28.6 & \textbf{43.1} & 32.4 \\
\textbf{Apple} & 2.1 & 26.5 & 17.9 & 24.7 & 24.5 & 23.4 & \textbf{30.6} & 15.5 & 26.0 & \underline{27.6} \\
\textbf{Acad} & 8.9 & 25.5 & 25.2 & 28.4 & \underline{30.8} & 20.3 & 24.3 & 22.0 & \textbf{30.9} & 26.0 \\
\textbf{Quant} & 3.7 & 17.4 & 11.6 & 21.9 & 21.1 & 17.4 & 22.3 & \underline{25.6} & 25.0 & \textbf{27.3} \\
\midrule
\textbf{Avg.} & 9.8 & 22.9 & 21.2 & \textbf{31.8} & 28.3 & 22.9 & 30.6 & 25.2 & \underline{31.4} & 29.1 \\
\bottomrule
\end{tabular}}
\end{table*}

\clearpage

\section{Image Type Breakdown by Domain}
\label{app:image_types}
\subsection{Image type taxonomy and classifier setup}
\label{app:image_types_taxonomy}
\begin{figure*}[p]
  \centering
  
  \begin{subfigure}[b]{0.32\textwidth}
    \includegraphics[width=\textwidth]{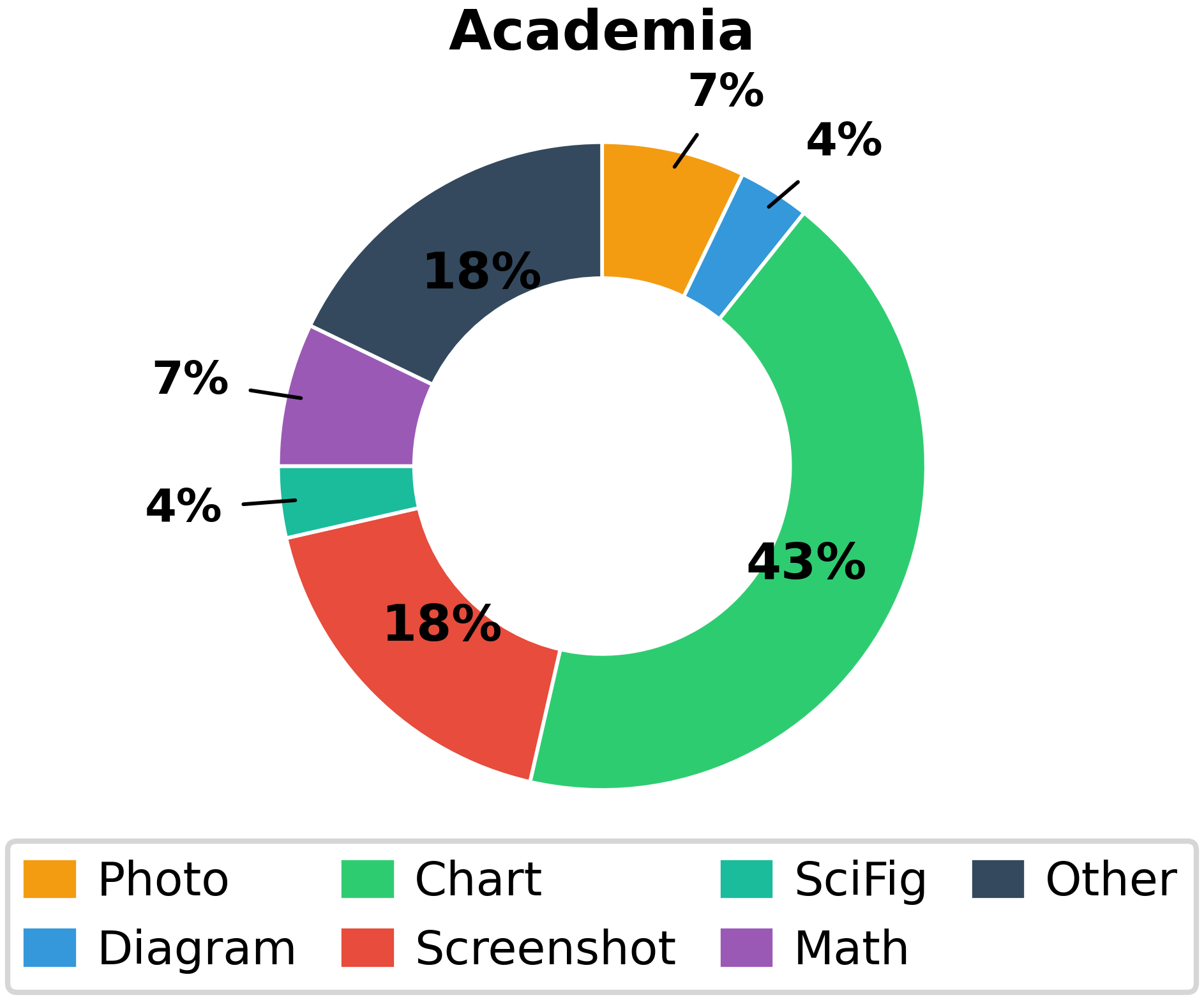}
    \caption{Academia}
  \end{subfigure}
  \hfill
  \begin{subfigure}[b]{0.32\textwidth}
    \includegraphics[width=\textwidth]{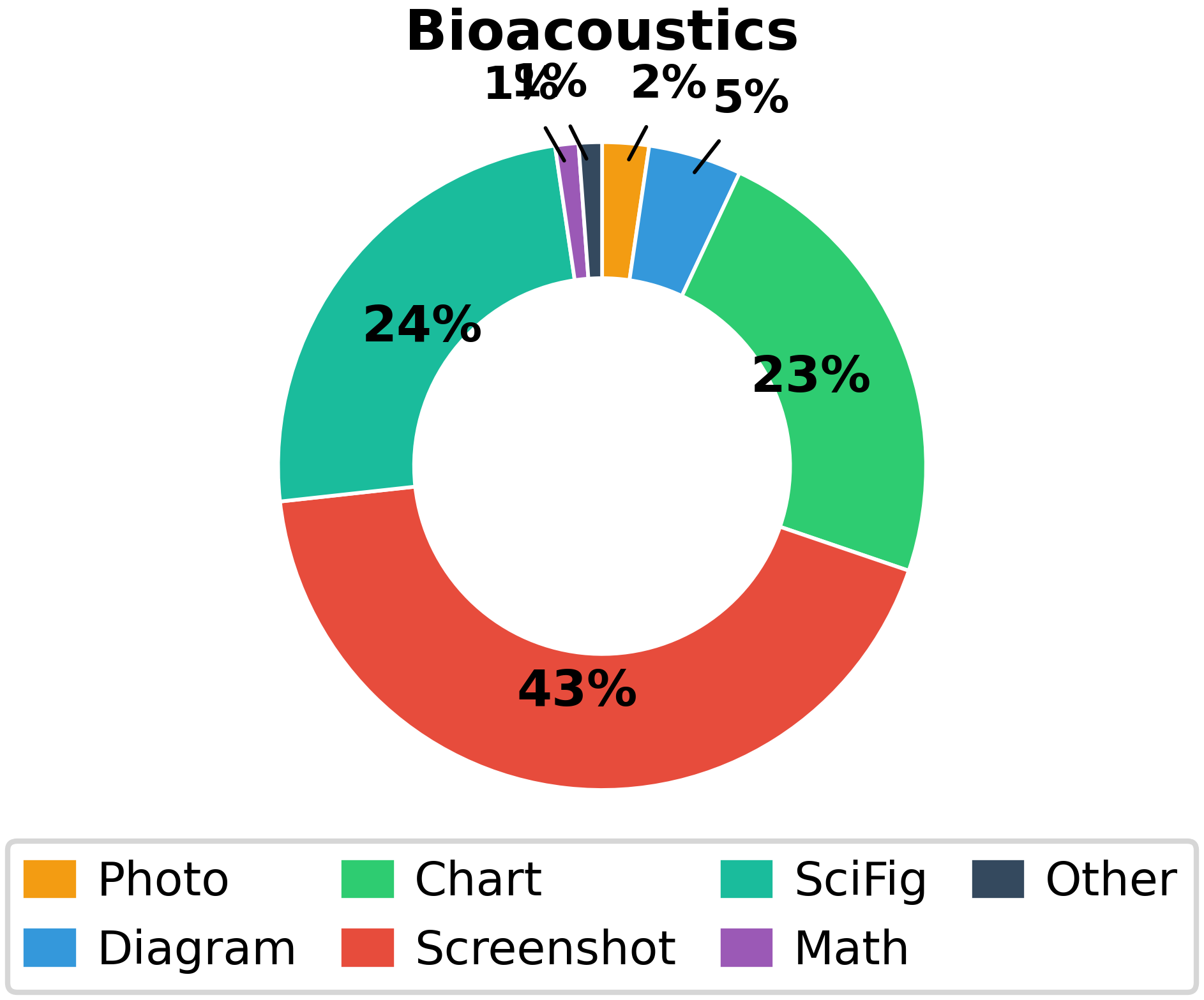}
    \caption{Bioacoustics}
  \end{subfigure}
  \hfill
  \begin{subfigure}[b]{0.32\textwidth}
    \includegraphics[width=\textwidth]{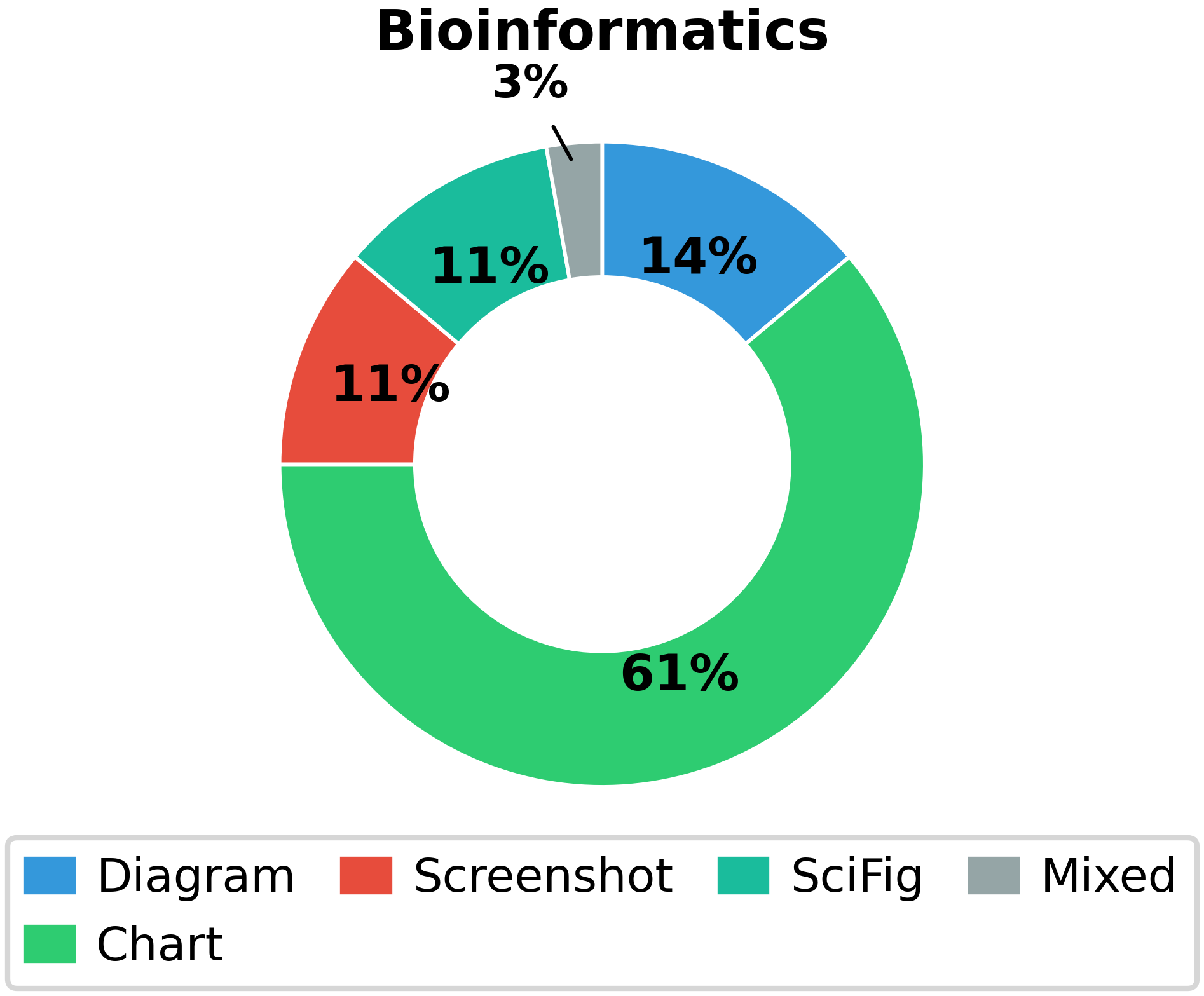}
    \caption{Bioinformatics}
  \end{subfigure}
  
  \vspace{0.5cm}
  
  \begin{subfigure}[b]{0.32\textwidth}
    \includegraphics[width=\textwidth]{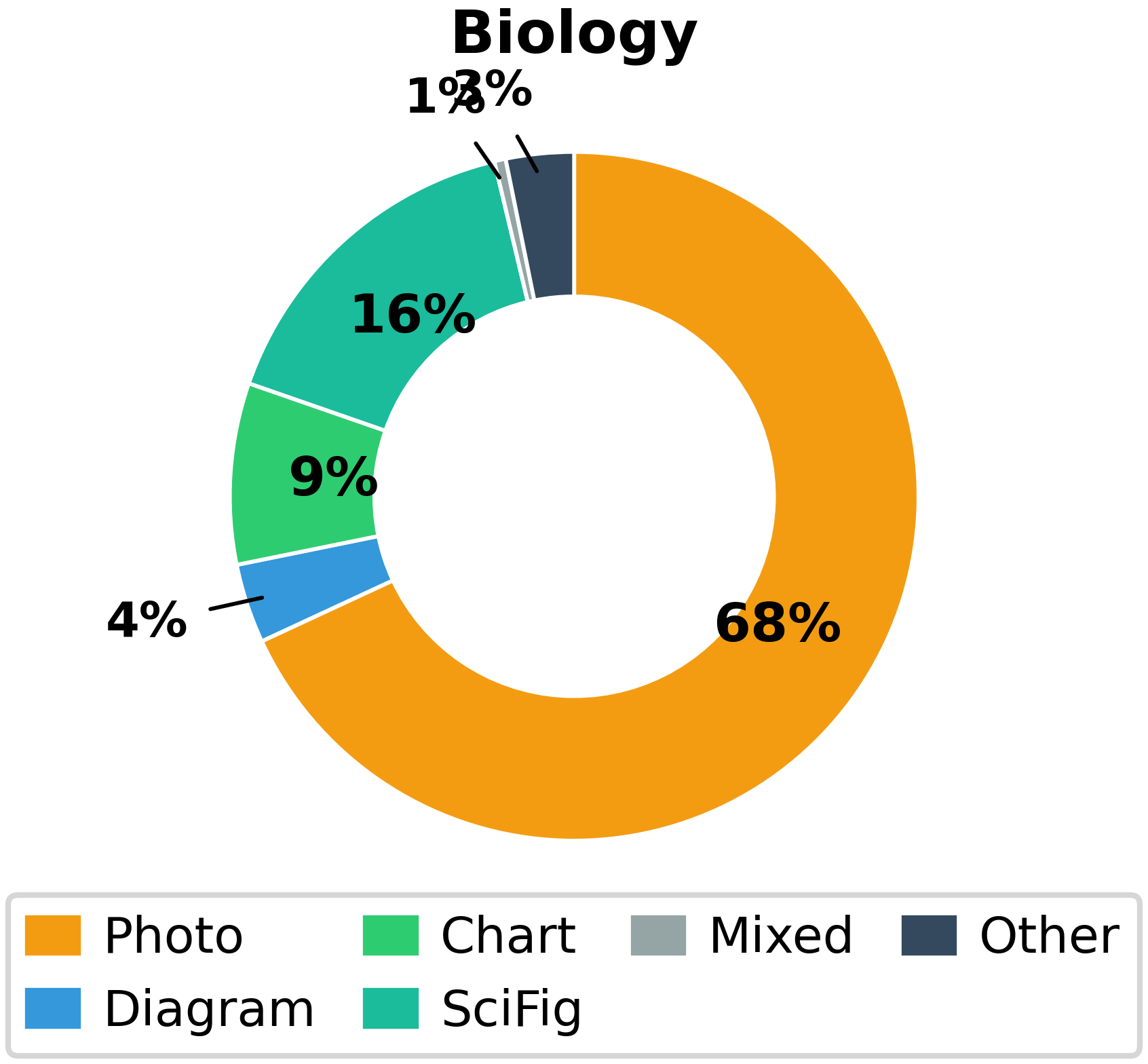}
    \caption{Biology}
  \end{subfigure}
  \hfill
  \begin{subfigure}[b]{0.32\textwidth}
    \includegraphics[width=\textwidth]{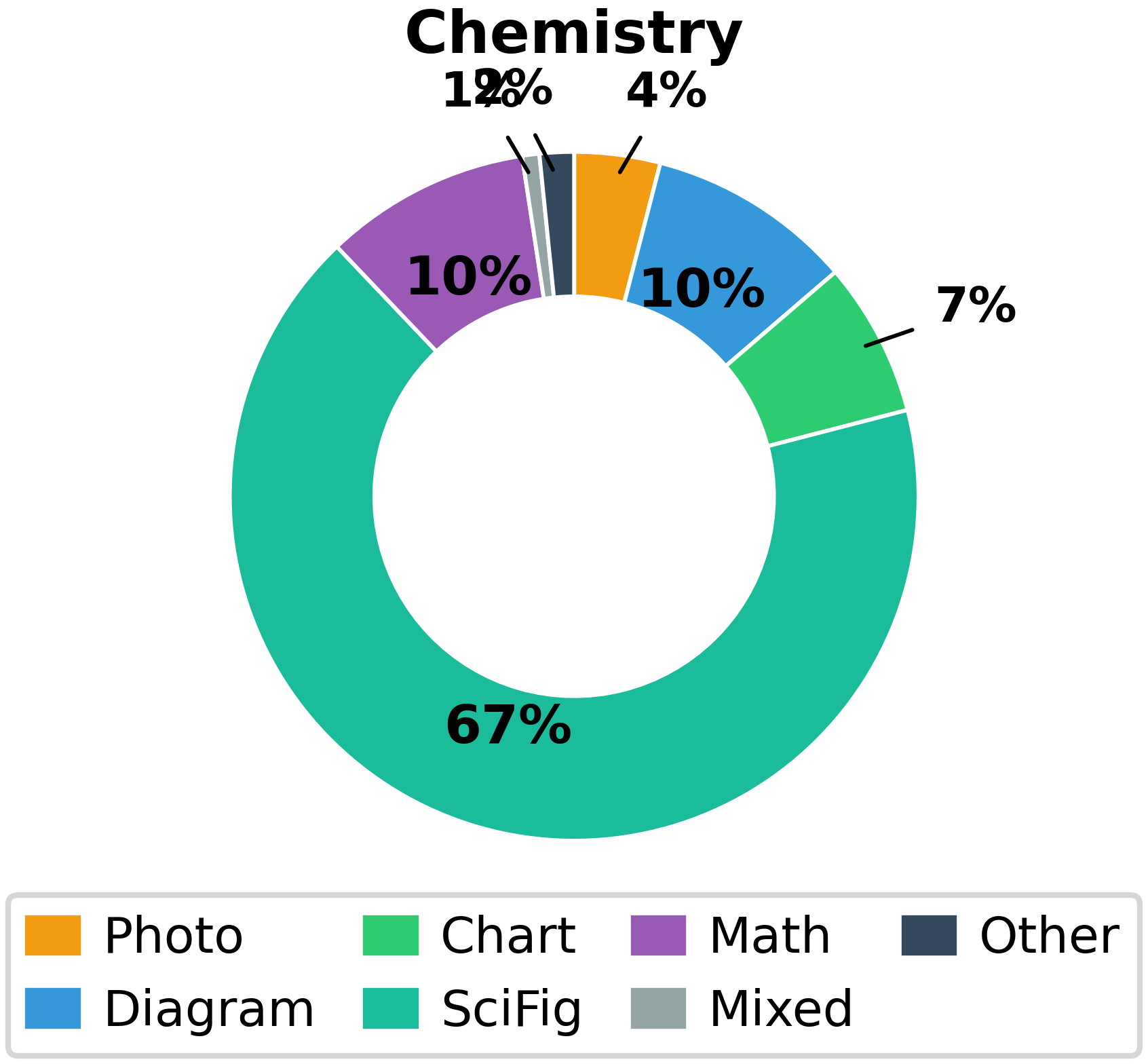}
    \caption{Chemistry}
  \end{subfigure}
  \hfill
  \begin{subfigure}[b]{0.32\textwidth}
    \includegraphics[width=\textwidth]{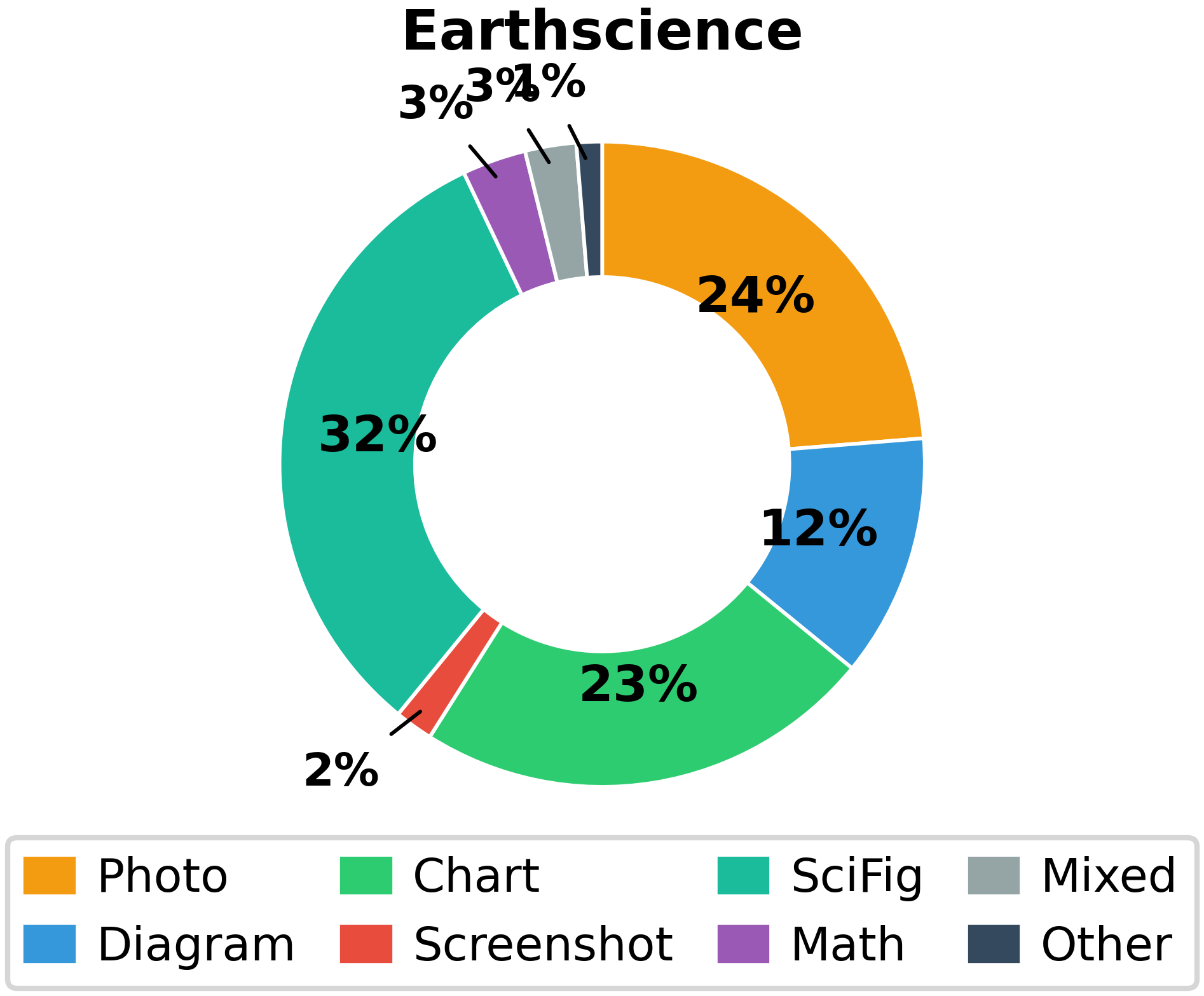}
    \caption{Earth Science}
  \end{subfigure}
  
  \vspace{0.5cm}
  
  \begin{subfigure}[b]{0.32\textwidth}
    \includegraphics[width=\textwidth]{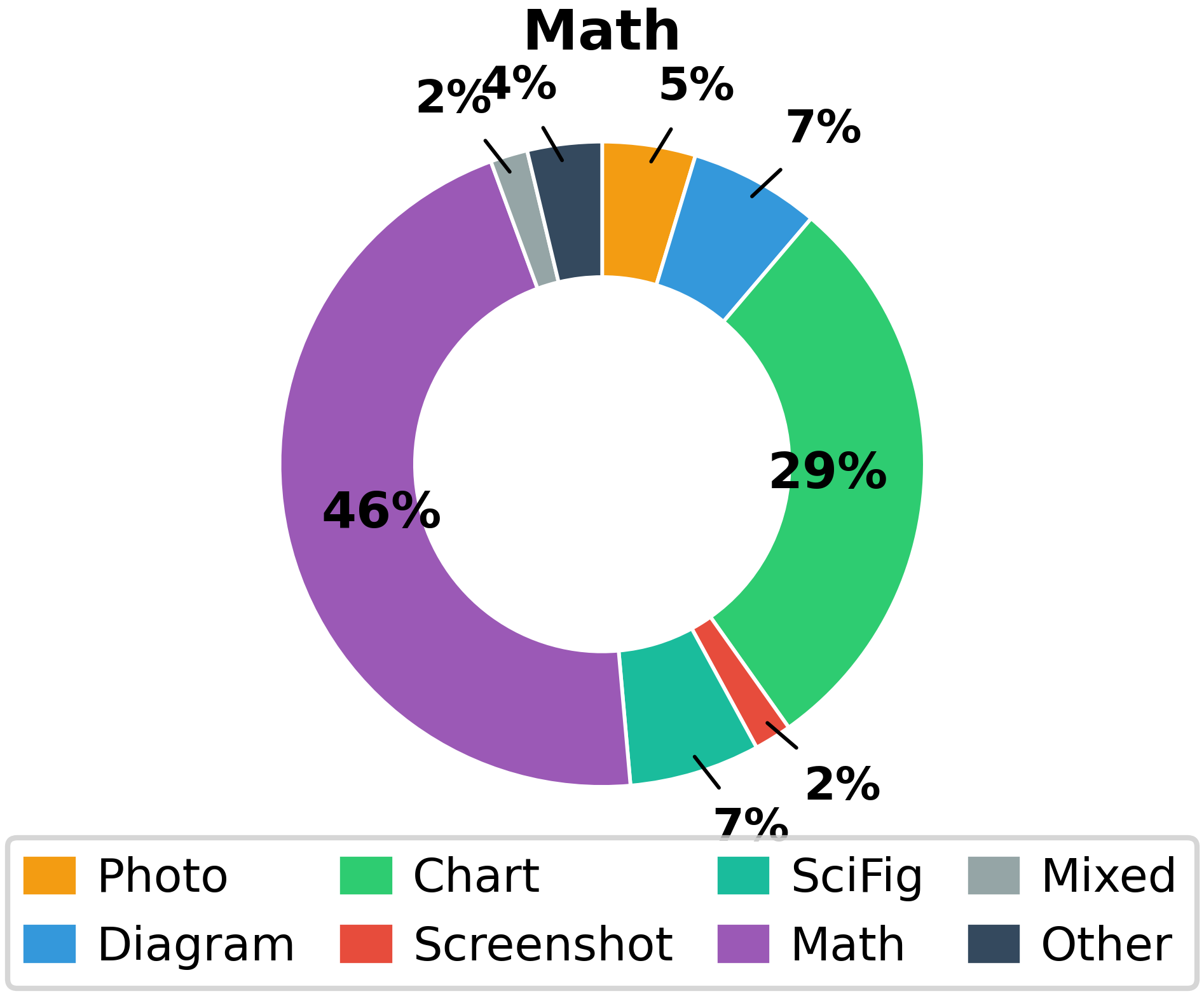}
    \caption{Mathematics}
  \end{subfigure}
  \hfill
  \begin{subfigure}[b]{0.32\textwidth}
    \includegraphics[width=\textwidth]{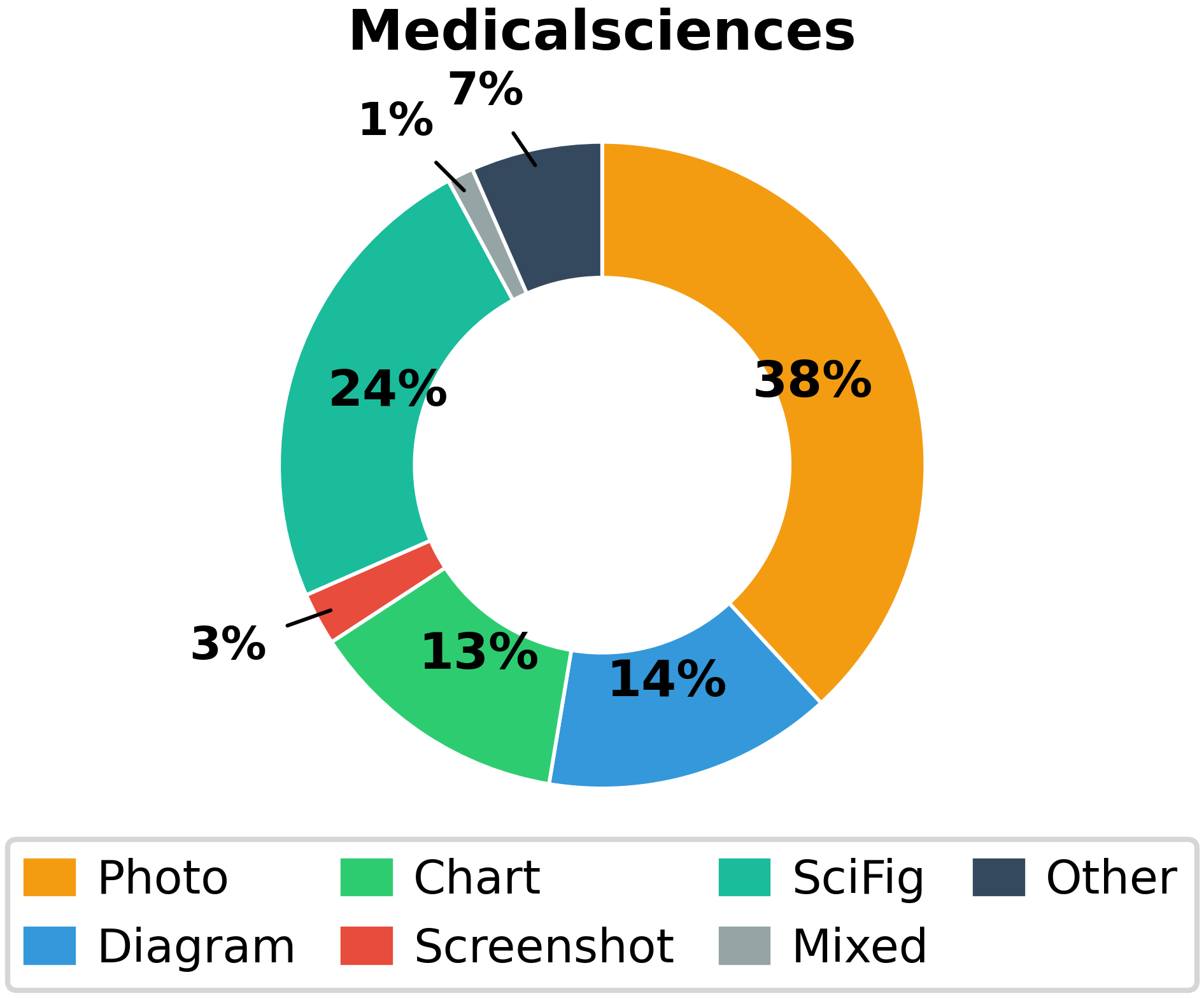}
    \caption{Medical Sciences}
  \end{subfigure}
  \hfill
  \begin{subfigure}[b]{0.32\textwidth}
    \includegraphics[width=\textwidth]{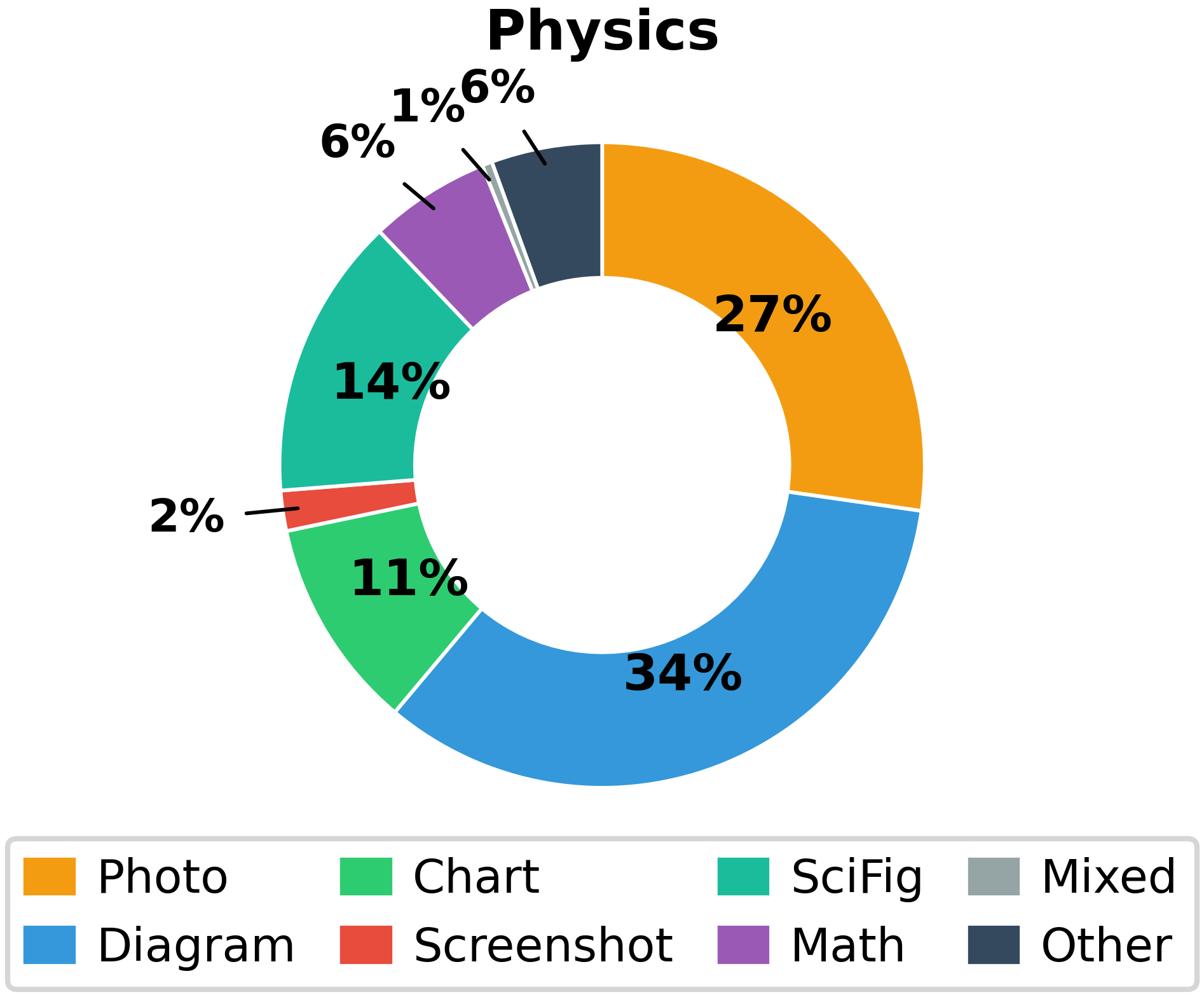}
    \caption{Physics}
  \end{subfigure}
  
  \caption{
  \textbf{Image type distribution for STEM \& Life Sciences domains.}
  STEM domains show high diversity in image types. 
  Biology (68.1\% photos) focuses on organism identification, 
  Chemistry (66.9\% scientific figures) contains molecular structures, 
  Physics (33.8\% diagrams, 27.3\% photos) balances theoretical schematics with experimental observations,
  and Mathematics (45.8\% mathematical notation) requires symbolic reasoning.
  Earth Science combines scientific figures (32.1\%), photos (23.7\%), and charts (23.1\%) for geological and atmospheric phenomena.
  }
  \label{fig:image_types_stem}
\end{figure*}

\begin{figure*}[p]
  \centering
  
  \begin{subfigure}[b]{0.32\textwidth}
    \includegraphics[width=\textwidth]{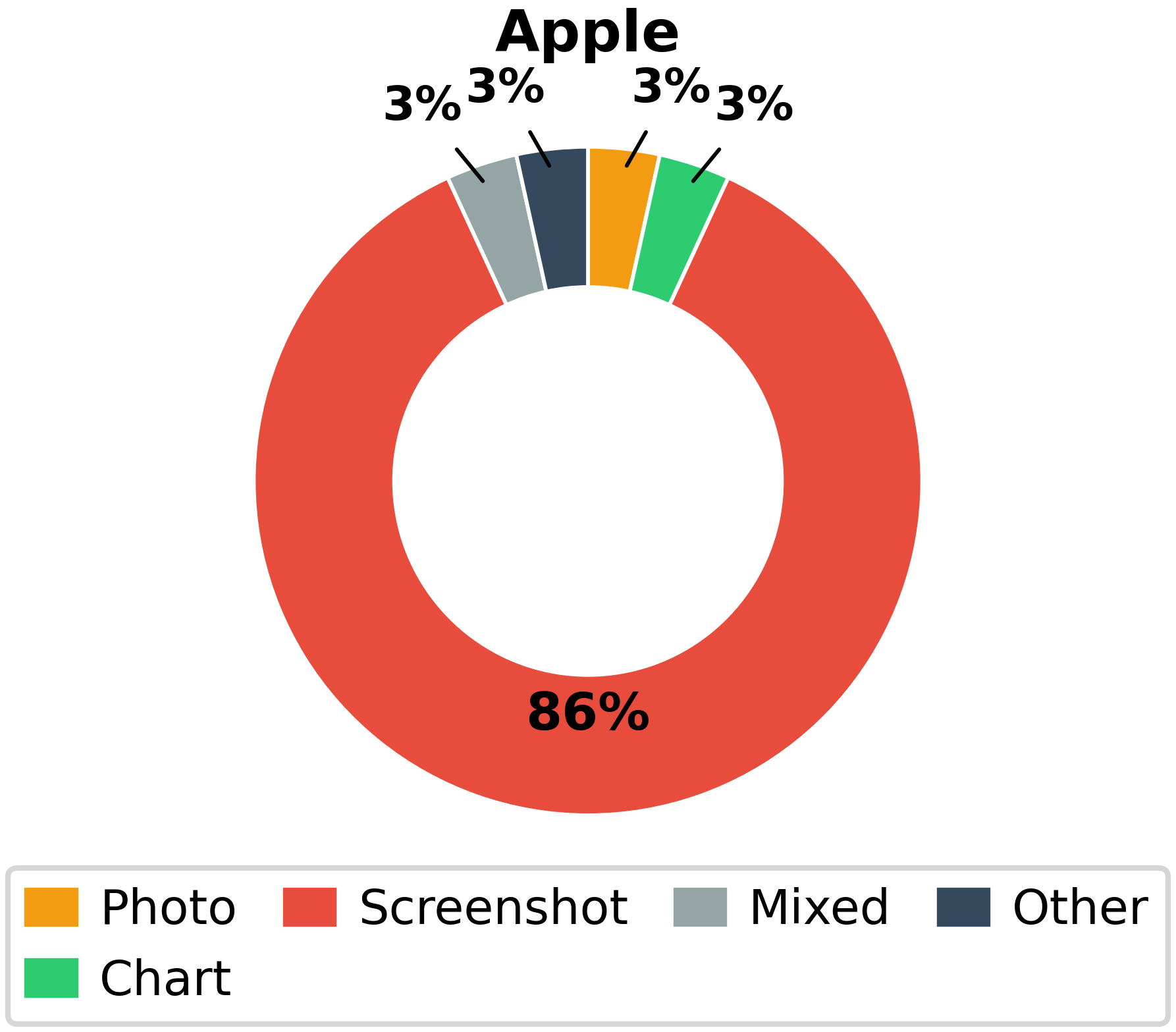}
    \caption{Apple}
  \end{subfigure}
  \hfill
  \begin{subfigure}[b]{0.32\textwidth}
    \includegraphics[width=\textwidth]{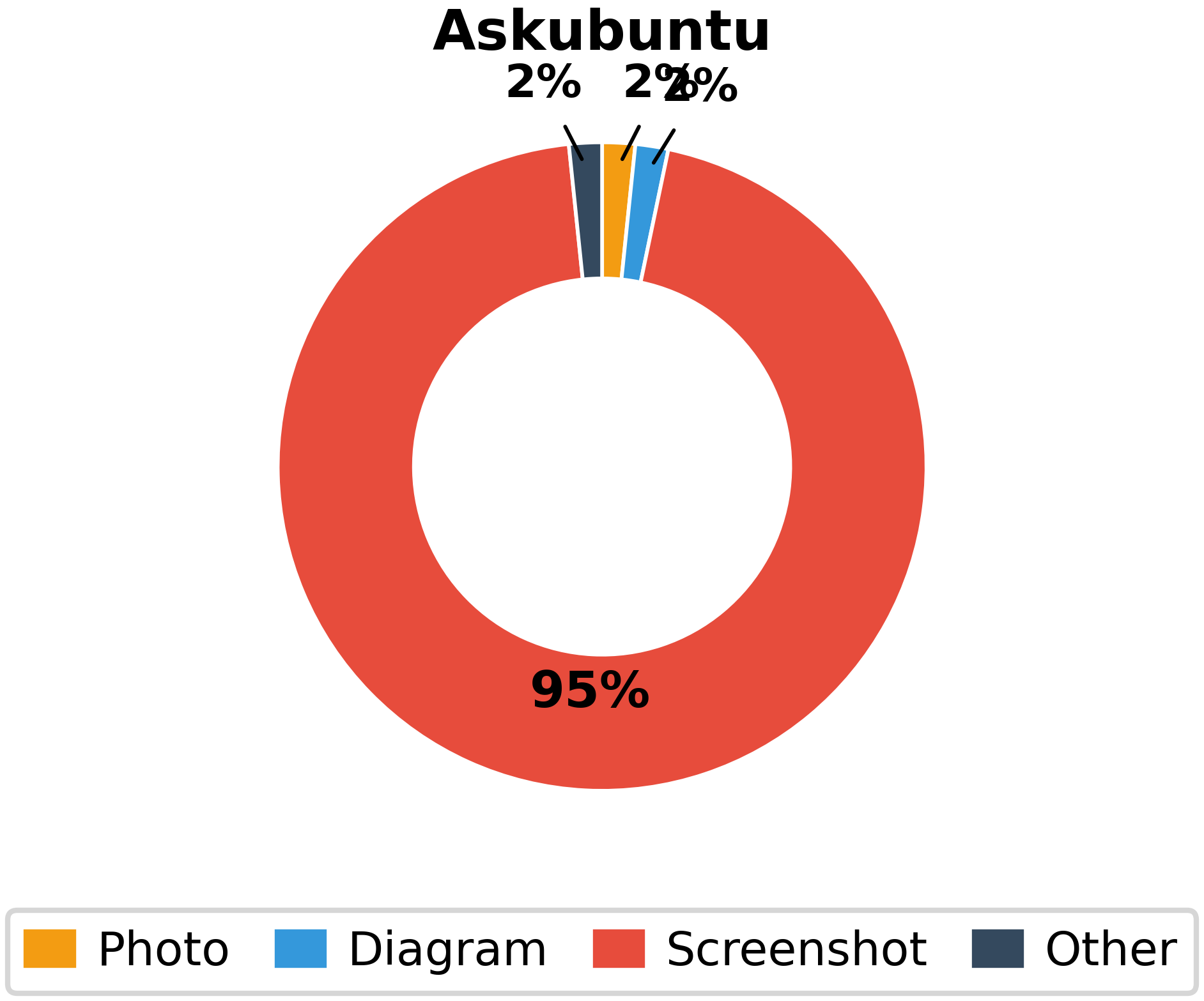}
    \caption{Ask Ubuntu}
  \end{subfigure}
  \hfill
  \begin{subfigure}[b]{0.32\textwidth}
    \includegraphics[width=\textwidth]{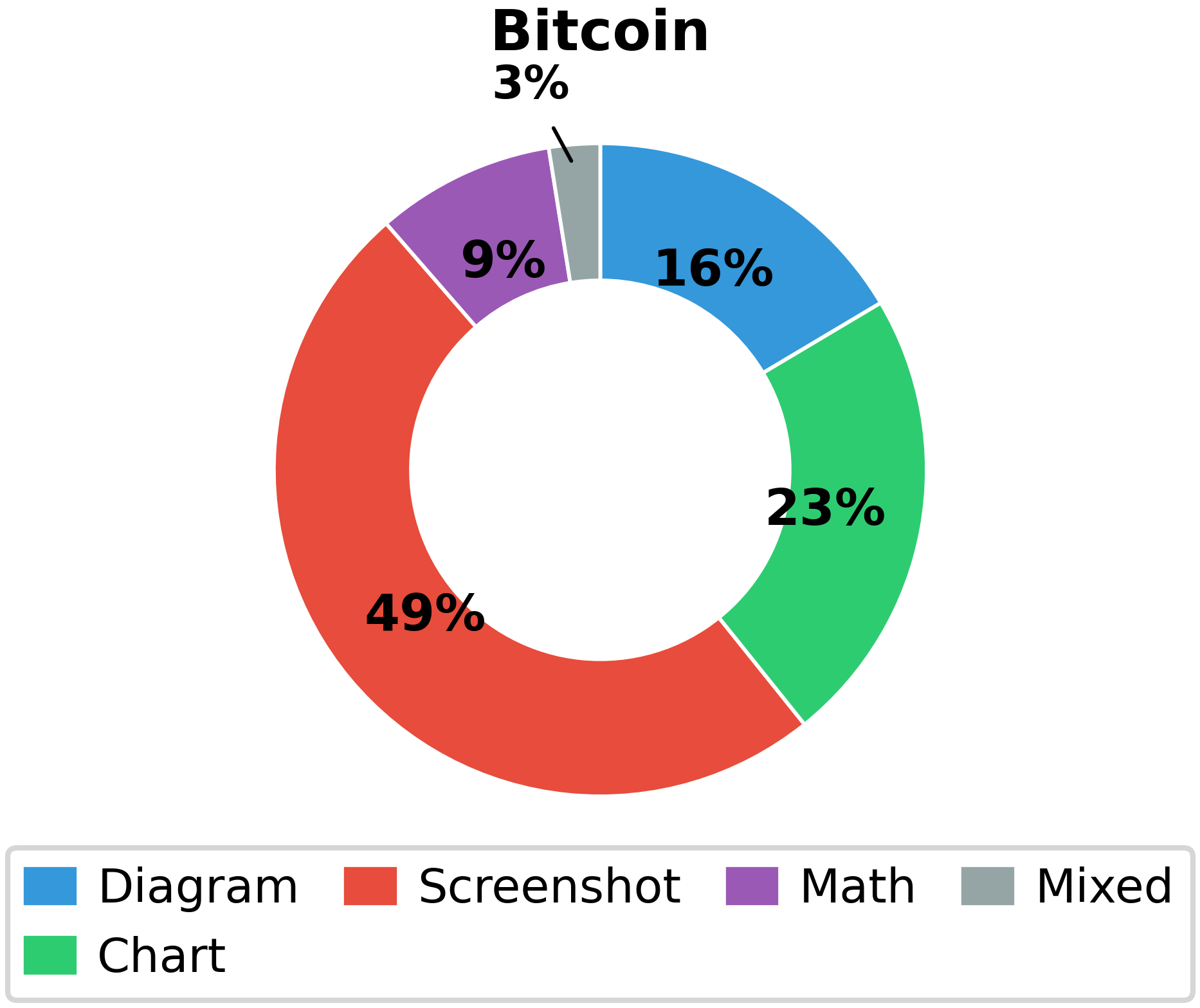}
    \caption{Bitcoin}
  \end{subfigure}
  
  \vspace{0.5cm}
  
  \begin{subfigure}[b]{0.32\textwidth}
    \includegraphics[width=\textwidth]{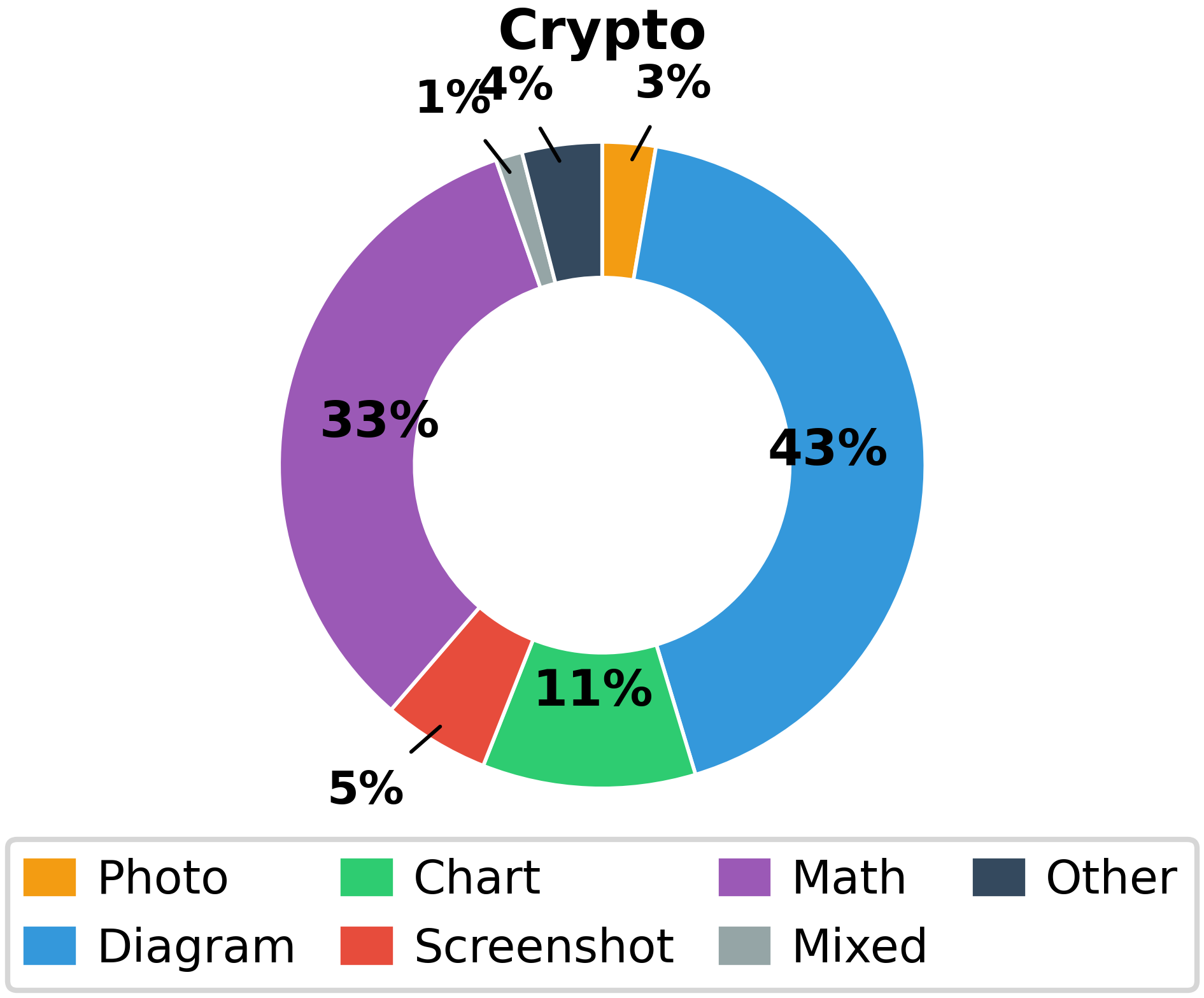}
    \caption{Cryptography}
  \end{subfigure}
  \hfill
  \begin{subfigure}[b]{0.32\textwidth}
    \includegraphics[width=\textwidth]{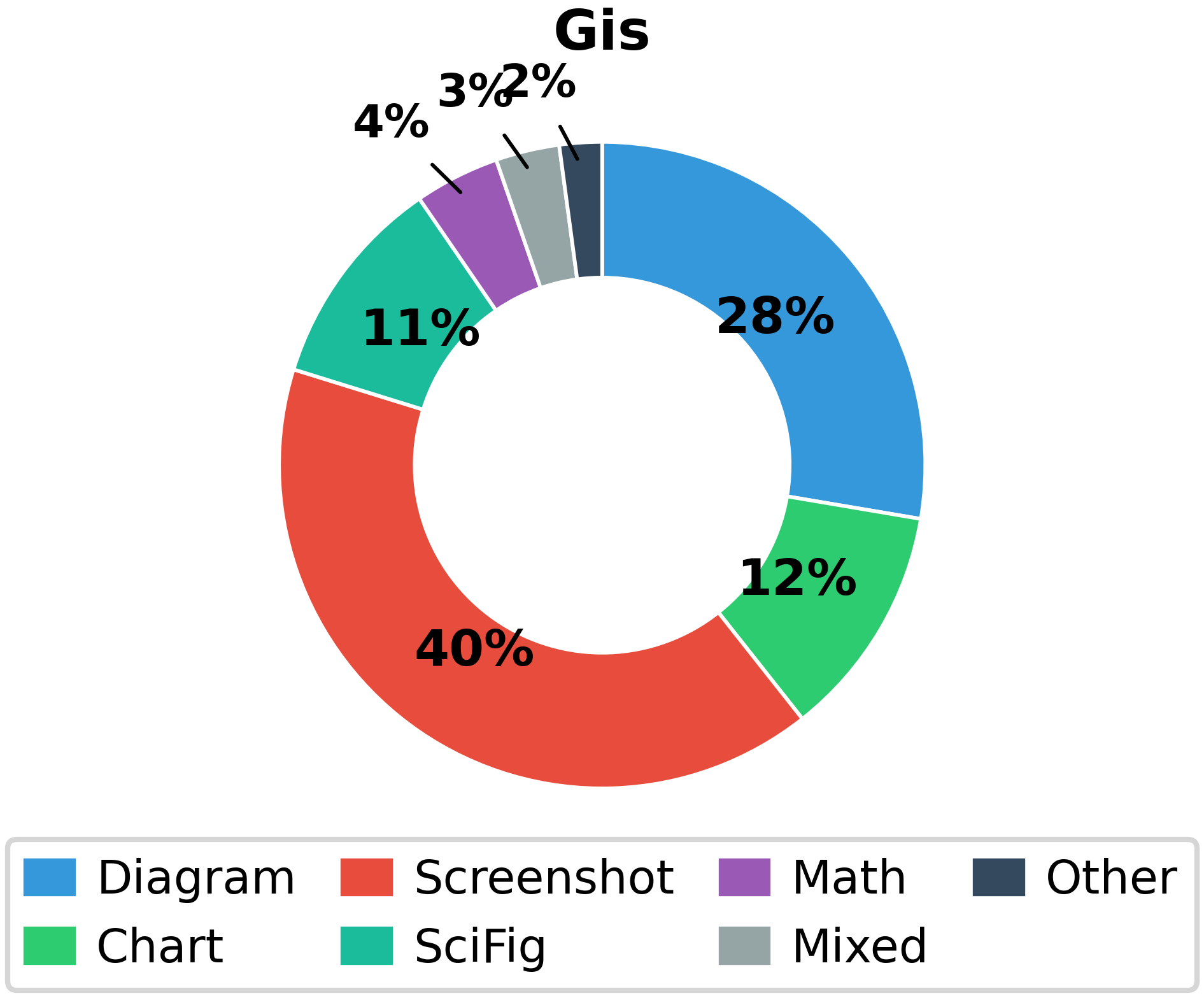}
    \caption{GIS}
  \end{subfigure}
  \hfill
  \begin{subfigure}[b]{0.32\textwidth}
    \includegraphics[width=\textwidth]{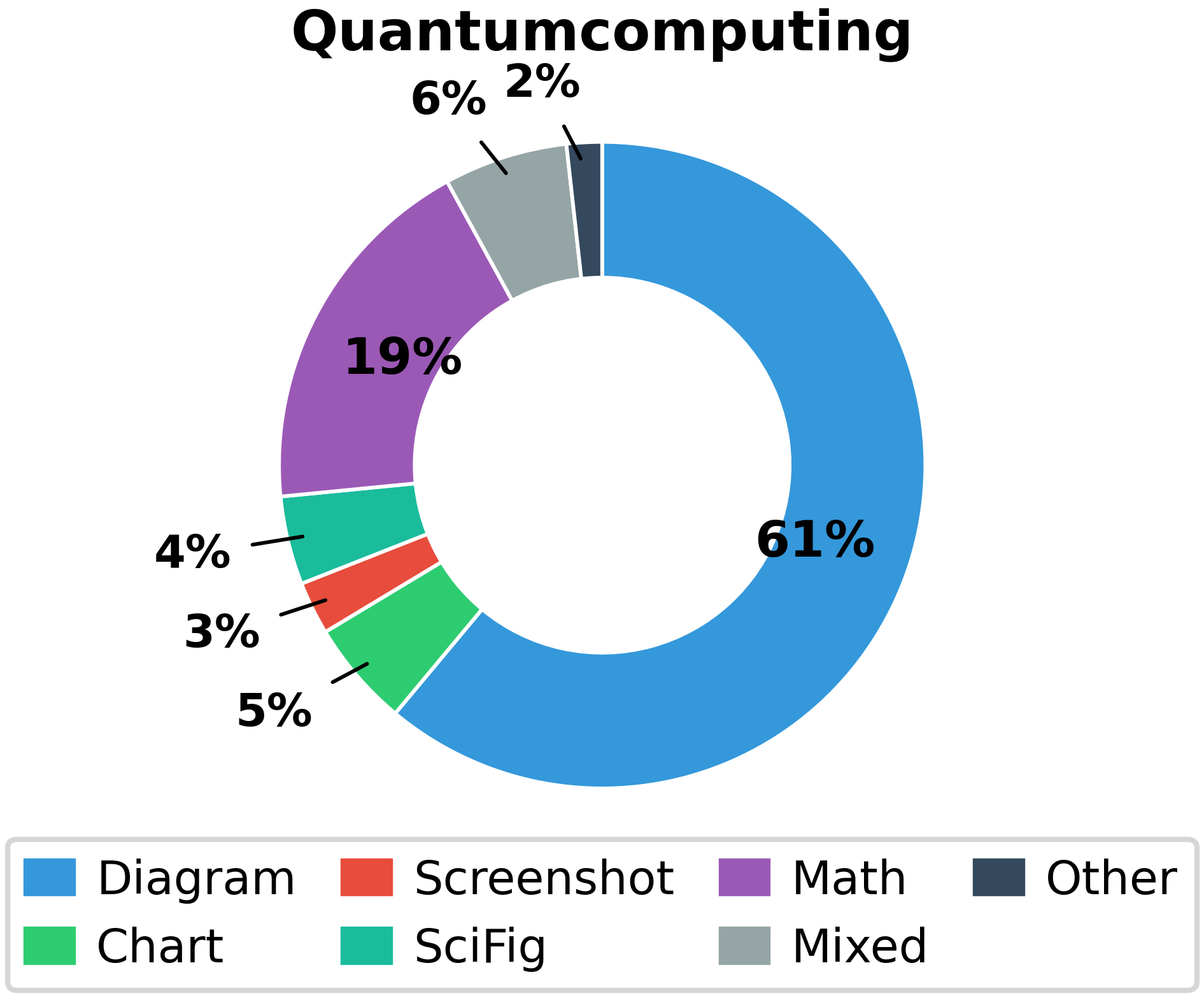}
    \caption{Quantum Computing}
  \end{subfigure}
  
  \vspace{0.5cm}
  
  \begin{subfigure}[b]{0.32\textwidth}
    \includegraphics[width=\textwidth]{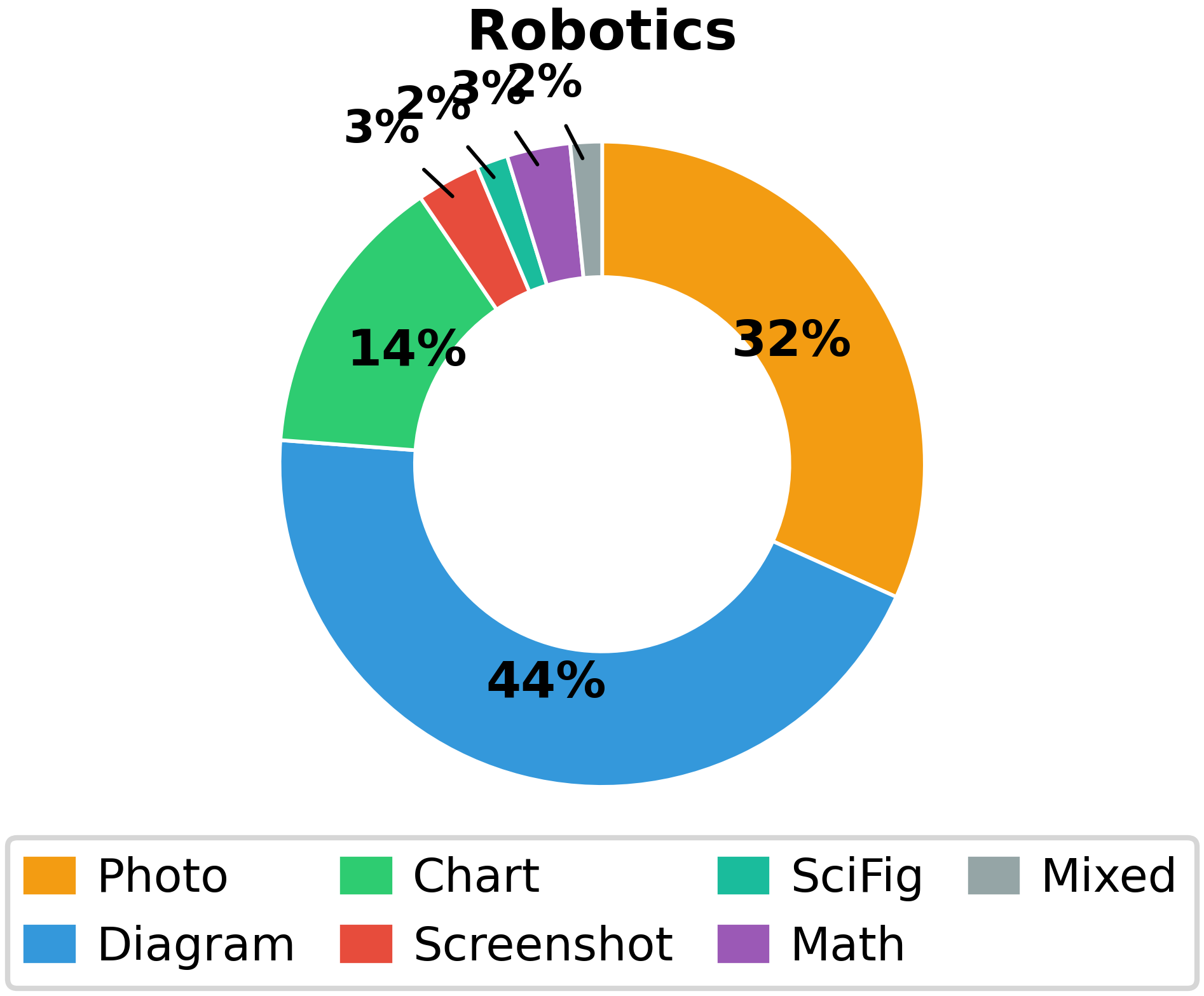}
    \caption{Robotics}
  \end{subfigure}
  \hfill
  \begin{subfigure}[b]{0.32\textwidth}
    \includegraphics[width=\textwidth]{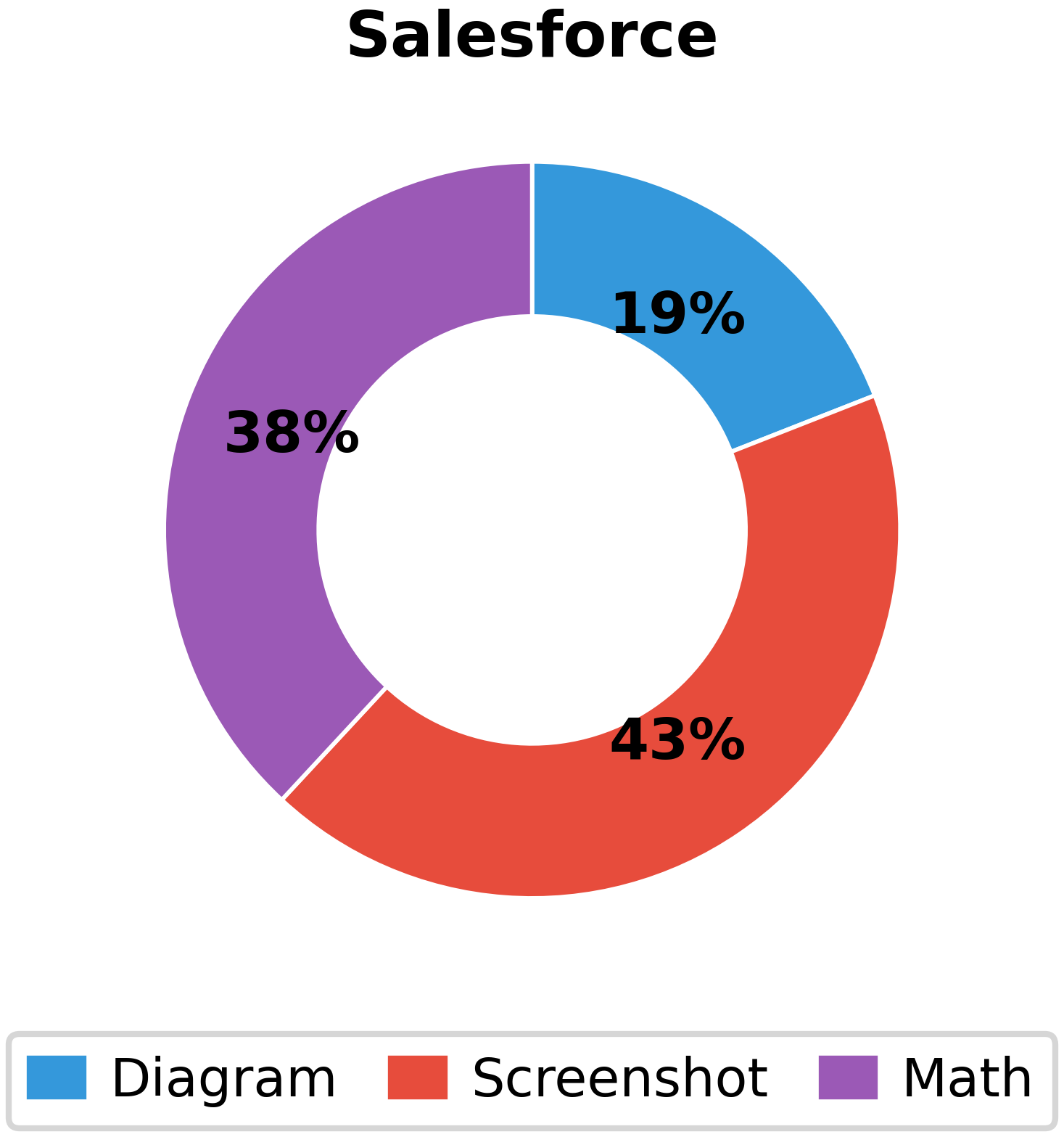}
    \caption{Salesforce}
  \end{subfigure}
  
  \caption{
  \textbf{Image type distribution for Software \& Technical Systems domains.}
  Computing domains are heavily dominated by screenshots and diagrams.
  Ask Ubuntu (95.1\% screenshots) and Apple (86.2\% screenshots) focus on debugging and configuration tasks.
  Quantum Computing (61.1\% diagrams) and Robotics (44.4\% diagrams) contain technical schematics.
  Cryptography balances diagrams (42.7\%) with mathematical notation (33.3\%).
  GIS combines screenshots (40.4\%) with diagrams (27.7\%) for geospatial analysis.
  }
  \label{fig:image_types_computing}
\end{figure*}

\begin{figure*}[p]
  \centering
  
  \begin{subfigure}[b]{0.32\textwidth}
    \includegraphics[width=\textwidth]{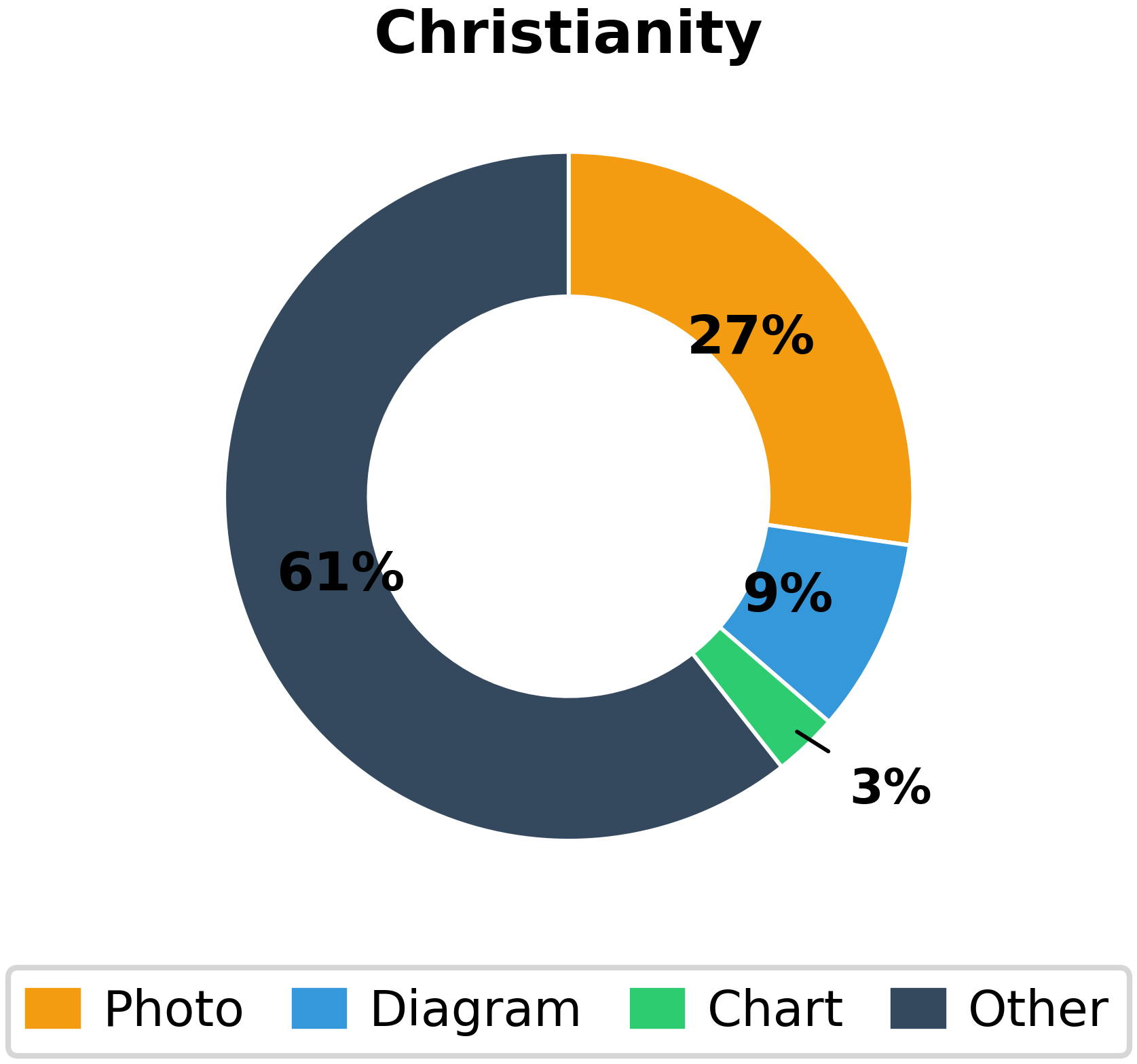}
    \caption{Christianity}
  \end{subfigure}
  \hfill
  \begin{subfigure}[b]{0.32\textwidth}
    \includegraphics[width=\textwidth]{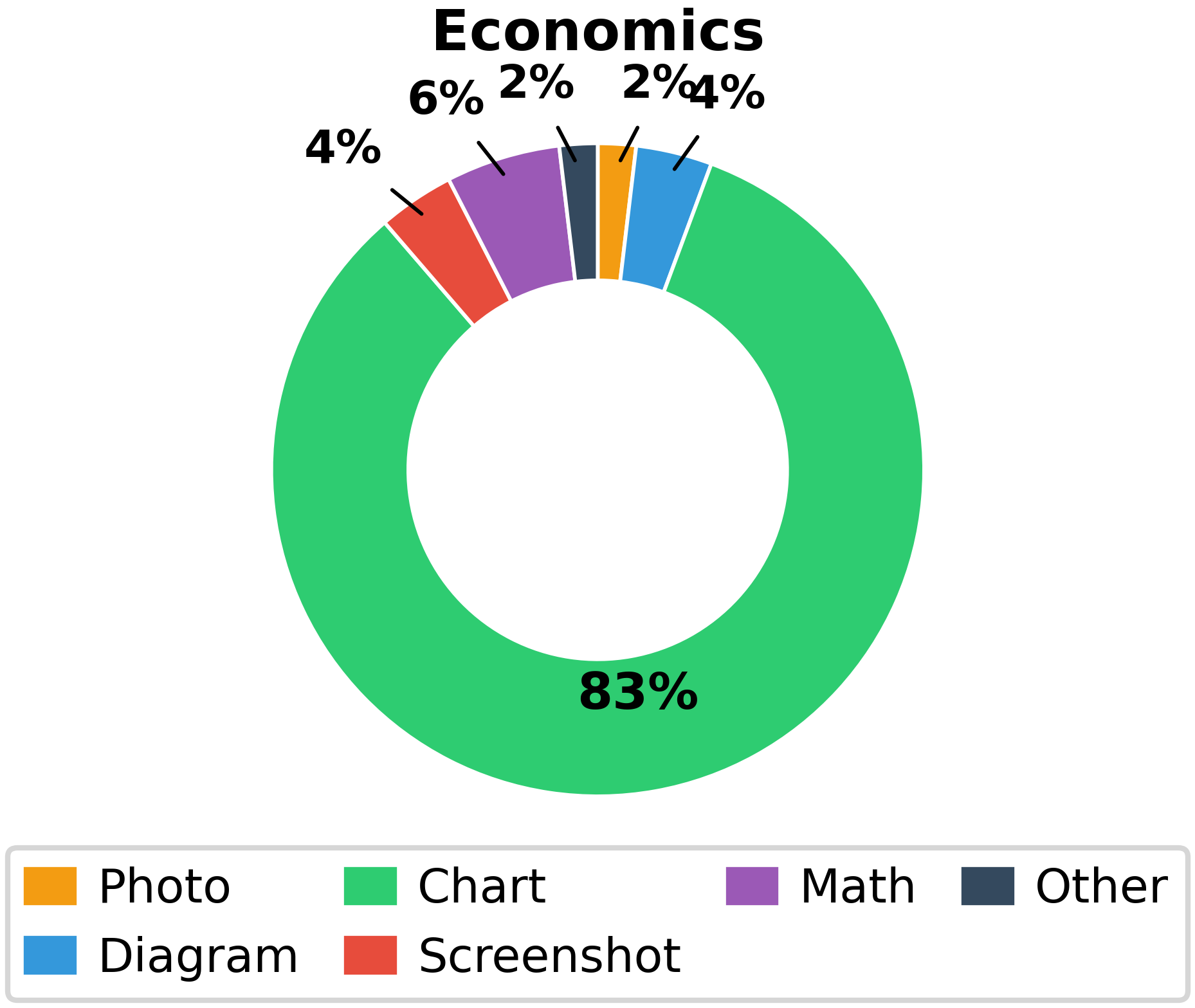}
    \caption{Economics}
  \end{subfigure}
  \hfill
  \begin{subfigure}[b]{0.32\textwidth}
    \includegraphics[width=\textwidth]{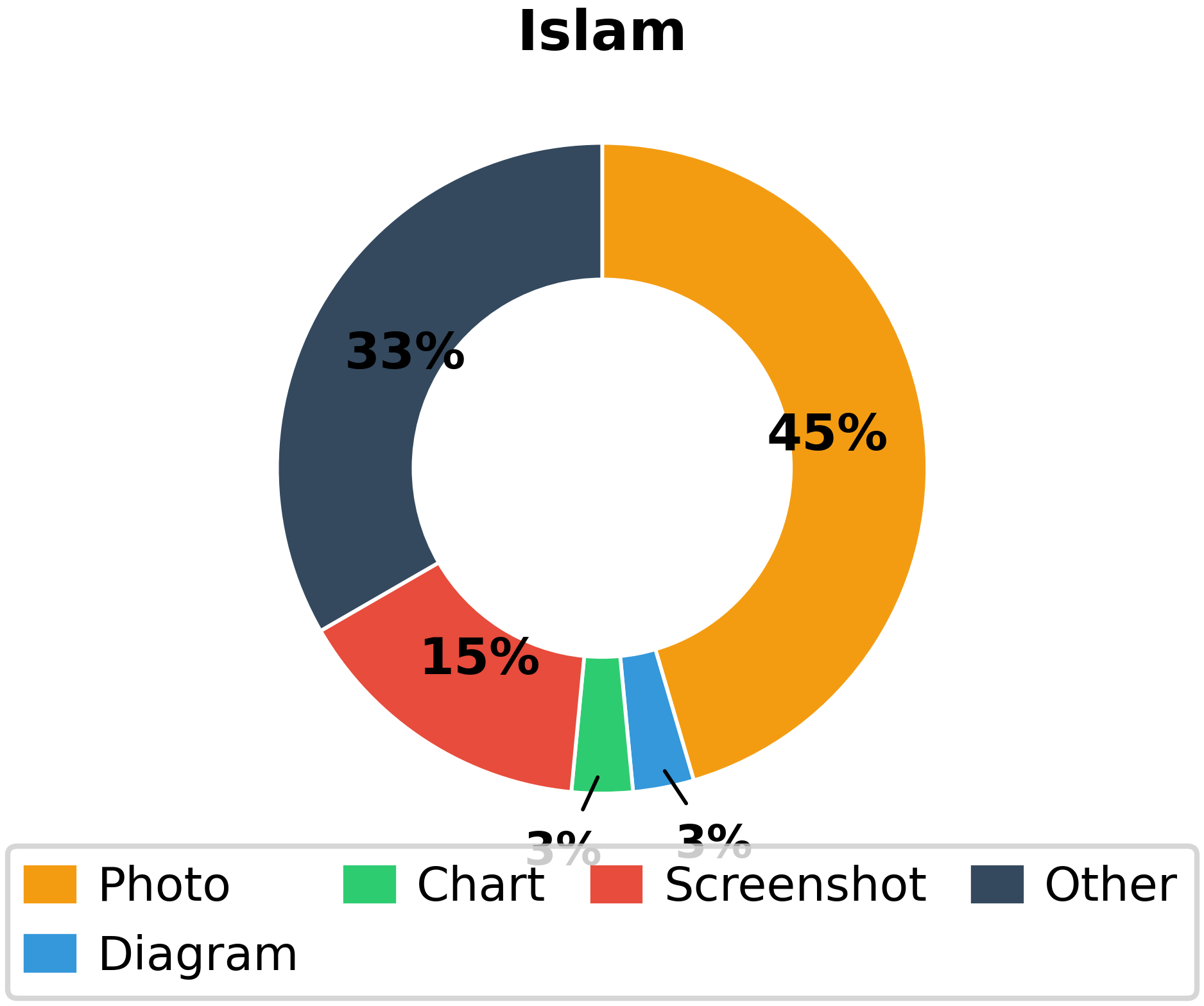}
    \caption{Islam}
  \end{subfigure}
  
  \vspace{0.5cm}
  
  \begin{subfigure}[b]{0.32\textwidth}
    \includegraphics[width=\textwidth]{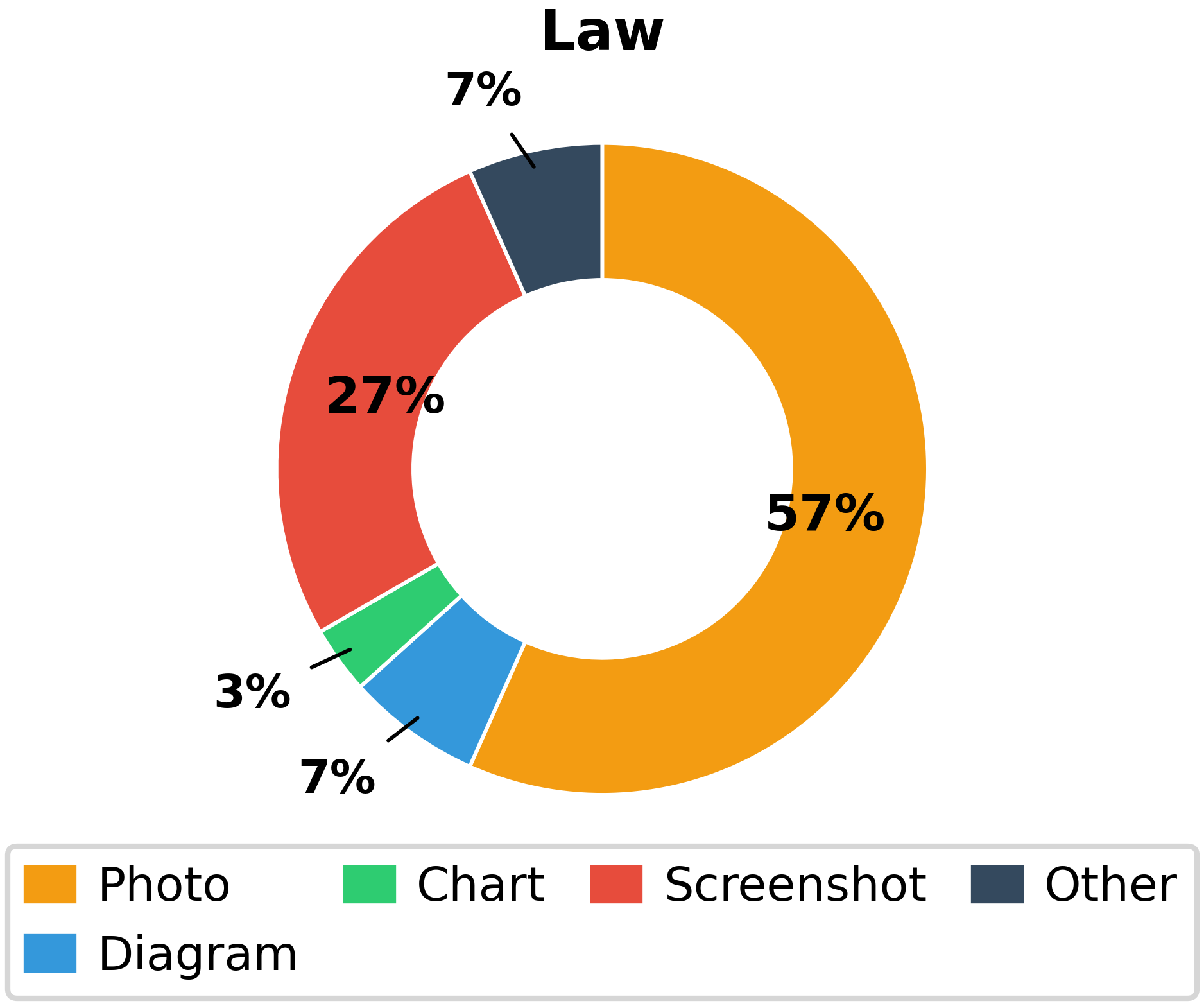}
    \caption{Law}
  \end{subfigure}
  \hfill
  \begin{subfigure}[b]{0.32\textwidth}
    \includegraphics[width=\textwidth]{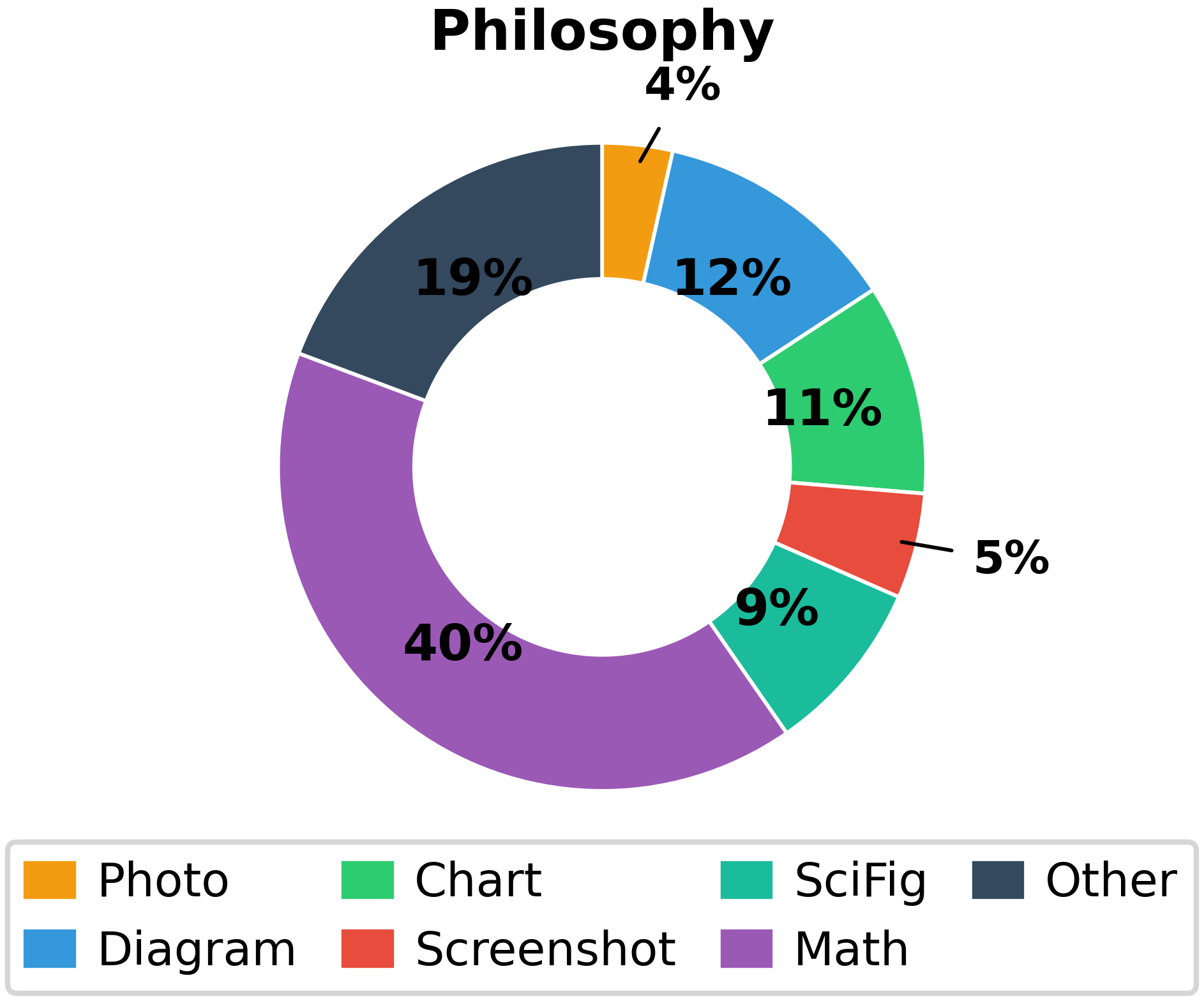}
    \caption{Philosophy}
  \end{subfigure}
  \hfill
  \begin{subfigure}[b]{0.32\textwidth}
    \includegraphics[width=\textwidth]{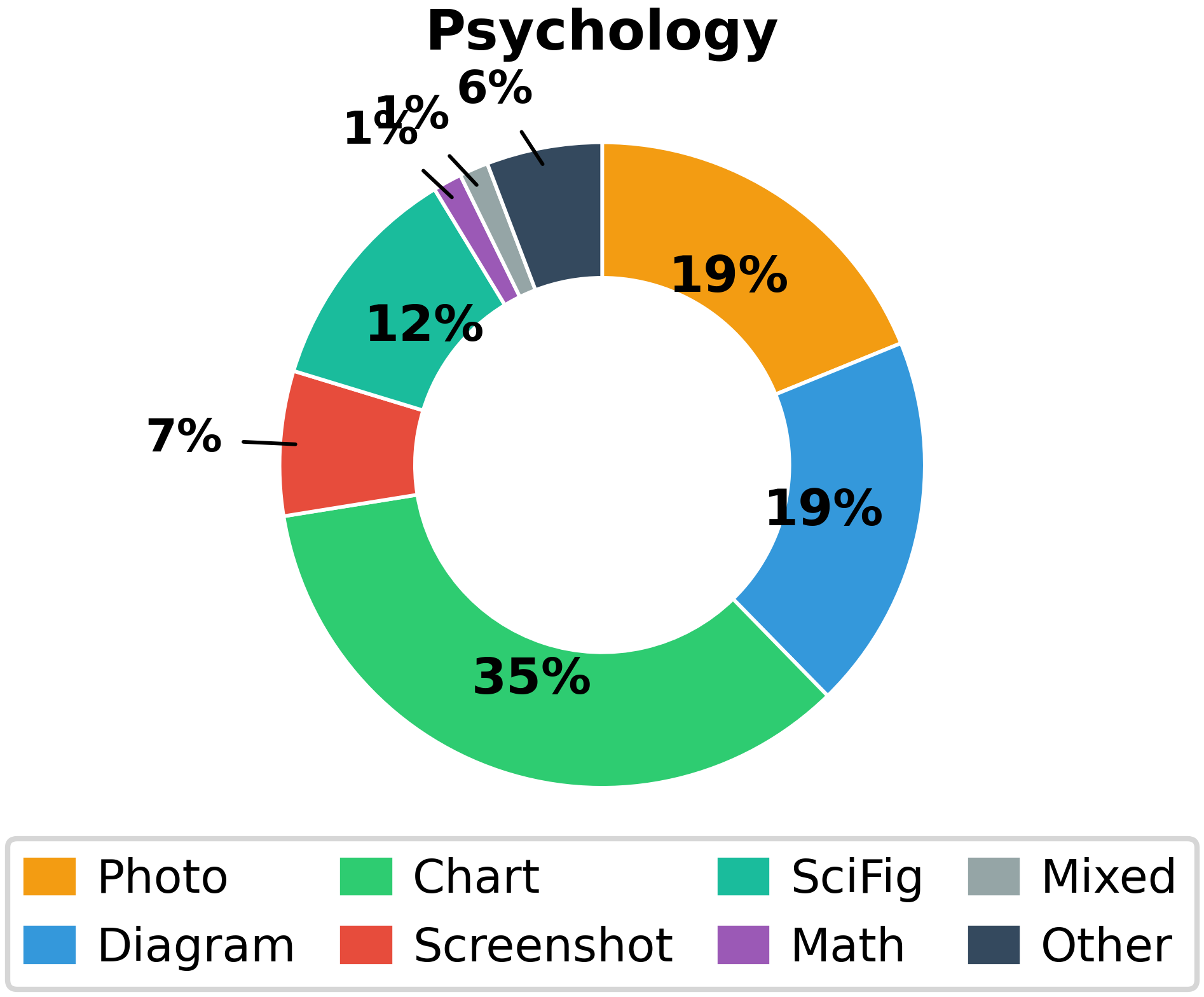}
    \caption{Psychology}
  \end{subfigure}
  
  \caption{
  \textbf{Image type distribution for Social Sciences \& Humanities domains.}
  Social science domains show varied patterns.
  Economics (83.0\% charts) and Psychology (34.8\% charts) focus heavily on data interpretation.
  Philosophy has high mathematical notation (40.4\%) for logical formalisms.
  Law (56.7\% photos) and Christianity (60.6\% other) contain illustrative images and religious artwork.
  Islam balances photos (45.5\%) with other content (33.3\%).
  }
  \label{fig:image_types_social}
\end{figure*}

\begin{figure*}[p]
  \centering
  
  \begin{subfigure}[b]{0.32\textwidth}
    \includegraphics[width=\textwidth]{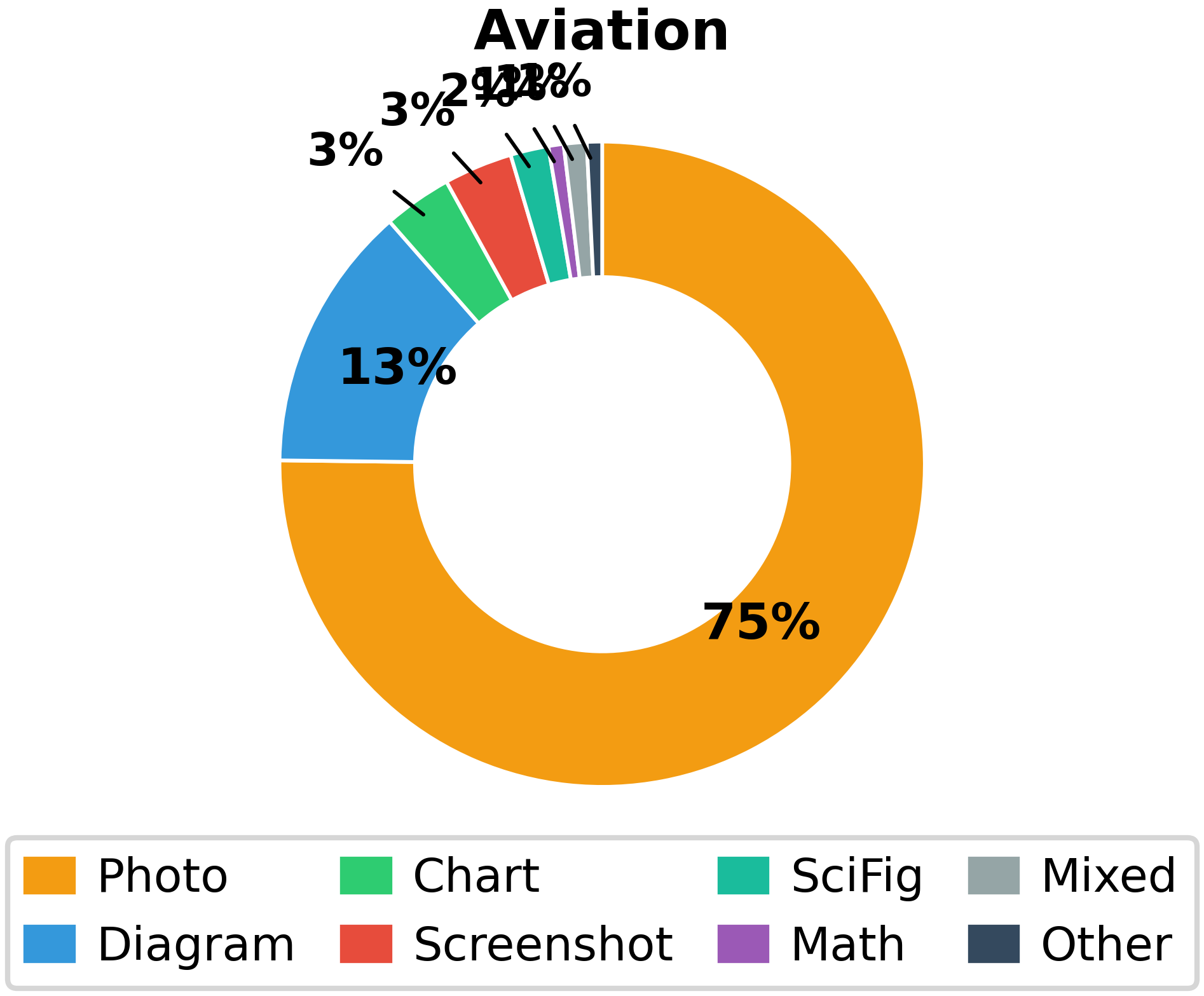}
    \caption{Aviation}
  \end{subfigure}
  \hfill
  \begin{subfigure}[b]{0.32\textwidth}
    \includegraphics[width=\textwidth]{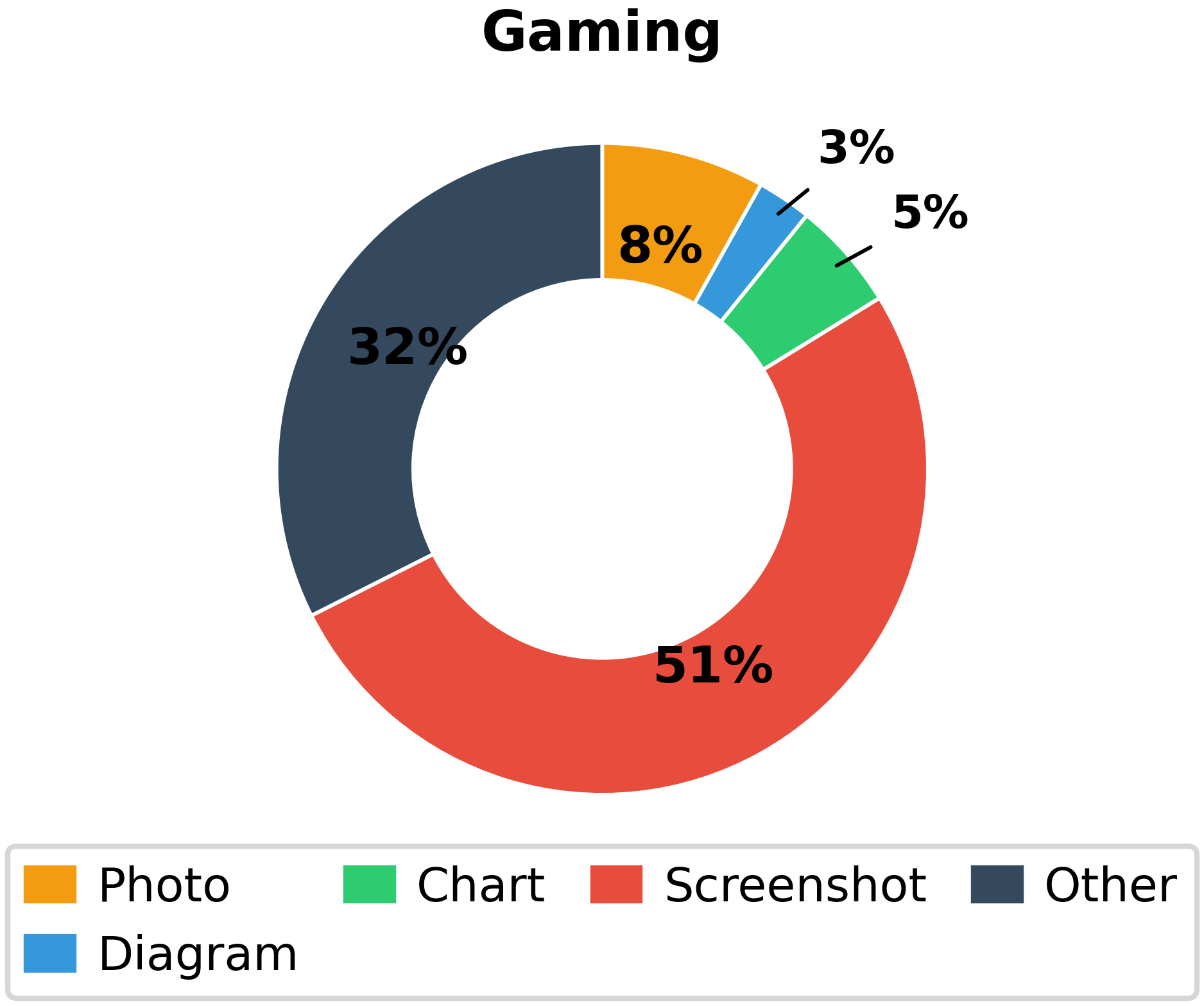}
    \caption{Gaming}
  \end{subfigure}
  \hfill
  \begin{subfigure}[b]{0.32\textwidth}
    \includegraphics[width=\textwidth]{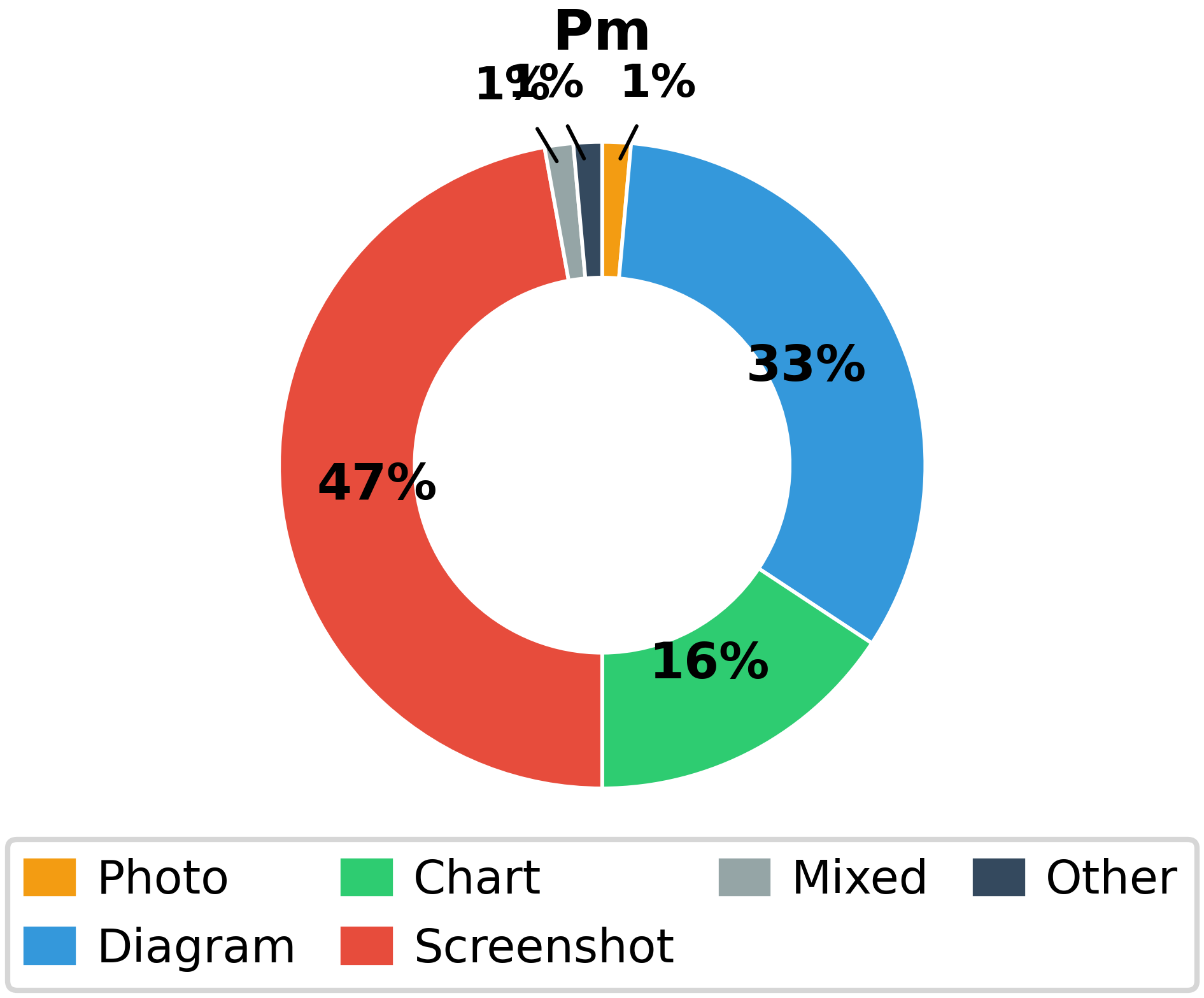}
    \caption{Project Management}
  \end{subfigure}
  
  \vspace{0.5cm}
  
  \begin{subfigure}[b]{0.32\textwidth}
    \includegraphics[width=\textwidth]{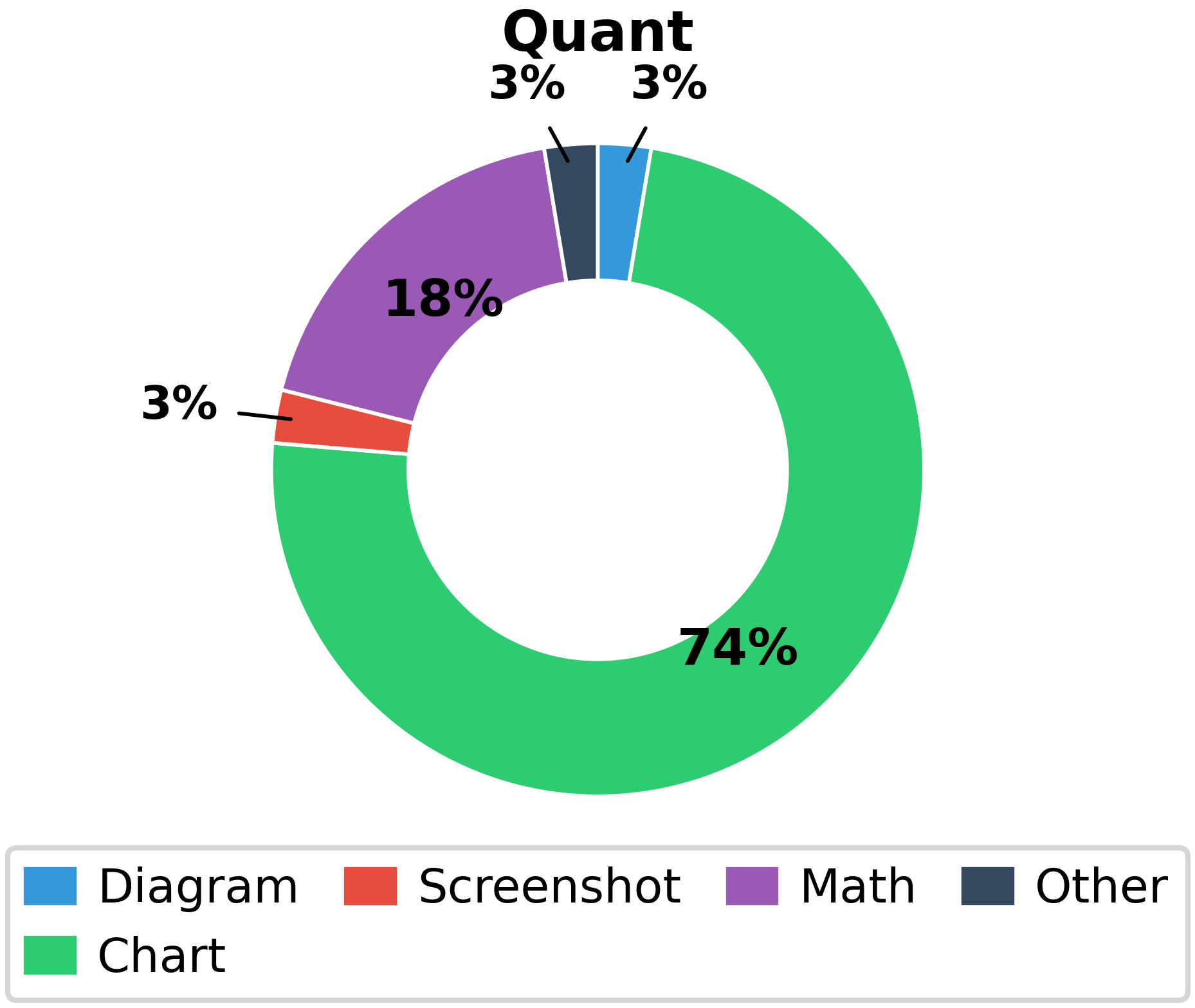}
    \caption{Quantitative Finance}
  \end{subfigure}
  \hfill
  \begin{subfigure}[b]{0.32\textwidth}
    \includegraphics[width=\textwidth]{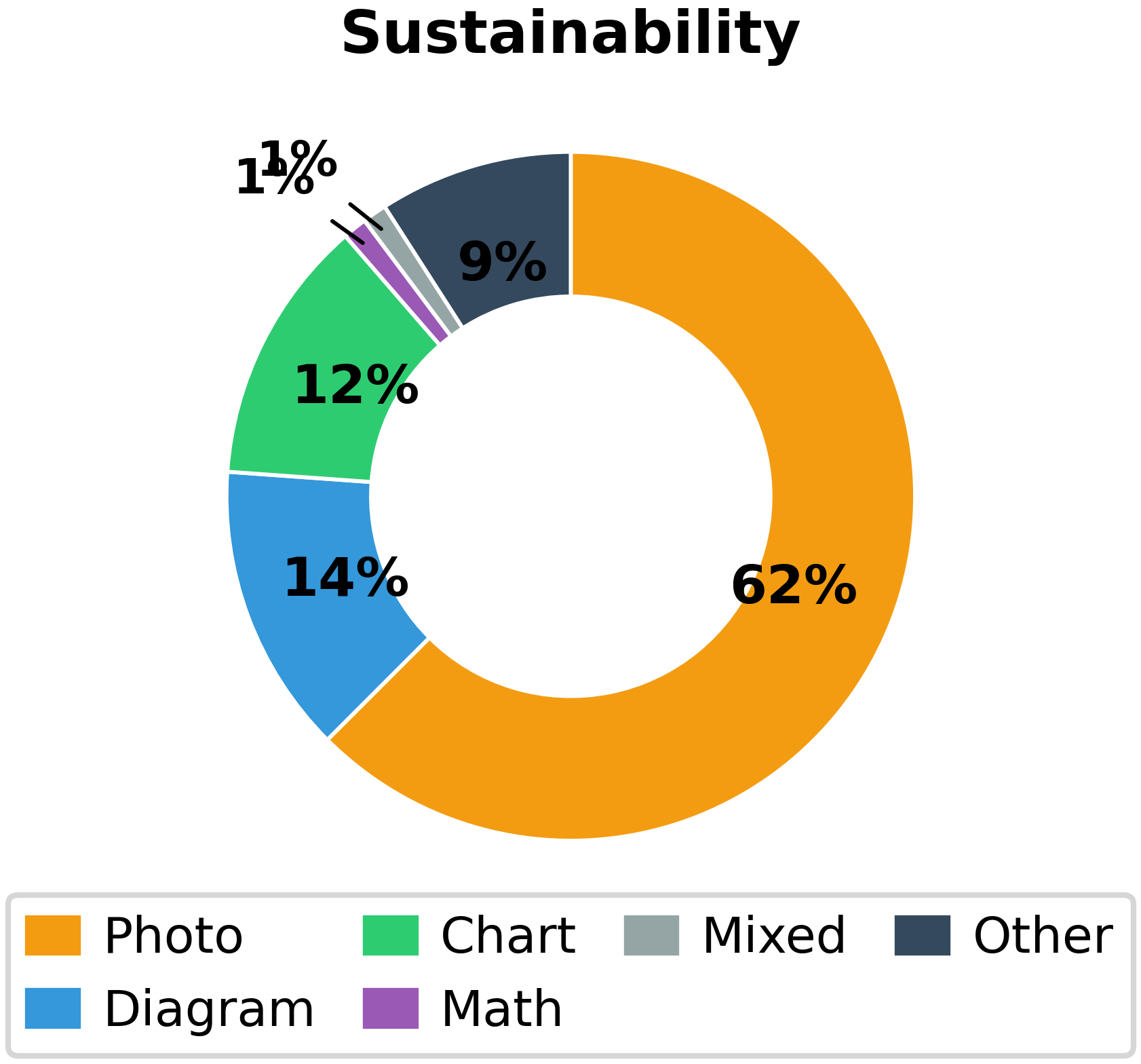}
    \caption{Sustainability}
  \end{subfigure}
  \hfill
  \begin{subfigure}[b]{0.32\textwidth}
    \includegraphics[width=\textwidth]{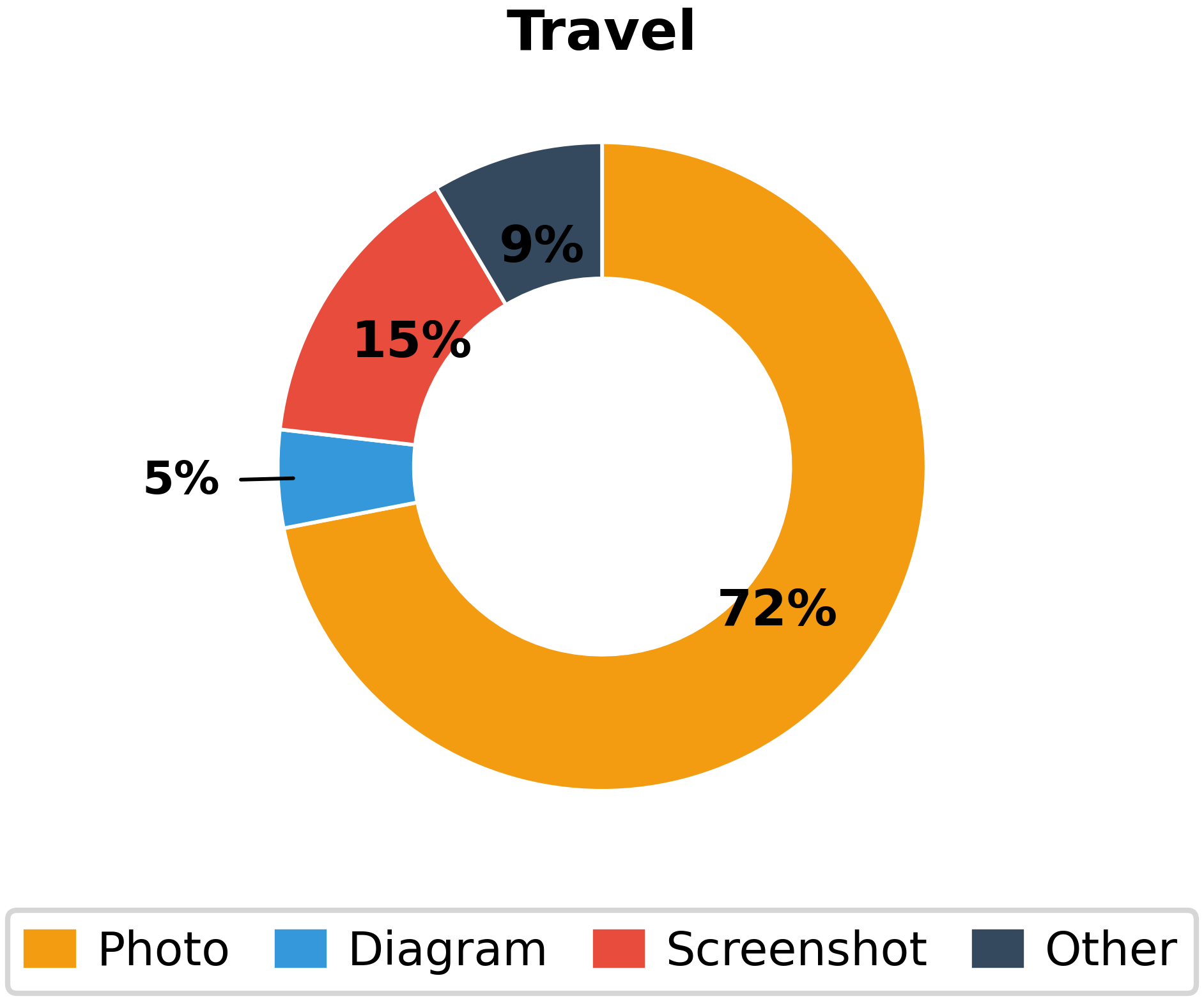}
    \caption{Travel}
  \end{subfigure}
  
  \caption{
  \textbf{Image type distribution for Applied domains.}
  Applied domains reflect their practical focus.
  Aviation (75.2\% photos) focuses on aircraft and equipment identification.
  Travel (72.0\% photos) contains location and landmark images.
  Sustainability (62.5\% photos) shows environmental and ecological subjects.
  Quantitative Finance (73.7\% charts) emphasizes financial data visualization.
  Project Management balances screenshots (47.1\%) with diagrams (32.9\%).
  Gaming is dominated by screenshots (51.4\%).
  }
  \label{fig:image_types_applied}
\end{figure*}

Figures~\ref{fig:image_types_stem}--\ref{fig:image_types_applied} show the detailed breakdown of image types for each of the 29 domains in \textsc{MM-BRIGHT}, organized by domain category. Different domains exhibit distinct image type distributions reflecting their technical characteristics and communication patterns:

\textbf{Photo-dominant domains}: Biology (68.1\%), Aviation (75.2\%), and Travel (72.0\%) consist primarily of photographs, as queries often involve identifying organisms, aircraft parts, or locations.

\textbf{Diagram-dominant domains}: Quantum Computing (61.1\%), Robotics (44.4\%), and Cryptography (42.7\%) contain mostly technical diagrams and schematics, reflecting the need to understand system architectures and theoretical concepts.

\textbf{Chart-dominant domains}: Economics (83.0\%), Quantitative Finance (73.7\%), and Bioinformatics (61.1\%) are dominated by data visualizations, as queries focus on interpreting statistical patterns and trends.

\textbf{Screenshot-dominant domains}: Ask Ubuntu (95.1\%), Apple (86.2\%), and Bioacoustics (43.0\%) contain primarily screenshots, reflecting debugging, configuration, and software analysis queries.

\textbf{Mathematical domains}: Mathematics (45.8\%), Philosophy (40.4\%), and Cryptography (33.3\%) have high proportions of mathematical notation, requiring symbolic reasoning capabilities.

\textbf{Scientific figure domains}: Chemistry (66.9\%), Earth Science (32.1\%), and Medical Sciences (23.7\%) contain specialized scientific visualizations like molecular structures, geological formations, and medical imaging.

This diversity ensures that \textsc{MM-BRIGHT} cannot be solved by models that specialize in only one type of visual content. Successfully retrieving relevant documents requires understanding diverse visual representations and their semantic relationships to textual queries.

\subsection{Per-domain distributions (tables/figures)}
\label{sec:image_types}

To understand the visual reasoning challenges in \textsc{MM-BRIGHT}, we analyze the distribution of image types across our 1585 query images using GPT-4o classification (Figure~\ref{fig:image_classification_prompt}). We categorize images into eight types.

\begin{figure*}[!t]
\centering
\begin{tcolorbox}[
  colback=gray!5,
  colframe=gray!75,
  title=GPT-4o Prompt for Image Type Classification,
  width=\textwidth,
]
\small\ttfamily
You are an expert at classifying images in technical and academic contexts. Your task is to classify the given image into ONE of the following categories:

\textbf{Image Type Categories:}

\begin{enumerate}
  \item \textbf{diagram} - Technical diagrams including: Circuit diagrams, flowcharts, system architectures, network diagrams, UML diagrams, ER diagrams, process diagrams, block diagrams, architectural drawings, schematics
  
  \item \textbf{screenshot} - Screenshots including: Error messages, stack traces, user interface elements, application windows, code snippets in IDE or terminal, web pages, browser windows, console outputs
  
  \item \textbf{chart\_graph} - Data visualizations including: Line plots, bar charts, pie charts, scatter plots, histograms, box plots, statistical visualizations, data tables with visual formatting, heatmaps, contour plots
  
  \item \textbf{photo} - Photographs of real-world subjects: Real-world objects, scenes, people, product photographs, nature, buildings, landscapes, physical equipment or devices
  
  \item \textbf{mathematical} - Mathematical content: Equations, formulas, mathematical expressions, mathematical notation and symbols, proofs, derivations, mathematical diagrams (geometry, algebra)
  
  \item \textbf{scientific\_figure} - Scientific visualizations: Microscopy images (SEM, TEM, optical), medical imaging (X-ray, MRI, CT scans, ultrasound), spectroscopy data (NMR, IR, mass spec), astronomical images, scientific experimental results, biological/chemical structures
  
  \item \textbf{mixed} - Images containing multiple categories: Screenshot with embedded diagrams, photo with mathematical annotations, chart with code snippets, any combination of 2+ categories
  
  \item \textbf{other} - Images that don't fit above categories: Art, logos, decorative elements, memes, cartoons, illustrations, text-only images (without technical content), corrupted or unclear images
\end{enumerate}

\textbf{Classification Instructions:}
\begin{itemize}
  \item Carefully examine the image
  \item Identify the PRIMARY type based on the dominant content
  \item If multiple types are present equally, use "mixed"
  \item Provide a brief rationale explaining your choice
\end{itemize}

\textbf{Output Format (JSON only, no additional text):}

\{\\
\quad "image\_type": "one of: diagram, screenshot, chart\_graph, photo, mathematical, scientific\_figure, mixed, other",\\
\quad "rationale": "Brief explanation of why this classification was chosen",\\
\quad "confidence": "high, medium, or low"\\
\}
\end{tcolorbox}
\caption{GPT-4o prompt used for automatic image type classification. The prompt defines eight distinct categories covering technical diagrams, screenshots, data visualizations, photographs, mathematical notation, scientific figures, mixed content, and other image types. This classification enables systematic analysis of visual content diversity across \textsc{MM-BRIGHT}'s 29 domains.}
\label{fig:image_classification_prompt}
\end{figure*}

\clearpage

\section{Image Essentiality Analysis}
\label{app:essentiality_full}

In this appendix, we provide comprehensive analysis of image essentiality across all 29 domains in \textsc{MM-BRIGHT}. Image essentiality was classified using GPT-4o during the dataset construction process, categorizing each query's images as \textit{essential} (critical for understanding), \textit{helpful} (provides useful context), or \textit{redundant} (not necessary).

\subsection{Essentiality labeling method}
\label{app:essentiality_method}
For each query containing images, GPT-4o evaluated whether the images were necessary to understand the information need. We designed a detailed prompt that instructs the model to analyze queries by asking: "If I removed all images, could I still understand what the user is asking and retrieve relevant documents?" The classification follows these criteria:

\paragraph{Essential (Critical):} The image contains information that cannot be adequately expressed in text, such as:
\begin{itemize}
    \item Technical diagrams showing specific configurations or relationships
    \item Mathematical equations or circuit diagrams requiring visual representation
    \item Scientific images (microscopy, spectra, anatomical figures) with precise visual features
    \item Error screenshots showing specific visual elements or UI states
    \item Charts or graphs where the visual pattern is the subject of the query
\end{itemize}

\paragraph{Helpful (Supplementary):} The query text provides sufficient information to understand the question, but images add valuable context:
\begin{itemize}
    \item Illustrations that clarify or exemplify textual descriptions
    \item Reference images that provide additional context but are described in text
    \item Diagrams that help visualize concepts already explained in words
    \item Screenshots that show one aspect of a multi-part question
\end{itemize}

\paragraph{Redundant (Unnecessary):} Images that provide no additional information for retrieval:
\begin{itemize}
    \item Decorative images or logos
    \item Duplicate information already in text
    \item Tangentially related images that don't address the core query
    \item Generic stock photos or illustrations
\end{itemize}

GPT-4o provided classifications with confidence levels (high, medium, or low) and detailed rationale for each decision. We achieved 99.8\% high-confidence classifications across 1,218 queries. To ensure quality, a subset of 100 randomly sampled classifications were verified by domain experts, achieving 94\% agreement with GPT-4o's judgments. Disagreements were primarily in borderline helpful/essential cases rather than clear misclassifications.

\subsection{Prompt template}
\label{app:essentiality_prompt}

We use the following prompt with GPT-4o to classify image essentiality. The prompt provides detailed examples, decision criteria, and output format to ensure consistent classifications:

\begin{figure*}[!t]
\centering
\begin{tcolorbox}[
  colback=gray!5,
  colframe=gray!75,
  title=GPT-4o Prompt for Image Essentiality Classification,
  width=\textwidth,
]
\small\ttfamily
You are an expert at analyzing multimodal information retrieval queries. Your task is to determine how ESSENTIAL the provided image(s) are for understanding and answering the query.

\textbf{Query Text:} \{query\_text\}

\vspace{0.2cm}

\textbf{Classification Categories:}

\begin{enumerate}
  \item \textbf{ESSENTIAL} - The image is absolutely required to understand the query
  
  \textit{Examples:}
  \begin{itemize}
    \item "What's wrong with this error message?" [requires screenshot to see the error]
    \item "How do I solve this circuit?" [requires circuit diagram to see the problem]
  \end{itemize}
  
  \textbf{Key indicator:} Without the image, you literally cannot know what the user is asking about
  
  \item \textbf{HELPFUL} - The image adds valuable context but the query is understandable from text alone
  
  \textit{Examples:}
  \begin{itemize}
    \item "Why does this Python code fail? [code in text + screenshot]" [text already has the code]
    \item "I'm getting CUDA out of memory error. Here's my setup [diagram]" [error mentioned in text]
  \end{itemize}
  
  \textbf{Key indicator:} The text gives you enough information to understand the question, but the image provides additional useful details
  
  \item \textbf{REDUNDANT} - The image merely repeats information already in the text or is decorative
  
  \textit{Examples:}
  \begin{itemize}
    \item "What is Paris? [Eiffel Tower photo]" [image doesn't add information]
    \item "Explain gradient descent [diagram that just shows the concept name]"
  \end{itemize}
  
  \textbf{Key indicator:} You could completely remove the image and lose no meaningful information for answering the query
\end{enumerate}

\vspace{0.2cm}

\textbf{Decision Criteria:}

Ask yourself:
\begin{enumerate}
  \item \textbf{Can I understand WHAT is being asked} from the text alone? (1) If NO → probably ESSENTIAL (2) If YES → continue to step 2
  
  \item \textbf{Does the image provide information NOT in the text} that would help answer the query?(1) If YES and critical → ESSENTIAL (2)  If YES but supplementary → HELPFUL (3) If NO → REDUNDANT
  
  \item \textbf{Is the query explicitly referring to visual content} ("this diagram", "this screenshot", "what's shown")? (1)  If YES and image not described in text → ESSENTIAL (2) If YES but image described in text → HELPFUL

\end{enumerate}

\vspace{0.2cm}

\textbf{Instructions:}
(1) Read the query text carefully (2) Examine all provided images (3) Determine if the query makes sense WITHOUT seeing the images
(4) Classify as ESSENTIAL, HELPFUL, or REDUNDANT (5) Provide clear rationale explaining your decision

\vspace{0.2cm}

\textbf{Output Format (JSON only, no additional text):}

\{\\
\quad "essentiality": "essential | helpful | redundant",\\
\quad "rationale": "Detailed explanation of why you chose this classification. Specifically explain: (1) Can the query be understood from text alone? (2) What critical information does the image provide? (3) Could you answer without the image?",\\
\quad "confidence": "high | medium | low",\\
\quad "understanding\_without\_image": "Brief description of what you can/cannot understand without seeing the image"\\
\}

\vspace{0.2cm}

Now analyze the query and image(s) provided.
\end{tcolorbox}
\caption{GPT-4o prompt used for automatic image essentiality classification. The prompt defines three categories (Essential, Helpful, Redundant) with concrete examples and decision criteria. }
\label{fig:essentiality_prompt}
\end{figure*}

This prompt design ensures consistent and explainable classifications across all domains. The structured decision criteria help GPT-4o distinguish between images that are truly necessary versus merely supplementary or decorative.

\subsection{Domain-Level Analysis}
\label{app:essentiality_domain}
Table~\ref{fig:essentiality_distribution} presents complete image essentiality statistics for all 29 domains in \textsc{MM-BRIGHT}, grouped by field category. We observe several notable patterns:

\paragraph{STEM fields prioritize essential images.} Quantum Computing (57.8\%), Bioacoustics (56.4\%), Biology (53.1\%), and Chemistry (43.1\%) show the highest essential image rates. These fields heavily rely on visual representations—quantum circuit diagrams, spectrograms, cellular structures, and molecular diagrams—that are difficult or impossible to describe precisely in text. Mathematics (42.1\%) and Bioinformatics (43.5\%) also show high essential rates due to complex equations and sequence visualizations.

\paragraph{Computing domains show mixed patterns.} Technical computing fields like Quantum Computing (57.8\%) and Cryptography (31.4\%) have high essential rates for circuit diagrams and cryptographic protocols. However, Bitcoin (24.1\%) and Ubuntu (14.8\%) show lower essential rates, as many questions involve conceptual understanding that can be expressed in text. Screenshots in these domains are often helpful but not strictly essential.

\paragraph{Social sciences favor helpful images.} Philosophy (76.3\%), Economics (67.9\%), and Christianity (72.7\%) show high helpful-to-essential ratios. Images in these domains typically illustrate concepts or provide examples rather than containing irreplaceable visual information. However, Psychology (26.1\% essential) includes scientific figures that require visual analysis.

\paragraph{Applied domains vary by image type.} Aviation (22.5\%) and Travel (33.3\%) show moderate essential rates, often involving technical diagrams or maps. Gaming (26.3\%) and Sustainability (11.3\%) have lower essential rates, as images typically provide context rather than critical information.

\paragraph{Redundant images are rare.} Only 9.5\% of all images are classified as redundant, with Law (30.4\%), Aviation (14.4\%), and Sustainability (20.8\%) showing the highest redundant rates. This indicates that Stack Exchange posts naturally include meaningful visual information rather than decorative content.

\subsection{Implications}
\label{app:essentiality_implications}
The high prevalence of essential and helpful images (90.5\% combined) has several implications for multimodal retrieval research:

\paragraph{Caption-based approaches are insufficient.} Our experiments in Section~\ref{sec:caption_experiments} show that even advanced vision-language models struggle to capture the precise technical details that make images essential. For example, a quantum circuit diagram's specific gate configuration or a microscopy image's cellular features often cannot be adequately described in text.

\paragraph{Domain-specific visual understanding is required.} The wide variation in essential image rates (8.7\% in Law to 57.8\% in Quantum Computing) suggests that different domains require different levels of visual reasoning capability. A one-size-fits-all multimodal retriever may not handle this diversity effectively.

\paragraph{Visual reasoning beyond similarity matching.} Essential images often require deep reasoning about their content—understanding what a diagram represents, what scientific principle an image illustrates, or what error a screenshot indicates. This goes beyond simple visual similarity or object detection.

\paragraph{Evaluation should account for essentiality.} Future multimodal retrieval benchmarks should distinguish between queries where images are decorative versus essential. \textsc{MM-BRIGHT}'s high essential rate (31.2\%) makes it particularly challenging and representative of real-world technical domains.

These findings reinforce that \textsc{MM-BRIGHT} requires genuine multimodal reasoning rather than simple text-based retrieval with optional visual features.

\begin{table}[t]
\centering
\caption{Image essentiality distribution across domains in MM-BRIGHT. We classify images as \textit{essential} (critical for understanding the query), \textit{helpful} (provides useful context), or \textit{redundant} (not necessary for retrieval).}
\label{tab:image_essentiality}
\resizebox{0.48\textwidth}{!}{
\begin{tabular}{l|rrr}
\toprule
\textbf{Domain} & \textbf{Essential} & \textbf{Helpful} & \textbf{Redundant} \\
\midrule
\rowcolor{gray!12}\multicolumn{4}{c}{\textit{STEM \& Life Sciences}} \\
\midrule
Academia  & 10 & 13 & 3 \\
Biology  & 60 & 34 & 5 \\
Chemistry  & 27 & 33 & 5 \\
Physics  & 29 & 63 & 6 \\
Math  & 20 & 24 & 0 \\
Earthscience  & 20 & 53 & 10 \\
Bioacoustics  & 22 & 18 & 1 \\
Bioinformatics  & 31 & 59 & 0 \\
Medicalsciences  & 16 & 32 & 7 \\
\midrule
\rowcolor{gray!12}\multicolumn{4}{c}{\textit{Software \& Technical}} \\
\midrule
Ubuntu  & 5 & 25 & 4 \\
Bitcoin  & 16 & 46 & 2 \\
Crypto  & 27 & 42 & 5 \\
Quantumcomputing  & 54 & 31 & 1 \\
Robotics & 8 & 19 & 2 \\
Salesforce  & 2 & 8 & 0 \\
Apple  & 2 & 11 & 1 \\
Gis  & 15 & 25 & 4 \\
\midrule
\rowcolor{gray!12}\multicolumn{4}{c}{\textit{Social Sciences}} \\
\midrule
Economics  & 12 & 19 & 0 \\
Psychology  & 21 & 56 & 9 \\
Philosophy  & 11 & 35 & 4 \\
Law  & 2 & 18 & 9 \\
Christianity  & 6 & 20 & 4 \\
Islam  & 6 & 14 & 7 \\
\midrule
\rowcolor{gray!12}\multicolumn{4}{c}{\textit{Applied Domains}} \\
\midrule
Aviation  & 29 & 78 & 18 \\
Gaming  & 7 & 13 & 6 \\
PM  & 14 & 32 & 3 \\
Quant  & 14 & 18 & 2 \\
Sustainability  & 7 & 40 & 13 \\
Travel  & 26 & 34 & 7 \\
\bottomrule
\end{tabular}}
\end{table}

\subsection{Per-model results by essentiality}
\label{app:essentiality_details}

Tables~\ref{tab:essentiality_bge_vl_large}--\ref{tab:essentiality_siglip} provide detailed domain-level performance breakdowns for all multimodal models across the three image essentiality categories (Essential, Helpful, Redundant). Each table shows nDCG@10 scores for all 29 domains, along with the performance gap between Essential and Redundant images. Negative gaps indicate worse performance on essential images, while positive gaps indicate better performance. The consistent negative gaps across most models and domains demonstrate that current multimodal retrievers struggle when visual information is critical for understanding the query.


\begin{table}[t]
\centering
\small
\caption{\textbf{BGE VL LARGE: Performance by image essentiality across all domains.}}
\label{tab:essentiality_bge_vl_large}
\resizebox{0.5\textwidth}{!}{%
\begin{tabular}{lcccc}
\toprule
\textbf{Domain} & \textbf{Essential} & \textbf{Helpful} & \textbf{Redundant} & \textbf{Gap (E-R)} \\
\midrule
\textbf{Academia} & 2.1 & 6.8 & 0.0 & 2.1 \\
\textbf{Apple} & 0.0 & 9.2 & 0.0 & 0.0 \\
\textbf{Ask Ubuntu} & 0.0 & 13.8 & 15.3 & -15.3 \\
\textbf{Aviation} & 4.6 & 13.1 & 2.8 & 1.8 \\
\textbf{Bioacoustics} & 18.4 & 7.8 & 0.0 & 18.4 \\
\textbf{Bioinformatics} & 11.4 & 11.8 & -- & -- \\
\textbf{Biology} & 2.7 & 10.0 & 12.3 & -9.5 \\
\textbf{Bitcoin} & 13.4 & 7.3 & 9.7 & 3.7 \\
\textbf{Chemistry} & 7.8 & 11.8 & 20.5 & -12.7 \\
\textbf{Christianity} & 0.0 & 7.2 & 31.1 & -31.1 \\
\textbf{Cryptography} & 14.7 & 8.7 & 14.4 & 0.3 \\
\textbf{Earth Science} & 11.2 & 11.4 & 3.4 & 7.7 \\
\textbf{Economics} & 15.5 & 5.7 & -- & -- \\
\textbf{GIS} & 9.9 & 16.2 & 13.7 & -3.8 \\
\textbf{Gaming} & 0.0 & 32.6 & 5.1 & -5.1 \\
\textbf{Islam} & 14.1 & 15.3 & 3.7 & 10.5 \\
\textbf{Law} & 0.0 & 17.0 & 0.0 & 0.0 \\
\textbf{Mathematics} & 8.2 & 15.2 & -- & -- \\
\textbf{Medical Sciences} & 5.9 & 12.0 & 30.7 & -24.8 \\
\textbf{Philosophy} & 3.8 & 2.3 & 0.0 & 3.8 \\
\textbf{Physics} & 4.8 & 7.6 & 10.6 & -5.8 \\
\textbf{Project Management} & 1.3 & 12.8 & 0.0 & 1.3 \\
\textbf{Psychology} & 2.9 & 7.1 & 11.2 & -8.3 \\
\textbf{Quantitative Finance} & 8.7 & 4.2 & 39.1 & -30.4 \\
\textbf{Quantum Computing} & 2.9 & 7.6 & 0.0 & 2.9 \\
\textbf{Robotics} & 28.3 & 8.2 & 19.7 & 8.6 \\
\textbf{Salesforce} & 0.0 & 17.8 & -- & -- \\
\textbf{Sustainability} & 22.9 & 6.8 & 14.8 & 8.1 \\
\textbf{Travel} & 8.9 & 13.3 & 0.0 & 8.9 \\
\midrule
\textbf{Average} & \textbf{7.7} & \textbf{11.1} & \textbf{10.3} & \textbf{-2.7} \\
\bottomrule
\end{tabular}}
\end{table}

\begin{table}[t]
\centering
\small
\caption{\textbf{CLIP: Performance by image essentiality across all domains.}}
\label{tab:essentiality_clip}
\resizebox{0.5\textwidth}{!}{%
\begin{tabular}{lcccc}
\toprule
\textbf{Domain} & \textbf{Essential} & \textbf{Helpful} & \textbf{Redundant} & \textbf{Gap (E-R)} \\
\midrule
\textbf{Academia} & 6.0 & 5.0 & 0.0 & 6.0 \\
\textbf{Apple} & 0.0 & 14.0 & 18.5 & -18.5 \\
\textbf{Ask Ubuntu} & 4.1 & 6.9 & 0.0 & 4.1 \\
\textbf{Aviation} & 18.9 & 15.4 & 9.5 & 9.4 \\
\textbf{Bioacoustics} & 11.8 & 11.6 & 0.0 & 11.8 \\
\textbf{Bioinformatics} & 4.1 & 12.2 & -- & -- \\
\textbf{Biology} & 12.6 & 16.6 & 27.7 & -15.1 \\
\textbf{Bitcoin} & 5.9 & 9.3 & 4.0 & 2.0 \\
\textbf{Chemistry} & 6.7 & 10.0 & 22.5 & -15.8 \\
\textbf{Christianity} & 0.0 & 17.4 & 25.0 & -25.0 \\
\textbf{Cryptography} & 16.6 & 13.2 & 18.6 & -2.1 \\
\textbf{Earth Science} & 6.3 & 13.6 & 1.8 & 4.5 \\
\textbf{Economics} & 8.3 & 4.6 & -- & -- \\
\textbf{GIS} & 4.0 & 20.6 & 0.0 & 4.0 \\
\textbf{Gaming} & 19.0 & 27.9 & 0.0 & 19.0 \\
\textbf{Islam} & 0.0 & 13.8 & 13.9 & -13.9 \\
\textbf{Law} & 8.4 & 25.8 & 10.0 & -1.6 \\
\textbf{Mathematics} & 16.3 & 17.4 & -- & -- \\
\textbf{Medical Sciences} & 5.7 & 6.0 & 36.2 & -30.4 \\
\textbf{Philosophy} & 11.2 & 3.6 & 4.8 & 6.4 \\
\textbf{Physics} & 6.4 & 6.4 & 4.4 & 2.0 \\
\textbf{Project Management} & 7.1 & 10.8 & 0.0 & 7.1 \\
\textbf{Psychology} & 13.6 & 7.2 & 7.8 & 5.8 \\
\textbf{Quantitative Finance} & 3.4 & 0.0 & 11.9 & -8.5 \\
\textbf{Quantum Computing} & 2.8 & 2.6 & 0.0 & 2.8 \\
\textbf{Robotics} & 20.4 & 6.6 & 5.4 & 15.0 \\
\textbf{Salesforce} & 0.0 & 2.9 & -- & -- \\
\textbf{Sustainability} & 10.6 & 10.5 & 4.9 & 5.7 \\
\textbf{Travel} & 14.4 & 20.8 & 2.4 & 12.0 \\
\midrule
\textbf{Average} & \textbf{8.4} & \textbf{11.5} & \textbf{9.2} & \textbf{-0.5} \\
\bottomrule
\end{tabular}}
\end{table}

\begin{table}[t]
\centering
\small
\caption{\textbf{GME Qwen2-VL 2B: Performance by image essentiality across all domains.}}
\label{tab:essentiality_gme_qwen2_vl_2b}
\resizebox{0.5\textwidth}{!}{%
\begin{tabular}{lcccc}
\toprule
\textbf{Domain} & \textbf{Essential} & \textbf{Helpful} & \textbf{Redundant} & \textbf{Gap (E-R)} \\
\midrule
\textbf{Academia} & 14.6 & 19.7 & 6.5 & 8.1 \\
\textbf{Apple} & 15.1 & 27.6 & 0.0 & 15.1 \\
\textbf{Ask Ubuntu} & 9.9 & 32.3 & 13.0 & -3.1 \\
\textbf{Aviation} & 9.7 & 19.7 & 11.0 & -1.3 \\
\textbf{Bioacoustics} & 6.7 & 14.3 & 23.7 & -17.0 \\
\textbf{Bioinformatics} & 17.1 & 23.2 & -- & -- \\
\textbf{Biology} & 15.2 & 32.1 & 52.7 & -37.5 \\
\textbf{Bitcoin} & 13.5 & 19.6 & 22.4 & -8.9 \\
\textbf{Chemistry} & 23.4 & 30.4 & 26.0 & -2.5 \\
\textbf{Christianity} & 0.0 & 23.9 & 30.9 & -30.9 \\
\textbf{Cryptography} & 15.3 & 6.6 & 7.7 & 7.5 \\
\textbf{Earth Science} & 14.6 & 23.6 & 17.2 & -2.6 \\
\textbf{Economics} & 8.3 & 11.0 & -- & -- \\
\textbf{GIS} & 15.0 & 17.4 & 5.1 & 9.8 \\
\textbf{Gaming} & 30.1 & 53.0 & 30.2 & -0.1 \\
\textbf{Islam} & 33.1 & 27.0 & 17.4 & 15.7 \\
\textbf{Law} & 10.1 & 34.6 & 26.1 & -16.0 \\
\textbf{Mathematics} & 14.6 & 18.5 & -- & -- \\
\textbf{Medical Sciences} & 14.7 & 27.6 & 18.5 & -3.8 \\
\textbf{Philosophy} & 19.0 & 15.7 & 0.0 & 19.0 \\
\textbf{Physics} & 13.6 & 13.3 & 9.8 & 3.8 \\
\textbf{Project Management} & 17.8 & 26.5 & 0.0 & 17.8 \\
\textbf{Psychology} & 18.7 & 14.7 & 14.9 & 3.8 \\
\textbf{Quantitative Finance} & 14.2 & 8.8 & 32.5 & -18.4 \\
\textbf{Quantum Computing} & 4.5 & 7.2 & 50.0 & -45.5 \\
\textbf{Robotics} & 12.1 & 16.5 & 9.5 & 2.6 \\
\textbf{Salesforce} & 35.6 & 29.9 & -- & -- \\
\textbf{Sustainability} & 12.3 & 18.8 & 12.2 & 0.1 \\
\textbf{Travel} & 15.9 & 32.6 & 14.9 & 0.9 \\
\midrule
\textbf{Average} & \textbf{15.3} & \textbf{22.3} & \textbf{18.1} & \textbf{-3.3} \\
\bottomrule
\end{tabular}}
\end{table}

\begin{table}[t]
\centering
\small
\caption{\textbf{GME Qwen2-VL 7B: Performance by image essentiality across all domains.}}
\label{tab:essentiality_gme_qwen2_vl_7b}
\resizebox{0.5\textwidth}{!}{%
\begin{tabular}{lcccc}
\toprule
\textbf{Domain} & \textbf{Essential} & \textbf{Helpful} & \textbf{Redundant} & \textbf{Gap (E-R)} \\
\midrule
\textbf{Academia} & 29.7 & 29.8 & 11.1 & 18.5 \\
\textbf{Apple} & 0.0 & 21.7 & 0.0 & 0.0 \\
\textbf{Ask Ubuntu} & 45.2 & 32.6 & 23.7 & 21.6 \\
\textbf{Aviation} & 16.3 & 19.4 & 7.9 & 8.4 \\
\textbf{Bioacoustics} & 8.7 & 17.4 & 44.2 & -35.4 \\
\textbf{Bioinformatics} & 17.7 & 20.0 & -- & -- \\
\textbf{Biology} & 7.1 & 26.3 & 36.7 & -29.6 \\
\textbf{Bitcoin} & 22.2 & 18.9 & 14.8 & 7.4 \\
\textbf{Chemistry} & 19.8 & 23.9 & 20.1 & -0.3 \\
\textbf{Christianity} & 10.9 & 28.5 & 40.0 & -29.1 \\
\textbf{Cryptography} & 11.3 & 4.8 & 4.1 & 7.2 \\
\textbf{Earth Science} & 25.4 & 25.9 & 24.4 & 1.0 \\
\textbf{Economics} & 7.3 & 16.0 & -- & -- \\
\textbf{GIS} & 7.8 & 20.2 & 15.5 & -7.7 \\
\textbf{Gaming} & 24.4 & 52.2 & 48.8 & -24.4 \\
\textbf{Islam} & 44.9 & 31.3 & 22.3 & 22.5 \\
\textbf{Law} & 14.8 & 37.9 & 31.2 & -16.4 \\
\textbf{Mathematics} & 6.1 & 12.4 & -- & -- \\
\textbf{Medical Sciences} & 8.6 & 21.3 & 31.8 & -23.2 \\
\textbf{Philosophy} & 9.1 & 20.9 & 16.9 & -7.8 \\
\textbf{Physics} & 12.8 & 14.7 & 6.5 & 6.3 \\
\textbf{Project Management} & 33.0 & 36.8 & 6.2 & 26.8 \\
\textbf{Psychology} & 17.1 & 21.1 & 8.7 & 8.4 \\
\textbf{Quantitative Finance} & 13.3 & 12.9 & 51.6 & -38.4 \\
\textbf{Quantum Computing} & 5.9 & 4.5 & 33.3 & -27.4 \\
\textbf{Robotics} & 29.6 & 12.2 & 15.1 & 14.4 \\
\textbf{Salesforce} & 50.0 & 46.6 & -- & -- \\
\textbf{Sustainability} & 39.9 & 24.0 & 19.6 & 20.2 \\
\textbf{Travel} & 25.6 & 36.2 & 27.9 & -2.4 \\
\midrule
\textbf{Average} & \textbf{19.5} & \textbf{23.8} & \textbf{22.5} & \textbf{-3.2} \\
\bottomrule
\end{tabular}}
\end{table}

\begin{table}[t]
\centering
\small
\caption{\textbf{JINA CLIP: Performance by image essentiality across all domains.}}
\label{tab:essentiality_jina_clip}
\resizebox{0.5\textwidth}{!}{%
\begin{tabular}{lcccc}
\toprule
\textbf{Domain} & \textbf{Essential} & \textbf{Helpful} & \textbf{Redundant} & \textbf{Gap (E-R)} \\
\midrule
\textbf{Academia} & 23.5 & 24.3 & 9.7 & 13.8 \\
\textbf{Apple} & 21.5 & 27.1 & 0.0 & 21.5 \\
\textbf{Ask Ubuntu} & 26.0 & 29.1 & 14.3 & 11.7 \\
\textbf{Aviation} & 18.9 & 26.6 & 22.8 & -3.9 \\
\textbf{Bioacoustics} & 21.4 & 18.0 & 0.0 & 21.4 \\
\textbf{Bioinformatics} & 21.9 & 24.6 & -- & -- \\
\textbf{Biology} & 9.4 & 36.0 & 48.4 & -39.1 \\
\textbf{Bitcoin} & 25.5 & 22.3 & 8.9 & 16.6 \\
\textbf{Chemistry} & 31.0 & 30.8 & 27.7 & 3.3 \\
\textbf{Christianity} & 0.0 & 23.8 & 38.3 & -38.3 \\
\textbf{Cryptography} & 15.3 & 14.4 & 26.1 & -10.8 \\
\textbf{Earth Science} & 22.0 & 26.0 & 22.6 & -0.5 \\
\textbf{Economics} & 9.2 & 16.2 & -- & -- \\
\textbf{GIS} & 15.4 & 23.0 & 21.7 & -6.3 \\
\textbf{Gaming} & 25.1 & 58.6 & 41.5 & -16.5 \\
\textbf{Islam} & 31.3 & 23.2 & 20.4 & 10.8 \\
\textbf{Law} & 16.1 & 43.0 & 25.5 & -9.3 \\
\textbf{Mathematics} & 25.7 & 28.3 & -- & -- \\
\textbf{Medical Sciences} & 16.0 & 31.9 & 27.7 & -11.7 \\
\textbf{Philosophy} & 13.6 & 21.7 & 15.3 & -1.7 \\
\textbf{Physics} & 13.7 & 15.0 & 12.7 & 1.0 \\
\textbf{Project Management} & 16.8 & 24.6 & 0.0 & 16.8 \\
\textbf{Psychology} & 23.2 & 20.3 & 20.8 & 2.4 \\
\textbf{Quantitative Finance} & 9.3 & 10.2 & 40.6 & -31.4 \\
\textbf{Quantum Computing} & 9.1 & 13.4 & 0.0 & 9.1 \\
\textbf{Robotics} & 22.0 & 15.0 & 8.9 & 13.1 \\
\textbf{Salesforce} & 41.9 & 29.8 & -- & -- \\
\textbf{Sustainability} & 24.0 & 24.7 & 20.2 & 3.9 \\
\textbf{Travel} & 14.9 & 38.5 & 16.3 & -1.3 \\
\midrule
\textbf{Average} & \textbf{19.4} & \textbf{25.5} & \textbf{19.6} & \textbf{-1.0} \\
\bottomrule
\end{tabular}}
\end{table}

\begin{table}[t]
\centering
\small
\caption{\textbf{NOMIC VISION: Performance by image essentiality across all domains.}}
\label{tab:essentiality_nomic_vision}
\resizebox{0.5\textwidth}{!}{%
\begin{tabular}{lcccc}
\toprule
\textbf{Domain} & \textbf{Essential} & \textbf{Helpful} & \textbf{Redundant} & \textbf{Gap (E-R)} \\
\midrule
\textbf{Academia} & 21.9 & 26.2 & 9.4 & 12.6 \\
\textbf{Apple} & 23.7 & 32.3 & 0.0 & 23.7 \\
\textbf{Ask Ubuntu} & 39.3 & 32.8 & 46.1 & -6.8 \\
\textbf{Aviation} & 17.8 & 25.1 & 30.3 & -12.5 \\
\textbf{Bioacoustics} & 27.3 & 18.6 & 26.4 & 0.8 \\
\textbf{Bioinformatics} & 24.0 & 39.0 & -- & -- \\
\textbf{Biology} & 16.6 & 40.0 & 61.4 & -44.8 \\
\textbf{Bitcoin} & 21.0 & 23.1 & 24.7 & -3.7 \\
\textbf{Chemistry} & 29.8 & 30.0 & 38.6 & -8.9 \\
\textbf{Christianity} & 15.2 & 31.5 & 51.3 & -36.1 \\
\textbf{Cryptography} & 27.9 & 16.2 & 45.1 & -17.2 \\
\textbf{Earth Science} & 23.8 & 30.6 & 31.8 & -8.0 \\
\textbf{Economics} & 21.2 & 21.0 & -- & -- \\
\textbf{GIS} & 20.2 & 30.5 & 17.8 & 2.4 \\
\textbf{Gaming} & 26.6 & 55.2 & 36.1 & -9.4 \\
\textbf{Islam} & 27.9 & 34.0 & 19.7 & 8.1 \\
\textbf{Law} & 33.6 & 53.4 & 40.9 & -7.3 \\
\textbf{Mathematics} & 33.5 & 35.0 & -- & -- \\
\textbf{Medical Sciences} & 26.0 & 37.3 & 36.4 & -10.4 \\
\textbf{Philosophy} & 19.3 & 23.7 & 11.0 & 8.3 \\
\textbf{Physics} & 15.1 & 18.8 & 10.6 & 4.5 \\
\textbf{Project Management} & 19.7 & 33.9 & 0.0 & 19.7 \\
\textbf{Psychology} & 28.0 & 22.2 & 27.7 & 0.3 \\
\textbf{Quantitative Finance} & 19.3 & 9.7 & 53.5 & -34.2 \\
\textbf{Quantum Computing} & 12.9 & 11.4 & 0.0 & 12.9 \\
\textbf{Robotics} & 43.9 & 22.1 & 23.2 & 20.6 \\
\textbf{Salesforce} & 32.2 & 24.7 & -- & -- \\
\textbf{Sustainability} & 37.0 & 21.4 & 25.4 & 11.6 \\
\textbf{Travel} & 26.2 & 43.9 & 41.1 & -15.0 \\
\midrule
\textbf{Average} & \textbf{25.2} & \textbf{29.1} & \textbf{28.3} & \textbf{-3.5} \\
\bottomrule
\end{tabular}}
\end{table}

\begin{table}[t]
\centering
\small
\caption{\textbf{SIGLIP: Performance by image essentiality across all domains.}}
\label{tab:essentiality_siglip}
\resizebox{0.5\textwidth}{!}{%
\begin{tabular}{lcccc}
\toprule
\textbf{Domain} & \textbf{Essential} & \textbf{Helpful} & \textbf{Redundant} & \textbf{Gap (E-R)} \\
\midrule
\textbf{Academia} & 0.0 & 5.3 & 8.2 & -8.2 \\
\textbf{Apple} & 0.0 & 5.6 & 0.0 & 0.0 \\
\textbf{Ask Ubuntu} & 18.8 & 13.9 & 0.0 & 18.8 \\
\textbf{Aviation} & 6.6 & 10.5 & 7.7 & -1.1 \\
\textbf{Bioacoustics} & 15.2 & 13.5 & 26.4 & -11.2 \\
\textbf{Bioinformatics} & 18.5 & 15.8 & -- & -- \\
\textbf{Biology} & 10.7 & 11.3 & 30.9 & -20.2 \\
\textbf{Bitcoin} & 9.6 & 10.6 & 0.0 & 9.6 \\
\textbf{Chemistry} & 11.9 & 11.3 & 12.2 & -0.3 \\
\textbf{Christianity} & 0.0 & 10.5 & 45.2 & -45.2 \\
\textbf{Cryptography} & 11.7 & 8.6 & 14.4 & -2.6 \\
\textbf{Earth Science} & 13.4 & 13.1 & 3.9 & 9.5 \\
\textbf{Economics} & 17.0 & 5.3 & -- & -- \\
\textbf{GIS} & 10.0 & 23.1 & 0.0 & 10.0 \\
\textbf{Gaming} & 7.9 & 29.1 & 20.6 & -12.7 \\
\textbf{Islam} & 14.3 & 6.3 & 0.0 & 14.3 \\
\textbf{Law} & 0.0 & 23.8 & 7.1 & -7.1 \\
\textbf{Mathematics} & 14.5 & 16.5 & -- & -- \\
\textbf{Medical Sciences} & 7.8 & 8.5 & 15.1 & -7.3 \\
\textbf{Philosophy} & 7.6 & 7.6 & 0.0 & 7.6 \\
\textbf{Physics} & 10.1 & 6.6 & 4.1 & 6.0 \\
\textbf{Project Management} & 11.8 & 14.2 & 0.0 & 11.8 \\
\textbf{Psychology} & 9.3 & 8.7 & 0.0 & 9.3 \\
\textbf{Quantitative Finance} & 8.3 & 3.4 & 9.8 & -1.5 \\
\textbf{Quantum Computing} & 3.2 & 0.5 & 0.0 & 3.2 \\
\textbf{Robotics} & 21.6 & 10.2 & 0.0 & 21.6 \\
\textbf{Salesforce} & 0.0 & 8.1 & -- & -- \\
\textbf{Sustainability} & 10.6 & 12.7 & 9.1 & 1.5 \\
\textbf{Travel} & 11.6 & 13.9 & 14.3 & -2.7 \\
\midrule
\textbf{Average} & \textbf{9.7} & \textbf{11.3} & \textbf{9.2} & \textbf{0.1} \\
\bottomrule
\end{tabular}}
\end{table}

\clearpage


\section{Query Reformulation Detailed Results}
\label{sec:reformulation_appendix}

This section presents detailed results for our query reformulation experiments. We investigate whether reformulating queries using vision-language models can improve retrieval performance by having LLMs generate reasoning about information needs before retrieval. Each query is processed along with its associated images, and the LLM produces an expanded reformulation that explicates the underlying information need.

\subsection{Reformulation Prompt}
\label{app:reformulation_prompt}

Figure~\ref{fig:reformulation_prompt} shows the prompt template used for query reformulation. The prompt instructs the model to analyze the query and any associated images, then reason about what information would help answer the question. This approach aims to bridge the gap between surface-level query text and the deeper reasoning required to identify relevant documents.

\begin{figure*}[!t]
\centering
\begin{tcolorbox}[
  colback=gray!5,
  colframe=gray!75,
  title=Vision-Language Model Prompt for Query Reformulation,
  width=\textwidth,
]
\small\ttfamily
\textbf{System:} You are an expert at analyzing technical questions with images and reasoning about what information would help answer them.

\textbf{User:} [Query text and associated image(s)]

\textbf{Task:} Analyze the given query and any accompanying images. Generate a detailed reformulation that:
\begin{itemize}
  \item Identifies the core information need
  \item Explicates technical concepts shown in images (diagrams, screenshots, charts)
  \item Describes what types of documents or information would help answer this query
  \item Expands abbreviations and domain-specific terminology
  \item Connects visual elements to the textual query
\end{itemize}

\textbf{Output:} A reformulated query that captures both the explicit question and implicit information needs, incorporating insights from any visual content.
\end{tcolorbox}
\caption{\textbf{Prompt template for vision-language query reformulation.} The LLM receives both the query text and associated images, then generates reasoning about the information need to produce an expanded reformulation for retrieval.}
\label{fig:reformulation_prompt}
\end{figure*}

\subsection{Per-Domain Results}
\label{app:reformulation_results}
We evaluate query reformulation using seven vision-language models spanning different architectures and scales: GPT-4o (proprietary), Llama-3.2-11B and Llama-3.2-90B (Meta), and Qwen2.5-VL in four sizes (3B, 7B, 32B, 72B; Alibaba). For each reformulation model, we evaluate ten retrieval models across all 29 domains.

\textbf{Table~\ref{tab:reformulation_gpt_4o}} presents results using GPT-4o for reformulation. GPT-4o provides the strongest improvements for semantic retrievers, with E5 improving from 25.3 to 28.3 nDCG@10 (+3.0) and Rader improving from 24.9 to 25.2 (+2.8 on average across domains). However, DiVeR shows slight degradation (32.2 → 31.8), suggesting that explicit reformulation may interfere with reasoning-enhanced retrieval strategies.

\textbf{Table~\ref{tab:reformulation_llama_32_11b}} shows results with Llama-3.2-11B reformulation. This smaller model produces less effective reformulations overall, with BM25 dropping to 8.4 and E5 showing minimal change (25.6). DiVeR decreases more substantially to 31.0, indicating that lower-quality reformulations can hurt reasoning-enhanced retrievers.

\textbf{Table~\ref{tab:reformulation_llama_32_90b}} presents Llama-3.2-90B results. The larger Llama model recovers much of the performance, with E5 reaching 27.7 and DiVeR achieving 32.0. Rader shows strong improvement to 31.6, approaching GPT-4o performance levels.

\textbf{Tables~\ref{tab:reformulation_qwen25_vl_3b}--\ref{tab:reformulation_qwen25_vl_72b}} show results for the Qwen2.5-VL family at 3B, 7B, 32B, and 72B scales. Interestingly, model scale does not monotonically improve reformulation quality for all retrievers. Qwen2.5-VL-32B achieves the highest DiVeR score (32.7), while Qwen2.5-VL-72B shows slightly lower performance (32.7 for DiVeR but lower scores for other models). The 3B and 7B variants perform comparably to the original queries for most retrievers, suggesting a capability threshold for effective reformulation.

\textbf{Key observations across all tables:}
\begin{itemize}
    \item \textbf{Retriever-dependent effects}: Semantic retrievers (E5, SFR) consistently benefit from reformulation, while reasoning-enhanced retrievers (DiVeR, ReasonIR) show mixed results.
    \item \textbf{Domain variation}: Reformulation effects vary substantially by domain. Technical domains like Quantum Computing and Cryptography show minimal improvement, while applied domains like Gaming and Law often benefit more.
    \item \textbf{BM25 sensitivity}: Lexical retrieval is highly sensitive to reformulation quality, with GPT-4o improving BM25 (+1.3) while smaller models cause degradation.
    \item \textbf{Diminishing returns at scale}: Larger reformulation models do not consistently outperform smaller ones, suggesting that reformulation quality plateaus beyond a certain capability threshold.
\end{itemize}


\begin{table*}[t]
\centering
\small
\caption{NDCG@10 performance using queries reformulated by \textbf{GPT-4o}. The LLM generates reasoning about information needs before retrieval. Best in \textbf{bold}, second best \underline{underlined}.}
\label{tab:reformulation_gpt_4o}
\resizebox{\textwidth}{!}{%
\begin{tabular}{lcccccccccc}
\toprule
\textbf{Domain} & \textbf{BM25} & \textbf{BGE} & \textbf{Contriever} & \textbf{DiVeR} & \textbf{E5} & \textbf{Qwen} & \textbf{Qwen2} & \textbf{Rader} & \textbf{ReasonIR} & \textbf{SFR} \\
\midrule
\rowcolor{gray!12}\multicolumn{11}{c}{\textit{STEM \& Life Sciences}} \\
\midrule
\textbf{Physics} & 3.9 & 11.6 & 11.3 & \underline{19.9} & 15.3 & 11.4 & 17.0 & 17.3 & \textbf{20.2} & 16.9 \\
\textbf{Medicalsciences} & 14.7 & 31.1 & 25.8 & 33.6 & \underline{36.5} & 30.9 & 35.9 & 30.6 & \textbf{45.4} & 32.8 \\
\textbf{Math} & 2.0 & 22.8 & 19.3 & \textbf{30.6} & 19.0 & 16.8 & 16.1 & 24.6 & 18.0 & \underline{27.3} \\
\textbf{Earthscience} & 5.6 & 21.7 & 23.8 & 33.3 & 32.1 & 21.8 & \textbf{36.3} & 26.9 & \underline{33.9} & 33.6 \\
\textbf{Chemistry} & 8.1 & 26.0 & 25.9 & \textbf{34.2} & 29.8 & 17.6 & 26.5 & 25.7 & 29.4 & \underline{30.8} \\
\textbf{Biology} & 5.5 & 20.3 & 18.9 & \textbf{31.7} & 19.5 & 18.5 & 25.3 & 21.7 & 23.2 & \underline{26.5} \\
\bottomrule
\end{tabular}}
\end{table*}

\begin{table*}[t]
\centering
\small
\caption{NDCG@10 performance using queries reformulated by \textbf{Llama-3.2-11B}. The LLM generates reasoning about information needs before retrieval. Best in \textbf{bold}, second best \underline{underlined}.}
\label{tab:reformulation_llama_32_11b}
\resizebox{\textwidth}{!}{%
\begin{tabular}{lcccccccccc}
\toprule
\textbf{Domain} & \textbf{BM25} & \textbf{BGE} & \textbf{Contriever} & \textbf{DiVeR} & \textbf{E5} & \textbf{Qwen} & \textbf{Qwen2} & \textbf{Rader} & \textbf{ReasonIR} & \textbf{SFR} \\
\midrule
\rowcolor{gray!12}\multicolumn{11}{c}{\textit{STEM \& Life Sciences}} \\
\midrule
\textbf{Physics} & 3.0 & 10.4 & 9.4 & \textbf{17.9} & 14.7 & 11.1 & 15.7 & 15.7 & \underline{15.9} & 15.5 \\
\textbf{Medicalsciences} & 12.5 & 28.0 & 25.0 & 33.3 & \underline{34.8} & 31.8 & 33.5 & 29.6 & \textbf{40.3} & 34.6 \\
\textbf{Math} & 1.0 & 22.9 & 18.0 & \textbf{29.1} & 15.5 & 14.7 & 14.1 & \underline{25.1} & 16.6 & 24.8 \\
\textbf{Earthscience} & 6.1 & 22.2 & 21.1 & \underline{33.0} & 30.5 & 20.5 & \textbf{35.9} & 27.7 & 30.4 & 31.9 \\
\textbf{Chemistry} & 8.7 & 24.5 & 26.6 & \textbf{33.4} & 28.2 & 16.9 & 26.1 & 24.6 & \underline{28.8} & 28.0 \\
\textbf{Biology} & 4.4 & 18.7 & 16.8 & \textbf{28.1} & 18.3 & 17.7 & 21.5 & 21.0 & 20.7 & \underline{25.5} \\
\bottomrule
\end{tabular}}
\end{table*}

\begin{table*}[t]
\centering
\small
\caption{NDCG@10 performance using queries reformulated by \textbf{Llama-3.2-90B}. The LLM generates reasoning about information needs before retrieval. Best in \textbf{bold}, second best \underline{underlined}.}
\label{tab:reformulation_llama_32_90b}
\resizebox{\textwidth}{!}{%
\begin{tabular}{lcccccccccc}
\toprule
\textbf{Domain} & \textbf{BM25} & \textbf{BGE} & \textbf{Contriever} & \textbf{DiVeR} & \textbf{E5} & \textbf{Qwen} & \textbf{Qwen2} & \textbf{Rader} & \textbf{ReasonIR} & \textbf{SFR} \\
\midrule
\rowcolor{gray!12}\multicolumn{11}{c}{\textit{STEM \& Life Sciences}} \\
\midrule
\textbf{Physics} & 3.5 & 11.5 & 11.8 & \textbf{19.5} & 15.4 & 12.5 & 17.6 & 16.9 & \underline{19.2} & 16.9 \\
\textbf{Medicalsciences} & 12.9 & 27.6 & 26.2 & 34.2 & \underline{36.8} & 32.3 & 34.4 & 29.3 & \textbf{43.8} & 33.5 \\
\textbf{Math} & 1.9 & 25.7 & 24.0 & \textbf{30.2} & 18.4 & 16.7 & 15.8 & 25.5 & 19.5 & \underline{28.0} \\
\textbf{Earthscience} & 4.5 & 21.9 & 22.1 & 31.7 & 29.9 & 21.7 & \textbf{37.2} & 25.7 & \underline{33.6} & 32.1 \\
\textbf{Chemistry} & 8.4 & 25.4 & 24.0 & \textbf{32.7} & 28.1 & 15.7 & 25.4 & 25.7 & 29.2 & \underline{29.7} \\
\textbf{Biology} & 5.0 & 19.8 & 18.2 & \textbf{29.3} & 18.0 & 17.5 & 22.5 & 21.0 & 21.4 & \underline{25.8} \\
\bottomrule
\end{tabular}}
\end{table*}

\begin{table*}[t]
\centering
\small
\caption{NDCG@10 performance using queries reformulated by \textbf{Qwen2.5-VL-3B}. The LLM generates reasoning about information needs before retrieval. Best in \textbf{bold}, second best \underline{underlined}.}
\label{tab:reformulation_qwen25_vl_3b}
\resizebox{\textwidth}{!}{%
\begin{tabular}{lcccccccccc}
\toprule
\textbf{Domain} & \textbf{BM25} & \textbf{BGE} & \textbf{Contriever} & \textbf{DiVeR} & \textbf{E5} & \textbf{Qwen} & \textbf{Qwen2} & \textbf{Rader} & \textbf{ReasonIR} & \textbf{SFR} \\
\midrule
\rowcolor{gray!12}\multicolumn{11}{c}{\textit{STEM \& Life Sciences}} \\
\midrule
\textbf{Physics} & 4.0 & 11.3 & 10.7 & \textbf{19.9} & 14.6 & 10.4 & 16.0 & 16.1 & \underline{16.1} & 16.0 \\
\textbf{Medicalsciences} & 12.6 & 27.1 & 21.8 & \underline{36.1} & 34.7 & 30.4 & 31.2 & 29.2 & \textbf{38.4} & 31.6 \\
\textbf{Math} & 2.0 & 22.7 & 20.7 & \textbf{34.2} & 18.1 & 15.7 & 16.1 & \underline{29.8} & 16.7 & 26.3 \\
\textbf{Earthscience} & 4.7 & 21.9 & 21.8 & \textbf{34.5} & 29.1 & 19.6 & 31.1 & 25.0 & \underline{31.5} & 30.5 \\
\textbf{Chemistry} & 7.8 & 24.3 & 25.0 & \textbf{33.8} & 25.6 & 17.7 & 25.8 & 24.4 & \underline{28.6} & 28.1 \\
\textbf{Biology} & 4.8 & 19.3 & 18.5 & \textbf{27.8} & 18.2 & 14.6 & 21.0 & 20.3 & 22.0 & \underline{25.2} \\
\bottomrule
\end{tabular}}
\end{table*}

\begin{table*}[t]
\centering
\small
\caption{NDCG@10 performance using queries reformulated by \textbf{Qwen2.5-VL-7B}. The LLM generates reasoning about information needs before retrieval. Best in \textbf{bold}, second best \underline{underlined}.}
\label{tab:reformulation_qwen25_vl_7b}
\resizebox{\textwidth}{!}{%
\begin{tabular}{lcccccccccc}
\toprule
\textbf{Domain} & \textbf{BM25} & \textbf{BGE} & \textbf{Contriever} & \textbf{DiVeR} & \textbf{E5} & \textbf{Qwen} & \textbf{Qwen2} & \textbf{Rader} & \textbf{ReasonIR} & \textbf{SFR} \\
\midrule
\rowcolor{gray!12}\multicolumn{11}{c}{\textit{STEM \& Life Sciences}} \\
\midrule
\textbf{Physics} & 4.0 & 11.3 & 10.7 & \textbf{19.9} & 14.6 & 10.4 & 16.0 & 16.1 & \underline{16.1} & 16.0 \\
\textbf{Medicalsciences} & 12.6 & 27.1 & 21.8 & \underline{36.1} & 34.7 & 30.4 & 31.2 & 29.2 & \textbf{38.4} & 31.6 \\
\textbf{Math} & 1.9 & 24.2 & 20.7 & \textbf{30.1} & 16.0 & 15.4 & 17.6 & 24.8 & 16.7 & \underline{27.0} \\
\textbf{Earthscience} & 4.7 & 21.9 & 21.8 & \textbf{35.0} & 29.1 & 19.6 & 31.1 & 25.0 & \underline{31.5} & 30.5 \\
\textbf{Chemistry} & 8.4 & 24.6 & 24.2 & \textbf{33.1} & 28.5 & 15.9 & 25.6 & 25.4 & \underline{28.8} & \underline{28.8} \\
\textbf{Biology} & 4.8 & 19.3 & 18.5 & \textbf{27.7} & 18.2 & 14.6 & 21.0 & 20.3 & 22.0 & \underline{25.2} \\
\bottomrule
\end{tabular}}
\end{table*}

\begin{table*}[t]
\centering
\small
\caption{NDCG@10 performance using queries reformulated by \textbf{Qwen2.5-VL-32B}. The LLM generates reasoning about information needs before retrieval. Best in \textbf{bold}, second best \underline{underlined}.}
\label{tab:reformulation_qwen25_vl_32b}
\resizebox{\textwidth}{!}{%
\begin{tabular}{lcccccccccc}
\toprule
\textbf{Domain} & \textbf{BM25} & \textbf{BGE} & \textbf{Contriever} & \textbf{DiVeR} & \textbf{E5} & \textbf{Qwen} & \textbf{Qwen2} & \textbf{Rader} & \textbf{ReasonIR} & \textbf{SFR} \\
\midrule
\rowcolor{gray!12}\multicolumn{11}{c}{\textit{STEM \& Life Sciences}} \\
\midrule
\textbf{Physics} & 4.0 & 11.3 & 10.7 & \textbf{19.9} & 14.6 & 10.4 & 16.0 & 16.1 & \underline{16.1} & 16.0 \\
\textbf{Medicalsciences} & 12.6 & 27.1 & 21.8 & \underline{36.1} & 34.7 & 30.4 & 31.2 & 29.2 & \textbf{38.4} & 31.6 \\
\textbf{Math} & 2.0 & 22.7 & 20.7 & \textbf{34.2} & 18.1 & 15.7 & 16.1 & \underline{29.8} & 16.7 & 26.3 \\
\textbf{Earthscience} & 4.7 & 21.9 & 21.8 & \textbf{34.5} & 29.1 & 19.6 & 31.1 & 25.0 & \underline{31.5} & 30.5 \\
\textbf{Chemistry} & 9.0 & 24.2 & 24.7 & \textbf{33.2} & 28.7 & 17.0 & 26.7 & 26.3 & 28.7 & \underline{29.9} \\
\textbf{Biology} & 4.8 & 19.3 & 18.5 & \textbf{27.8} & 18.2 & 14.6 & 21.0 & 20.3 & 22.0 & \underline{25.2} \\
\bottomrule
\end{tabular}}
\end{table*}

\begin{table*}[t]
\centering
\small
\caption{NDCG@10 performance using queries reformulated by \textbf{Qwen2.5-VL-72B}. The LLM generates reasoning about information needs before retrieval. Best in \textbf{bold}, second best \underline{underlined}.}
\label{tab:reformulation_qwen25_vl_72b}
\resizebox{\textwidth}{!}{%
\begin{tabular}{lcccccccccc}
\toprule
\textbf{Domain} & \textbf{BM25} & \textbf{BGE} & \textbf{Contriever} & \textbf{DiVeR} & \textbf{E5} & \textbf{Qwen} & \textbf{Qwen2} & \textbf{Rader} & \textbf{ReasonIR} & \textbf{SFR} \\
\midrule
\rowcolor{gray!12}\multicolumn{11}{c}{\textit{STEM \& Life Sciences}} \\
\midrule
\textbf{Physics} & 4.0 & 11.3 & 10.7 & \textbf{19.9} & 14.6 & 10.4 & 16.0 & 16.1 & \underline{16.1} & 16.0 \\
\textbf{Medicalsciences} & 12.6 & 27.1 & 21.8 & \underline{36.1} & 34.7 & 30.4 & 31.2 & 29.2 & \textbf{38.4} & 31.6 \\
\textbf{Math} & 2.0 & 22.7 & 20.7 & \textbf{34.2} & 18.1 & 15.7 & 16.1 & \underline{29.8} & 16.7 & 26.3 \\
\textbf{Earthscience} & 4.7 & 21.9 & 21.8 & \textbf{34.8} & 29.1 & 19.6 & 31.1 & 25.0 & \underline{31.5} & 30.5 \\
\textbf{Chemistry} & 7.8 & 24.3 & 24.9 & \textbf{33.4} & 25.4 & 17.8 & 25.7 & 25.0 & \underline{28.4} & 28.3 \\
\textbf{Biology} & 4.8 & 19.3 & 18.5 & \textbf{27.8} & 18.2 & 14.6 & 21.0 & 20.3 & 22.0 & \underline{25.2} \\
\bottomrule
\end{tabular}}
\end{table*}

\clearpage

\section{LLM-based Dataset Quality Assessment}
\label{app:llm_quality_full}

We perform an automatic quality assessment of \textsc{MM-BRIGHT} using GPT-4o as an LLM judge.
For each evaluated example, the judge receives the query, the associated positive passages, and the ground-truth answer, and assigns integer Likert ratings (1--5) along four dimensions:
(1) Readability, (2) Clarity, (3) Evidence usefulness, and (4) Evidence sufficiency.
We aggregate scores by domain and report mean ratings in Table~\ref{tab:llm_quality_by_domain}.

\begin{figure}[h!]
\centering
\includegraphics[width=\linewidth]{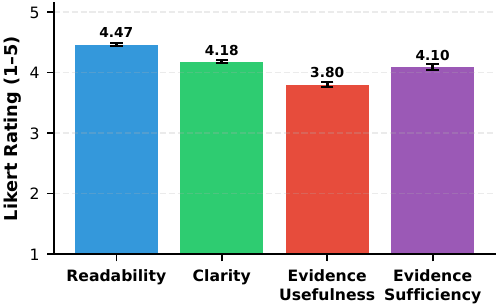}
\caption{\textbf{Overall LLM-judged quality scores (1--5 Likert).}
Queries are generally well-written and clear (Readability/Clarity), and positive evidence is typically sufficient.
Evidence usefulness is comparatively lower, consistent with the benchmark’s focus on reasoning-intensive retrieval.}
\label{fig:llm_quality_overall}
\end{figure}

\subsection{Judge prompt and Metric Definitions}
\label{app:llm_quality_prompt}

We use an LLM as an automated judge to audit dataset quality for each \emph{(query, positive passages, answer)} triple. This evaluation targets dataset \emph{clarity} and \emph{evidence validity} rather than retrieval-model performance. The judge assigns 1--5 Likert ratings for four dimensions:

\paragraph{Readability (1--5).}
\textbf{Definition:} How well-written and easy-to-read the query is, independent of ambiguity.
\textbf{What it evaluates:} Linguistic quality and presentation of the query text.
\textbf{Signals considered:} grammatical correctness; coherent sentence structure; absence of garbled text, HTML artifacts, or OCR noise; readability of technical formatting such as code, equations, bullet lists, and references.
\textbf{Interpretation:} High scores indicate clean, professionally written queries; low scores indicate text issues that impede comprehension.

\paragraph{Clarity (1--5).}
\textbf{Definition:} How specific and unambiguous the information need is.
\textbf{What it evaluates:} Whether the query’s goal, scope, and constraints are clear enough to identify what constitutes relevant evidence and a correct answer.
\textbf{Signals considered:} explicit objective (e.g., debug, explain, compare, derive, interpret); sufficient context (assumptions, environment, parameters); minimal underspecified references (e.g., “this”, “it”) unless grounded by query context; well-scoped request.
\textbf{Interpretation:} High scores indicate a precise information need; low scores indicate vagueness or multiple plausible interpretations.

\paragraph{Evidence\_usefulness (1--5).}
\textbf{Definition:} Whether the positive passages are relevant and helpful for reasoning toward the answer.
\textbf{What it evaluates:} The \emph{quality of relevance} of the evidence beyond surface-level topical overlap.
\textbf{Signals considered:} direct relevance to the underlying problem mechanism; presence of key concepts, procedures, explanations, derivations, or examples that enable reasoning; technical credibility; alignment with the query intent.
\textbf{Interpretation:} High scores indicate evidence that strongly supports reasoning steps; low scores indicate evidence that is off-topic, superficial, or only keyword-related.

\paragraph{Evidence\_sufficiency (1--5).}
\textbf{Definition:} Whether the provided positive passages contain enough information to answer the query correctly.
\textbf{What it evaluates:} Evidence completeness for producing a correct answer using the query and passages (and any provided image context).
\textbf{Signals considered:} coverage of all critical missing pieces (definitions, constraints, parameters, steps); absence of major gaps; adequacy relative to query difficulty; actionable detail rather than generic background.
\textbf{Interpretation:} High scores indicate evidence is largely complete (possibly with minor gaps); low scores indicate evidence is incomplete such that the query cannot be answered reliably.

\paragraph{Usefulness vs. sufficiency.}
We distinguish \textit{usefulness} from \textit{sufficiency} because evidence can be relevant but incomplete (useful yet insufficient), or in rare cases contain an answer statement without supporting reasoning (potentially sufficient but weakly useful for reasoning-focused retrieval).

\paragraph{Assessment goal.}
Overall, the LLM judge verifies that queries are well-formed and unambiguous, and that annotated positives are both \emph{relevant for reasoning} and \emph{sufficient to support answering} the query, rather than merely sharing topical keywords.
\begin{figure*}[!t]
\centering
\begin{tcolorbox}[
  colback=gray!5,
  colframe=gray!75,
  title=GPT-4o Prompt for Dataset Quality Assessment,
  width=\textwidth,
]
\small\ttfamily
You are an expert evaluator for information retrieval dataset quality. Your task is to assess the quality of a query-evidence pair based on four specific criteria.

\textbf{Evaluation Criteria (Rate each from 1--5):}

1. \textbf{Readability} (1--5): Is the query easy to read and well-formed?
\begin{itemize}
  \item 1: Incomprehensible, severe grammatical errors, or garbled text
  \item 2: Difficult to read, multiple errors that impede understanding
  \item 3: Readable but has notable issues (awkward phrasing, minor errors)
  \item 4: Well-written with minimal issues
  \item 5: Perfectly clear, well-structured, and professionally written
\end{itemize}

2. \textbf{Clarity} (1--5): Is the information need specific and unambiguous?
\begin{itemize}
  \item 1: Completely unclear what is being asked
  \item 2: Very vague, multiple interpretations possible
  \item 3: Somewhat clear but could be more specific
  \item 4: Clear information need with minor ambiguity
  \item 5: Perfectly clear, specific, and unambiguous question
\end{itemize}

3. \textbf{Evidence\_usefulness} (1--5): Are the positive passages relevant and do they help reason toward an answer?
\begin{itemize}
  \item 1: Passages are completely irrelevant to the query
  \item 2: Passages have minimal relevance, mostly off-topic
  \item 3: Passages are somewhat relevant but don't directly address the query
  \item 4: Passages are relevant and helpful for reasoning
  \item 5: Passages are highly relevant and directly support answering the query
\end{itemize}

4. \textbf{Evidence\_sufficiency} (1--5): Do the provided positive passages contain enough information to answer the query?
\begin{itemize}
  \item 1: No useful information to answer the query
  \item 2: Very limited information, cannot reasonably answer
  \item 3: Partial information, can answer incompletely
  \item 4: Most information needed, can answer with minor gaps
  \item 5: Complete information to fully answer the query
\end{itemize}

\hrule
\vspace{0.35em}

\textbf{Query:}\\
\{query\}

\vspace{0.35em}

\textbf{Positive Passages (Evidence):}\\
\{positive\_passages\}

\vspace{0.35em}

\textbf{Answer:}\\
\{answer\_section\}

\vspace{0.35em}
\hrule
\vspace{0.35em}


\textbf{Output Format (JSON only, no additional text):}
\{
\newline\quad "Readability": <integer 1--5>,
\newline\quad "Clarity": <integer 1--5>,
\newline\quad "Evidence\_usefulness": <integer 1--5>,
\newline\quad "Evidence\_sufficiency": <integer 1--5>,
\newline\quad "rationale": "Brief explanation of your ratings"
\newline\}\end{tcolorbox}
\caption{\textbf{Prompt used for LLM-based dataset quality assessment.}
The judge rates Readability, Clarity, Evidence usefulness, and Evidence sufficiency on a 1--5 Likert scale.}
\label{fig:llm_quality_prompt_box}
\end{figure*}

\begin{table}[t]
\centering
\caption{LLM-based quality assessment by domain (mean Likert score; 1--5).}
\label{tab:llm_quality_by_domain}
\resizebox{\linewidth}{!}{
\begin{tabular}{l r c c c c}
\toprule
\textbf{Domain} & \textbf{N} & \textbf{Read.} & \textbf{Clar.} & \textbf{Usef.} & \textbf{Suff.} \\
\midrule
Aviation & 110 & 4.72 & 4.42 & 3.83 & 4.31 \\
Biology & 97 & 4.70 & 4.30 & 4.15 & 3.96 \\
Physics & 94 & 4.43 & 4.22 & 3.68 & 4.35 \\
Bioinformatics & 90 & 4.34 & 4.06 & 3.49 & 4.04 \\
Earthscience & 85 & 4.60 & 4.28 & 4.06 & 4.25 \\
Quantumcomputing & 83 & 4.25 & 4.11 & 3.86 & 4.51 \\
Psychology & 76 & 4.51 & 4.14 & 3.76 & 4.29 \\
Crypto & 74 & 4.14 & 3.95 & 3.23 & 3.82 \\
Travel & 68 & 4.68 & 4.35 & 3.94 & 3.90 \\
Bitcoin & 64 & 4.33 & 4.06 & 3.69 & 4.42 \\
Sustainability & 62 & 4.63 & 4.21 & 3.82 & 4.40 \\
Medicalsciences & 54 & 4.37 & 4.09 & 3.89 & 3.54 \\
Math & 45 & 4.38 & 4.18 & 3.51 & 4.22 \\
Pm & 44 & 4.27 & 4.05 & 4.00 & 3.68 \\
Gis & 44 & 4.52 & 4.25 & 3.34 & 3.98 \\
Philosophy & 41 & 4.05 & 3.71 & 3.90 & 4.41 \\
Chemistry & 40 & 4.55 & 4.33 & 4.00 & 3.77 \\
Bioacoustics & 39 & 4.54 & 4.10 & 3.72 & 4.24 \\
Askubuntu & 35 & 4.66 & 4.46 & 4.20 & 3.99 \\
Quant & 34 & 4.26 & 4.00 & 3.74 & 3.35 \\
Economics & 31 & 4.45 & 4.16 & 3.97 & 4.30 \\
Robotics & 30 & 4.43 & 4.03 & 3.90 & 4.43 \\
Law & 30 & 4.50 & 4.10 & 4.03 & 4.20 \\
Christianity & 30 & 4.67 & 4.40 & 4.03 & 3.97 \\
Academia & 26 & 4.54 & 4.19 & 3.92 & 3.50 \\
Gaming & 26 & 4.58 & 4.31 & 3.92 & 3.69 \\
Islam & 26 & 4.42 & 4.08 & 3.96 & 3.69 \\
Apple & 14 & 4.57 & 4.29 & 3.14 & 4.00 \\
Salesforce & 10 & 4.10 & 3.90 & 3.30 & 3.80 \\
\bottomrule
\end{tabular}}
\end{table}

\clearpage

\section{RAG answer evaluation prompt (GPT-4 judge)}
\label{app:rag_judge_prompt}

We evaluate answer correctness using an LLM-as-a-judge protocol following BRIGHT. The judge receives the query, the model-generated answer, and the reference answer, and returns (i) a brief rationale and (ii) a scalar score from 0 to 100 indicating coverage of the reference answer.

\begin{figure}[t]
\centering
\begin{tcolorbox}[
  colback=gray!5,
  colframe=gray!75,
  title=GPT-4 Prompt for RAG Answer Evaluation,
  width=0.8\textwidth
]
\small\ttfamily
You are a teacher to judge student's answer. \\

---------- PROBLEM START ---------- \\
\{query\} \\
---------- PROBLEM END ---------- \\

---------- STUDENT ANSWER START ---------- \\
\{pred\} \\
---------- STUDENT ANSWER END ---------- \\

---------- REFERENCE ANSWER START ---------- \\
\{gold\} \\
---------- REFERENCE ANSWER END ---------- \\

Criteria: \\
0 - The student's answer is completely irrelevant or blank. \\
10 - The student's answer addresses about 10\% of the reference content. \\
20 - The student's answer addresses about 20\% of the reference content. \\
30 - The student's answer addresses about 30\% of the reference content. \\
40 - The student's answer addresses about 40\% of the reference content. \\
50 - The student's answer addresses about 50\% of the reference content. \\
60 - The student's answer addresses about 60\% of the reference content. \\
70 - The student's answer addresses about 70\% of the reference content. \\
80 - The student's answer addresses about 80\% of the reference content. \\
90 - The student's answer addresses about 90\% of the reference content. \\
100 - The student's answer addresses about 100\% of the reference content. \\

Use the following format to give a score: \\
REASON: \\
Describe why you give a specific score \\
SCORE: \\
The score you give, e.g., 60 \\
Do not say anything after the score. \\
\end{tcolorbox}
\caption{Prompt used to evaluate RAG-generated answers with GPT-4 as a judge.}
\label{fig:rag_judge_prompt}
\end{figure}

\clearpage

 \section{Dataset Examples}
 \label{app:dataset_examples}

\begin{table*}[t]
\caption{\textbf{Bioinformatics example.} A randomly sampled query with one positive and one negative document from MM-BRIGHT.}
\centering
\small
\begin{tabular}{p{1\textwidth}}
\toprule
\textit{\textbf{Query}} \\
\midrule
How can I improve or otherwise investigate an unreliable genome tree?\\ 
Summary My genome tree doesn't agree with my gene trees and I get the feeling that my genome tree might be wrong, possibly due to long branch attraction, but I don't know how to check/fix it.\\ 
Background\\ 
I have a set of genes of interest recovered from genomes and metagenomes. I also have some bins built from the metagenomes of interest.\\ 
Previous work I've built multiple gene trees, which tend to agree with each other, as well as with other phylogenetic methods (16S rRNA, initial genome trees where available, conserved orientation of the genes of interest). So at this point I though I had a pretty good idea of how my genes / species of interest are linked to the species tree.\\ 
Concern\\ 
 I've assembled the metagenomes and recovered bins. I've refined bins of interest with anvio, based on sequence composition / differential coverage, and kept only the bins which appeared quite high quality (>70\% completion*,\\ 
  - with the exception of a bin of particular interest which I accepted with 62\%. I have de-replicated these bins, added some known genomes, and tried to build a species tree with Orthofinder (with the proteomes). There is a problem occurs with a clade of 3 bins I will call 'X'.\\ 
  image: enter image description here \\ 
  Trees Based on the gene-of-interest phylogenies, X looks like probably a sister-group to the rest of known genomes from order 'A'. Which is very interesting. With the 16S rRNA, it's not quite as clear (there are no 16S rRNA sequences in the bins); however, some 16S rRNA amplified from one of the original samples fit the bill as 'sister-clade to basically everything else in group A'). The first (100 or so) BLAST hits for this 16S sequence are all from group 'A', still. Tree conflict However, in the proteome tree, X as well as Y (Y being a sequenced genome that I suspected was related to X based on 16S data) end up in order 'B' instead of 'A'. GTDBtk also labels X as belonging to clade 'B', not 'A'. Now,I know that genes and genomes don't always have the same evolutionary trajectories. But multiple sets of data lead me to a different conclusion than the proteome tree, so I have to at least question whether the proteome tree is correct; what's likely to be the problem (I suspect long-branch attraction), and, if it is long-branch attraction (or anything else) how can I uncover the correct signal?\\ 
 Thank you for your time! \\
\midrule
\textit{\textbf{Query images}} \\
\midrule
\includegraphics[width=0.6\linewidth]{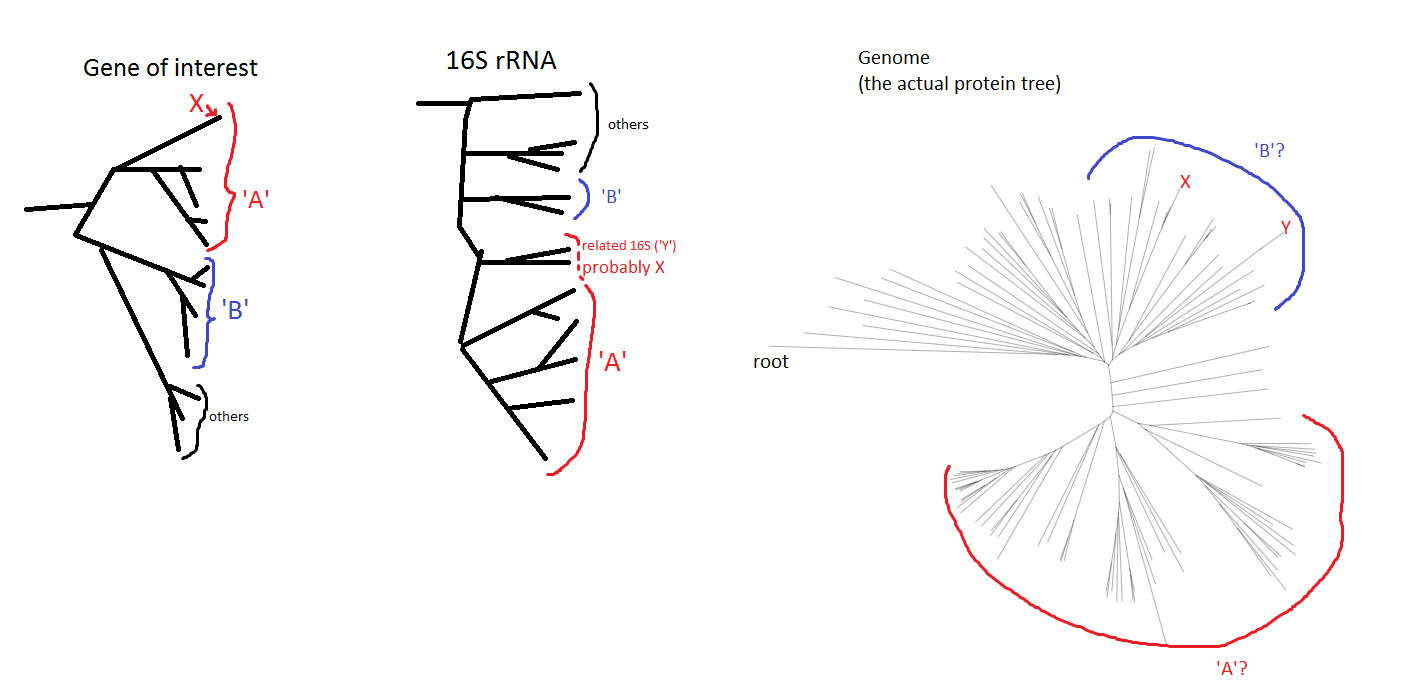}\\
\midrule
\textit{\textbf{Example positive document}} \\
\midrule
GBLOCKS is currently the most frequently used masking program that attempts to assess the quality of alignment position by position. It calculates the degree of conservation for each aligned position and then uses it to select conserved “blocks” for further analyses [27]. However, positions with low conservation scores could still be homologous (e.g., fast evolving sites). Such sites might contain useful phylogenetic information, sometimes more so than these highly conserved positions. To overcome this limitation, GBLOCKS tries to “rescue” these poorly conserved but potentially homologous positions as long as they belong to a block flanked by highly conserved columns at both ends and satisfy a set of ad hoc rules (e.g., the maximum number of contiguous nonconserved positions allowed is 8 and the minimum length of a block is 10). However, in real alignments, homologous regions are not always punctuated by highly conserved columns. In addition, these ad hoc rules are quite arbitrary with little theoretic support. \\
\midrule
\textit{\textbf{Example negative document}} \\
\midrule
Ammonium nitrate explosive systems DOEpatents Stinecipher, Mary M.; Coburn, Michael D. 1981-01-01 Novel explosives which comprise mixtures of ammonium nitrate and an ammonium salt of a nitroazole in desired ratios are disclosed. A preferred nitroazole is 3,5-dinitro-1,2,4-triazole. The explosive and physical properties of these explosives may readily be varied by the addition of other explosives and oxidizers. Certain of these mixtures have been found to act as ideal explosives. Ammonium nitrate explosive systems Stinecipher, Mary M.; Coburn, Michael D. Novel explosives which comprise mixtures of ammonium nitrate and an ammonium salt of a nitroazole in desired ratios are disclosed. A preferred nitroazole is 3,5-dinitro-1,2,4-triazole. The explosive and physical properties of these explosives may readily be varied by the addition of other explosives and oxidizers. Certain of these mixtures have been found to act as ideal explosives. 
 \\
\bottomrule
\end{tabular}
\label{tab:bioinformatics_example}
\end{table*}

\begin{table*}[t]
\caption{\textbf{Crypto example.} A randomly sampled query with one positive and one negative document from MM-BRIGHT.}
\centering
\small
\begin{tabular}{p{1\textwidth}}
\toprule
\textit{\textbf{Query}} \\
\midrule
Why do Feistel ciphers need round keys?\\ 
Looking at the design for Feistel ciphers, they use a list of round keys which are generated from the main key using the key schedule of the associated block cipher. Some block ciphers need this as to prevent repetition, but why does a Feistel network need it?\\ 
image: Feistel Network\\ 
If \$F\$ is a good PRF, then the output should be indistinguishable from random after the first few rounds. Under the random oracle model, one would expect that \$L\_0\$ is made pseudorandom after the first round, then \$R\_0\$ is made pseudorandom right after that. Continuing this should never return you back to the plaintext in a reasonable amount of time. So my question is why are round keys used in the Feistel network as opposed to making all the round keys the same? Did I get anything wrong in my reasoning? \\
\midrule
\textit{\textbf{Query images}} \\
\midrule
\includegraphics[height=0.6\linewidth]{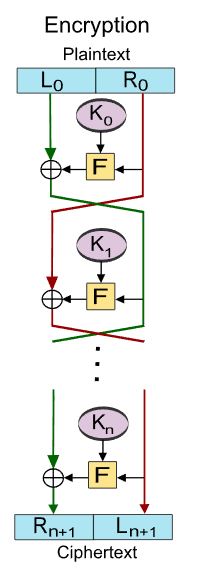}\\
\midrule
\textit{\textbf{Example positive document}} \\
\midrule
Key Features of Feistel Cipher\\ 
Works on blocks of data (not on individual characters like substitution ciphers).\\ 
The plaintext block is split into two halves: left (L) and right (R).\\ 
In each round, one half is modified using a function of the other half and a round key.\\ 
Multiple rounds make the cipher stronger.\\ 
Same process is used for both encryption and decryption. \\
\midrule
\textit{\textbf{Example negative document}} \\
\midrule
Before moving on to the online part of the sampling, we check for some bad events on \textbackslash{}(\textbackslash{}tau \textbackslash{}) itself. The event bad\textbackslash{}(\textbackslash{}tau \textbackslash{})-switch comes from the PRP-PRF switch we perform when we respond to the adversaries queries with replacement, instead of without replacement, as a permutation would do. The event bad\textbackslash{}(\textbackslash{}tau \textbackslash{})-\textbackslash{}(\textbackslash{}widehat\{Y\}\textbackslash{}) is the forced collision on \textbackslash{}(\textbackslash{}widehat\{Y\}\textbackslash{}) we mentioned earlier. bad\textbackslash{}(\textbackslash{}tau \textbackslash{})-3path involves a simultaneous 3-collision on R and S, which must involve a path of length 3. (For instance, one way to achieve this is as follows: an encryption query \textbackslash{}((L\_1, R)\textbackslash{}) giving \textbackslash{}((S, T\_1)\textbackslash{}); then a decryption query \textbackslash{}((S, T\_2)\textbackslash{}) yielding \textbackslash{}((L\_2, R)\textbackslash{}), making a path of length 2; and finally, a second encryption query with \textbackslash{}((L\_3, R)\textbackslash{}) giving \textbackslash{}((S, T\_3)\textbackslash{}), extending the path to length 3.) Finally, the event bad\textbackslash{}(\textbackslash{}tau \textbackslash{})-3coll involves a 3-collision on R or S where the last two come from oracle outputs. The precise definitions of these bad events are given in Fig. 3. \\
\bottomrule
\end{tabular}
\label{tab:crypto_example}
\end{table*}

\begin{table*}[t]
\caption{\textbf{Earthscience example.} A randomly sampled query with one positive and one negative document from MM-BRIGHT.}
\centering
\begin{tabular}{p{0.97\textwidth}}
\toprule
\textit{\textbf{Query}} \\
\midrule
How are subsurface wave speeds determined without subsurface sensors?\\ 
This is something I've never quite understood from a geology class I took years ago:\\ 
Consider the following picture (courtesy of wikipedia)\\ 
 image: enter image description here\\ 
Obviously, we can't possibly have sensors deep in the mantle (or core).  So, how exactly are these wave speeds determined? \\
\midrule
\textit{\textbf{Query images}} \\
\midrule
\begin{minipage}{0.49\linewidth}
  \centering
  \includegraphics[width=\linewidth]{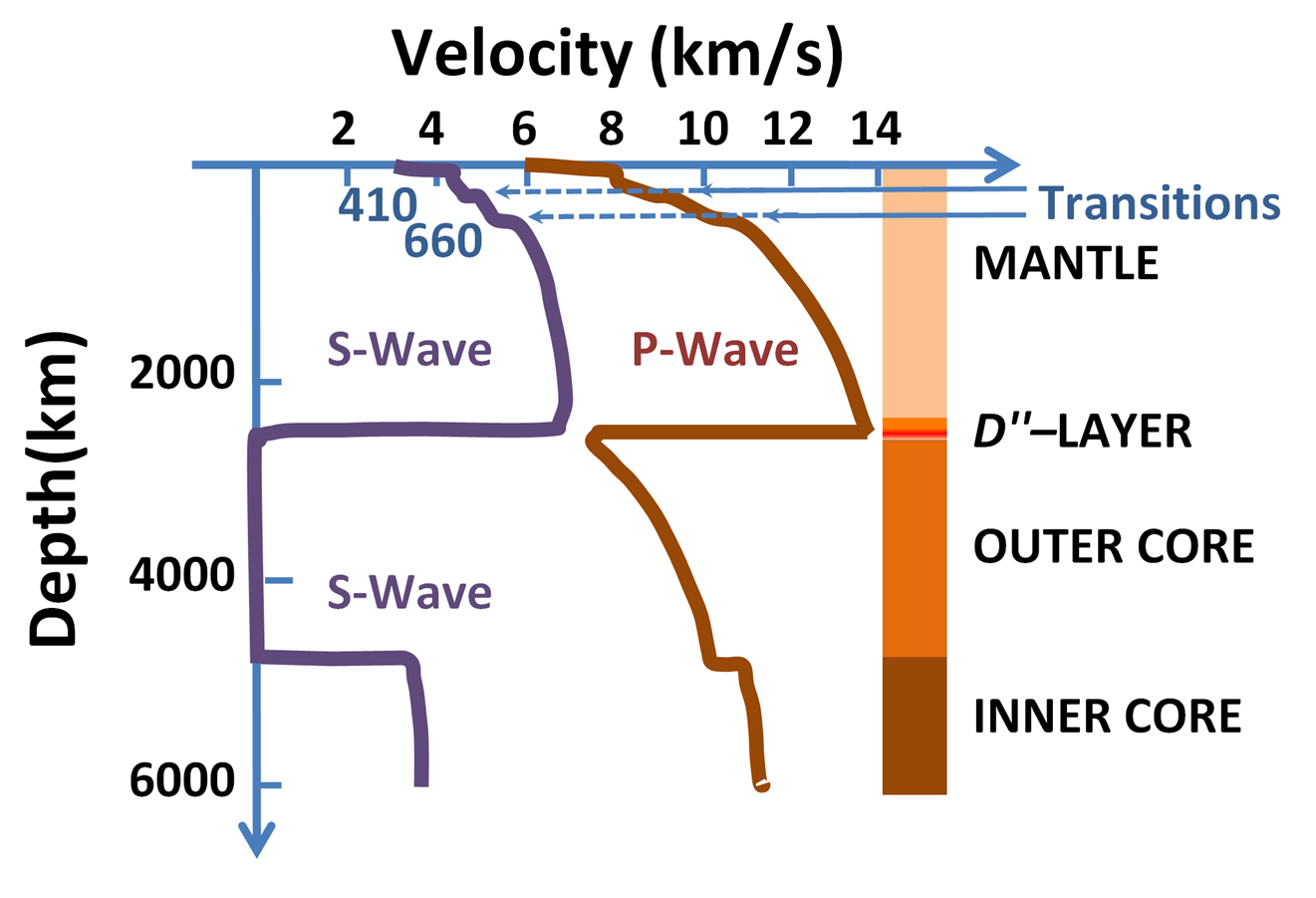}
\end{minipage} \\
\midrule
\textit{\textbf{Example positive document}} \\
\midrule
Seismic tomography or seismotomography is a technique for imaging the subsurface of the Earth using seismic waves.[2] The properties of seismic waves are modified by the material through which they travel. By comparing the differences in seismic waves recorded at different locations, it is possible to create a model of the subsurface structure. Most commonly, these seismic waves are generated by earthquakes or man-made sources such as explosions. Different types of waves, including P, S, Rayleigh, and Love waves can be used for tomographic images, though each comes with their own benefits and downsides and are used depending on the geologic setting, seismometer coverage, distance from nearby earthquakes, and required resolution. The model created by tomographic imaging is almost always a seismic velocity model, and features within this model may be interpreted as structural, thermal, or compositional variations. Geoscientists apply seismic tomography to a wide variety of settings in which the subsurface structure is of interest, ranging in scale from whole-Earth structure to the upper few meters below the surface. \\
\midrule
\textit{\textbf{Example negative document}} \\
\midrule
Earthquake energy is a function of magnitude. Both the magnitude and the seismic moment are related to the amount of energy that is radiated by an earthquake. \\
\bottomrule
\end{tabular}
\label{tab:earthscience_example}
\end{table*}

\begin{table*}[t]
\caption{\textbf{Medicalsciences example.} A randomly sampled query with one positive and one negative document from MM-BRIGHT.}
\centering
\small
\begin{tabular}{p{1\textwidth}}
\toprule
\textit{\textbf{Query}} \\
\midrule
Why does congenital duodenal obstruction cause a "double bubble" sign to appear on imaging?\\ 
Duodenal obstruction is caused by failure of recanalization of the duodenum during embryological organogenesis. It classically presents as a "double bubble" sign on x-ray or ultrasound. I understand that the more proximal (left-sided) dilation is due to gas in the stomach, but what is the cause of the more distal (right-sided) dilation?\\ 
image: enter image description here \\
\midrule
\textit{\textbf{Query images}} \\
\midrule
\includegraphics[width=0.4\linewidth]{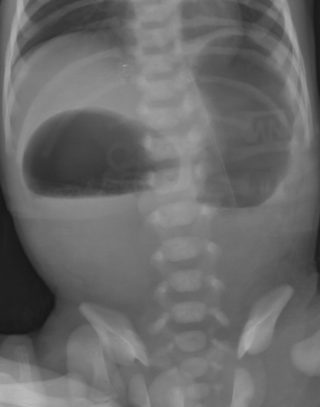}\\
\midrule
\textit{\textbf{Example positive document}} \\
\midrule
Antenatal imaging will show a double bubble, an echoless stomach filled with amniotic fluid, and a second nearby but the more distal fluid-filled, often circular (but blind-ended) structure (the second bubble) that is the obstructed portion of the duodenum.  The use of prenatal ultrasound has allowed for an earlier diagnosis of duodenal atresia.  An advantage of neonatal abdominal ultrasound is that it can be performed in the neonatal intensive care unit or nursery.  When antenatal ultrasound is performed, the duodenum is usually not filled with fluid, and the presence of a fluid-filled duodenum suggests duodenal atresia.  If a double-bubble sign is seen on antenatal ultrasound, then the sonographer needs to demonstrate a connection between the two fluid-filled structures because foregut duplication cyst, as well as other abdominal cysts, may simulate the appearance of a double-bubble sign.[7][8]\\ 
The initial postnatal radiographic evaluation for diagnosing duodenal atresia is a plain abdominal x-ray.  There is gas in the stomach and the proximal duodenum in duodenal atresia but an absence of gas distally in the small or large bowel.  A plain abdominal x-ray may reveal the double-bubble sign, which is seen postnatally as a large radiolucent (air-filled) stomach usually in the normal position to the left of the midline, and a smaller, more distal bubble to the right of the midline, which represents a dilated duodenum.  A double-bubble sign on an abdominal x-ray is a reliable indicator of duodenal atresia.  Other causes of intestinal obstruction may simulate a double-bubble sign.  The annular pancreas is the second most common cause of duodenal atresia.  Jejunal or more distal obstruction may dilate more distally, or more than two bubbles may be present.[8] \\
\midrule
\textit{\textbf{Example negative document}} \\
\midrule
Uterine fibroids are benign monoclonal tumors [Reference Bowden, Skorupski, Kovanci and Rajkovic1] that occur in up to 60\% of reproductive age women and 80\% of women during their lifetime [Reference Laughlin, Schroeder and Baird2]. While many women with fibroids are asymptomatic, the clinical presentation and symptoms can vary and include menorrhagia, pelvic pressure, bowel and urinary complaints, pain, and infertility. Fibroids are the leading cause of hysterectomy and are a significant public health care burden, estimated to exceed 34 billion dollars annually in the United States [Reference Cardozo, Clark, Banks, Henne, Stegmann and Segars3] (Figs 31.1 and 31.2). Fig. 31.1 Multiple fibroids on hysterectomy specimen. Fig. 31.2 Multiple fibroids during myomectomy showing submucous, intramural and subserous fibroids. \\
\bottomrule
\end{tabular}
\label{tab:medicalsciences_example}
\end{table*}

\begin{table*}[t]
\caption{\textbf{Quant example.} A randomly sampled query with one positive and one negative document from MM-BRIGHT.}
\centering
\small
\begin{tabular}{p{1\textwidth}}
\toprule
\textit{\textbf{Query}} \\
\midrule
Vanilla Option Prices from Local Vol Surface (using neither MC nor PDE)\\ 
There are numerous papers that describe the derivation of the Local-Vol equation using available market prices of options. For example:\\ 
 - Dupire's formula (see e.g. OpenGamma (2013)) gives us LV in terms of absolute strike \& price:\\ 
image: enter image description here\\ 
 - Gatheral (2003) gives us LV in terms of log-strike \& total BS-implied variance:\\ 
image: enter image description here\\ 
But what I am wondering is, given a pure Local Vol surface \$\textbackslash{}sigma\_\{LV\}\$, how does one recover Vanilla European market prices \$C\_\{Mkt\}\$?\\ 
I can think of Monte-Carlo or PDE methods, but are there standard (semi)-analytical techniques which are faster than the first two? That is, are there ways to do:\\ 
 - Convert local vols \$\textbackslash{}sigma\_\{LV\}\textbackslash{}rightarrow\$ BS-implied vols \$\textbackslash{}sigma\_\{BS\}\textbackslash{}rightarrow\$ market prices \$C\_\{Mkt\}\$ (3 steps)\\ 
 - Convert local vols \$\textbackslash{}sigma\_\{LV\}\textbackslash{}rightarrow\$ market prices \$C\_\{Mkt\}\$ (2 steps) \\
\midrule
\textit{\textbf{Query images}} \\
\midrule
\includegraphics[width=0.3\linewidth]{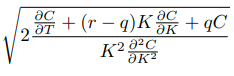}\\
\includegraphics[width=0.3\linewidth]{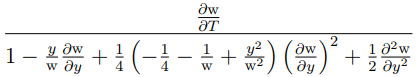}\\
\midrule
\textit{\textbf{Example positive document}} \\
\midrule
It is well known that, in the short-maturity limit, the implied volatility approaches the integral harmonic mean of the local volatility with respect to log-strike; see [H. Berestycki, Busca, and Florent, Quant. Finance, 2 (2002), pp. 61--69]. This short paper is dedicated to a complementary model-free result: An arbitrage-free implied volatility in fact is the harmonic mean of a positive function for any fixed maturity. We investigate the latter function, which is tightly linked to Fukasawa's invertible map 1/2\\ 
M. Fukasawa, Math. Finance, 22 (2012), pp. 753\-\-762, and its relation with the local volatility surface. It turns out that the log-strike transformation\\ 
defines a new coordinate system in which the short-dated implied volatility approaches the arithmetic (as opposed to harmonic) mean of the local volatility. \\
\midrule
\textit{\textbf{Example negative document}} \\
\midrule
Figure 39 shows plots of the first seven polynomials for the various families of orthogonal polynomials in our study. In our implementation, we employed the following recursive definitions: Chebyshev Polynomials of the Second Kind Legendre Polynomials Bessel Polynomials Laguerre Polynomials These recursive formulations were the simplest to implement the KAN layers. However, using these recursive definitions is not the most computationally efficient method, as using some of the series or closed-form expressions for the polynomials could be quicker. Moreover, it would be prudent to test the alternative of using min-max normalization instead of tanh to keep the inputs in the [-1, 1] range. Below is an example of how a Legendre KAN layer is implemented in our framework: This custom layer generates the KAN transformation for each layer based on the Legendre polynomials, while similar classes handle the other orthogonal polynomials. In , the network uses B-splines for the activation functions with additional grid extension mechanisms to adjust the grid dynamically during training. However, in our implementation, we rely purely on the intrinsic properties of the orthogonal polynomials, which provide stability and efficiency without the need for grid adjustments. In future work, it would be prudent to test whether combining layers of different polynomials increases performance. Also, from our experiments, it is clear that our KANs would benefit significantly from more regularization since dropout alone is not enough. We anticipate that the targeted pruning of nodes mechanism used in the original paper should work very well. \\
\bottomrule
\end{tabular}
\label{tab:quant_example}
\end{table*}

\begin{table*}[t]
\caption{\textbf{Quantumcomputing example.} A randomly sampled query with one positive and one negative document from MM-BRIGHT.}
\centering
\small
\begin{tabular}{p{1\textwidth}}
\toprule
\textit{\textbf{Query}} \\
\midrule
Devising "structured initial guesses" for random parametrized quantum circuits to avoid getting stuck in a flat plateau\\ 
The recent McClean et al. paper Barren plateaus in quantum neural network training landscapes shows that for a wide class of reasonable parameterized quantum circuits, the probability that the gradient along any reasonable direction is non-zero to some fixed precision is exponentially small as a function of the number of qubits.\\ 
This seems to affect Noisy Intermediate-Scale Quantum (NISQ) programs (as proposed by e.g. John Preskill) since they involve hybrid quantum-classical algorithms, ie training a parameterized quantum circuit with a classical optimization loop.\\ 
image: Fig 1 from the paper\\ 
My question: How do you avoid getting stranded on those barren plateaus? Concretely, how would one go about building one's Ansatz Haar states to avoid getting stuck in those plateaus? The paper proposes but does not elaborate:\\ 
  One approach to avoid these landscapes in the quantum setting is to\\ 
  use structured initial guesses, such as those adopted in quantum\\ 
  simulation. \\
\midrule
\textit{\textbf{Query images}} \\
\midrule
\includegraphics[width=0.6\linewidth]{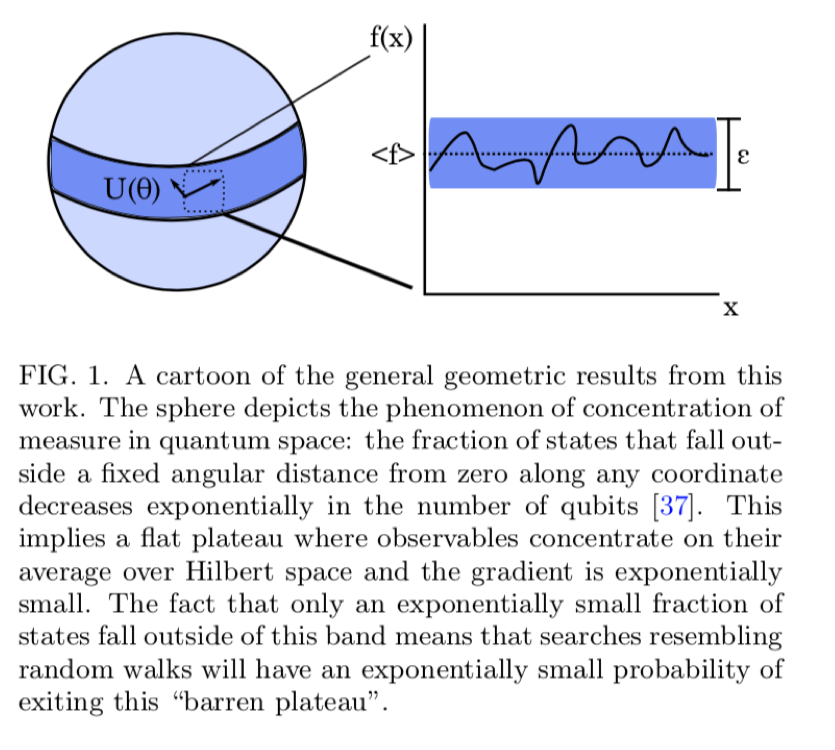}\\
\midrule
\textit{\textbf{Example positive document}} \\
\midrule
3.- The initial condition for the optimization is too far from the solution, which happens extremely often when is taken randomly. Having good initial conditions increase the probability to start the optimization with a non-zero gradient so that the protocol can converge to the optimum. For example, in [5] is proposed to classically pre-train the variational circuit with tensor networks to have a good initial condition for the quantum optimization.\\ 
4.- The problem of a lifetime: the noise [6]. The impact of the noise can be reduced by hardware-efficient circuits and local measurements.\\ 
In this project, we develop a Qiskit module that includes several proposals to reduce the impact of barren plateaus in variational quantum algorithms. We also provide an early implementation on Pennylane. For more details visit the introductory notebook or each individual tutorial notebook. \\
\midrule
\textit{\textbf{Example negative document}} \\
\midrule
Advances in quantum computing have led to the NISQ era. Physical quantum computers with upwards of hundreds of qubits are available. Certain quantum computers are also available as cloud services, such as IBM’s quantum computers [2]. These NISQ era quantum computers are noisy, meaning that the quantum states lack stability. As time progresses, the states decohere, resulting in a loss of accuracy. Furthermore, the gates used to manipulate the qubits often create slight deviations, resulting in incorrect solutions [8]. Despite these issues, the realization of such machines enables the implementation of a certain subclass of quantum and HQC algorithms. In order to successfully implement these algorithms, several considerations must be made during the implementation and testing processes. Several such considerations include how data are encoded in qubits (e.g., binary encoding or amplitude encoding) and how oracles are implemented (e.g., how many ancilla qubits are needed and how many gates are required to implement the required functionality) [8].\\
\bottomrule
\end{tabular}
\label{tab:quantumcomputing_example}
\end{table*}

\begin{table*}[t]
\caption{\textbf{Sustainability example.} A randomly sampled query with one positive and one negative document from MM-BRIGHT.}
\centering
\small
\begin{tabular}{p{1\textwidth}}
\toprule
\textit{\textbf{Query}} \\
\midrule
Do light fixtures with small solar panels require direct sunlight?\\ 
I purchased an Osram Ledvance DoorLED Solar product which charges its own batteries and turns on automatically at nighttime to provide light.  It requires a 7 hour minimum charge time.  When there is no motion, it dispenses light at 20\% of its maximum luminance.\\ 
I installed the product to receive daylight (not direct sunlight) and had it charge for 12 hours.  However, at night, I realized the product had stopped providing light after about 5 hours of use at 20\% intensity.  It is placed at a location where I am positive there is no motion to trigger full luminance.\\ 
Web research I have done indicates that direct sunlight is not required for solar panels and nowhere on the product packaging does it say that direct placement in sunlight is required.\\ 
Is direct sunlight a requirement for such solar-powered products?\\ 
image: enter image description here\\ 
image: enter image description here \\
\midrule
\textit{\textbf{Query images}} \\
\midrule
\includegraphics[width=0.3\linewidth]{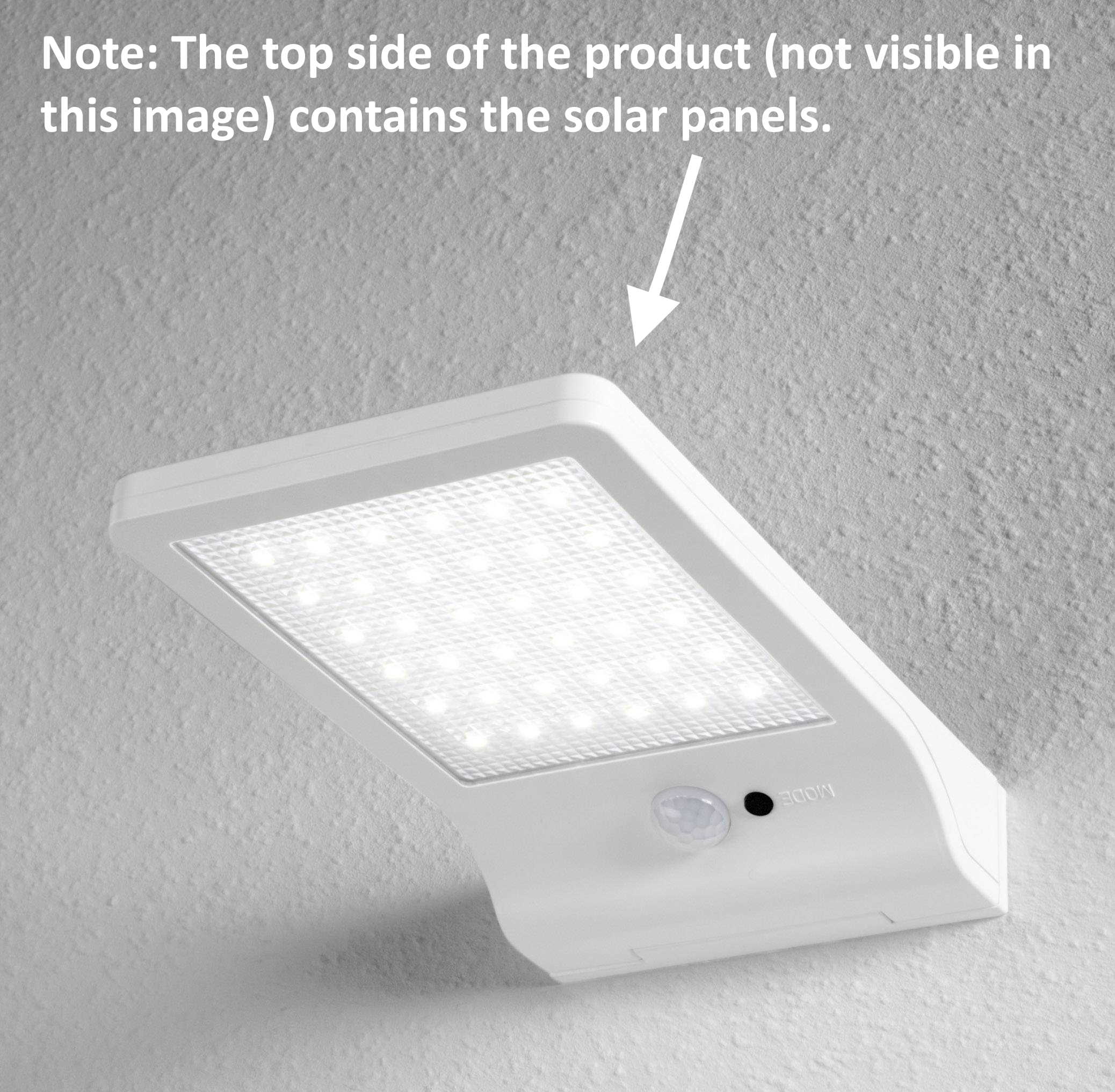}\\
\includegraphics[width=0.3\linewidth]{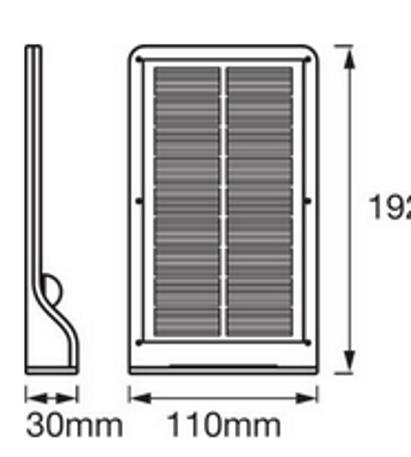}\\
\midrule
\textit{\textbf{Example positive document}} \\
\midrule
Do solar panels need direct sunlight to function?\\ 
Solar panels work even without direct sunlight. They can use both direct and indirect light, like on cloudy days. Yet, they work best with about four hours of direct sunlight daily.\\ 
How do photovoltaic cells in solar panels convert sunlight into electricity?\\ 
Solar panels have photovoltaic cells, usually made of silicon. These cells turn light photons into electricity by releasing electrons. The more light, the stronger the electric current. \\
\midrule
\textit{\textbf{Example negative document}} \\
\midrule
Dirt and debris can impact the performance of your solar LED lights. Create a regular cleaning schedule and remove any dust from your solar panels with a soft cloth. You should also replace the batteries in older solar models for maximum energy retention.Regular maintenance checks can also reveal loose wiring, faulty bulbs or prolific foliage obstructing light sources. Rectifying these issues at the onset ensures optimal brightness at night.\\
\bottomrule
\end{tabular}
\label{tab:sustainability_example}
\end{table*}

\begin{table*}[t]
\caption{\textbf{Philosophy example.} A randomly sampled query with one positive and one negative document from MM-BRIGHT.}
\centering
\small
\begin{tabular}{p{1\textwidth}}
\toprule
\textit{\textbf{Query}} \\
\midrule
How might aesthetics be radically Other?\\ 
I've read a handful of book by or about Levinas, but some time ago and without notes. IIRC his central ethical theme is that other people are not an aspect of the self, that our obligation to them exists in their being irreducible to ourselves.\\ 
E.g. you have this by Zahavi\\ 
image: Self awareness and Alterity\\ 
i.e. that alterity is originarily an ethical relation, so that its call to conscience is more important that any aspect of comprehension: and so (it seems to me) independent of how we seem to exist as individuals.\\ 
  Against these enigmas, every mode of comprehension runs aground....\\ 
  The other person is an event I can neither predict nor control.\\ 
Question:\\ 
Assuming that is a fair definition of "radical alterity", I wondered if there has been any attempt to apply a similar concept to aesthetics (rather than its usual place of obligation). \\
\midrule
\textit{\textbf{Query images}} \\
\midrule
\includegraphics[width=0.5\linewidth]{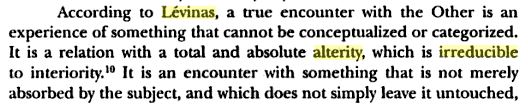}\\
\midrule
\textit{\textbf{Example positive document}} \\
\midrule
In his poetic travel report of Greece entitled Augenblicke in Griechenland, the Austrian writer Hugo von Hofmannsthal wrote the most beautiful description of his experience. The poet describes his feelings in front of the statues of five women covered with long outfits. In the philosophical tradition of Burke, Kant, Mendelssohn, Hegel, and Nietzsche, the sublime (das Erhabene) is first perceived as an addition or even a contradiction to the central concept of aesthetics as beauty. The difference is that beauty could be determined by what is sensitively present in appearance as an idea. This tradition goes from Plato to Hegel. Therefore, the sublime is understood by termination or cut concerning the appearance of the concept of beauty. Having recaptured the empty place of beauty, the sublime does not change the subject of our knowledge in an aesthetic sense. What should be changing refers to the whole body-soul relationship between humans and the world. With the sublime emerges a rare moment (Augenblick) of ascension over appearance and reality. \\
\midrule
\textit{\textbf{Example negative document}} \\
\midrule
In the preface to the massive 700+ pages Logical Investigations (1900–01), Husserl writes that his overarching aim is to provide a new foundation for logic and epistemology (Hua 18/6 [2001/I: 2]; for in depth discussions of the work, see, e.g., de Boer 1966 , Benoist 2001). The main task of the first part, Prolegomena to Pure Logic, is twofold: To offer a criticism of psychologism and to argue that scientific knowledge presupposes ideality. The position Husserl is targeting claims that the task of epistemology is to investigate the nature of our perceiving, believing, judging, and knowing. Since all of these phenomena are psychical processes, it may seem that only psychology can investigate them. This also holds true for our scientific and logical reasoning, and ultimately logic must therefore be regarded as part of psychology and the laws of logic as psycho-logical regularities, whose nature and validity must be empirically investigated and established (Hua 18/64, 89 [2001/I: 40, 56]). As Husserl argues, this line of reasoning commits the mistake of ignoring the fundamental difference between the domains of logic and psychology, “between ideal and real laws, between normative and causal regulation, between logical and real necessity, between logical and real grounds” (Hua 18/80 [2001/I: 50]). Logic is not an empirical science and is not concerned with the genesis of spatiotemporal objects or processes, but with the validity of ideal structures and laws. Psychology by contrast is an empirical science that investigates the empirical nature of consciousness. Whereas the domain of logic is characterized by certainty and exactness, the domain of psychology is characterized by the same mere probability as all empirical sciences (Hua 18/181 [2001: I/113–114]). A further mistake made by psychologism is that it doesn’t distinguish sufficiently between the object of knowledge and the act of knowing. Whereas the act of knowing is a subjective process that elapses in time and has a beginning and an end, the objects of logic, the logical truths, theories, principles, propositions, sentences, and proofs are not subjective experiences with temporal duration, but atemporal idealities. When I think of the theorem of Pythagoras and when you think of it, we must be able to think about the same theorem, even if our respective thought processes are different (Hua 19/49, 97–98 [2001/I: 195, 224–225]). 
\\
\bottomrule
\end{tabular}
\label{tab:philosophy_example}
\end{table*}

\begin{table*}[t]
\caption{\textbf{Physics example.} A randomly sampled query with one positive and one negative document from MM-BRIGHT.}
\centering
\small
\begin{tabular}{p{1\textwidth}}
\toprule
\textit{\textbf{Query}} \\
\midrule
Can I survive a free fall using a ramp and a rope?\\ 
Can I survive a free fall by carrying a very light and resistant ramp using a rope?\\ 
image: enter image description here\\ 
Note: lets assume the ramp is a little bit heavier at the bottom and I am very skilled at making it always land correctly, also I am wearing a ultra-oiled suit. \\
\midrule
\textit{\textbf{Query images}} \\
\midrule
\includegraphics[width=0.6\linewidth]{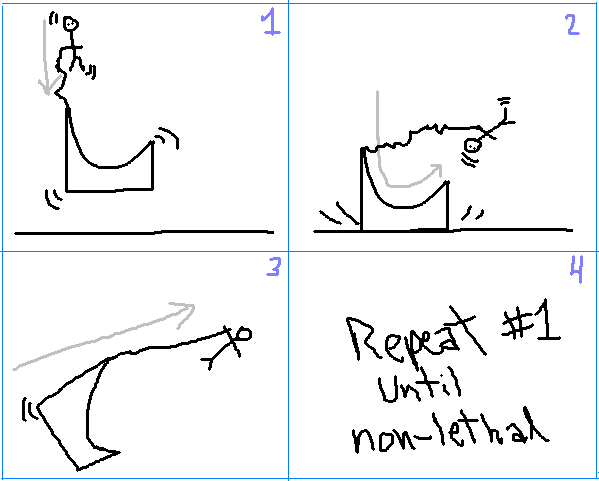}\\
\midrule
\textit{\textbf{Example positive document}} \\
\midrule
Free fall\\ 
From Wikipedia, the free encyclopedia\\ 
For other uses, see Free fall (disambiguation).\\ 
In classical mechanics, free fall is any motion of a body where gravity is the only force acting upon it. A freely falling object may not necessarily be falling down in the vertical direction. If the common definition of the word "fall" is used, an object moving upwards is not considered to be falling, but using scientific definitions, if it is subject to only the force of gravity, it is said to be in free fall. The Moon is thus in free fall around the Earth, though its orbital speed keeps it in very far orbit from the Earth's surface.\\ 
In a roughly uniform gravitational field gravity acts on each part of a body approximately equally. When there are no other forces, such as the normal force exerted between a body (e.g. an astronaut in orbit) and its surrounding objects, it will result in the sensation of weightlessness, a condition that also occurs when the gravitational field is weak (such as when far away from any source of gravity).\\ 
The term "free fall" is often used more loosely than in the strict sense defined above. Thus, falling through an atmosphere without a deployed parachute, or lifting device, is also often referred to as free fall. The aerodynamic drag forces in such situations prevent them from producing full weightlessness, and thus a skydiver's "free fall" after reaching terminal velocity produces the sensation of the body's weight being supported on a cushion of air. \\
\midrule
\textit{\textbf{Example negative document}} \\
\midrule
Essential for ecommerce, still-life, or prop-based storytelling. Matte black headphone — lying on white seamless, three-quarter angle, detailed reflections.Snippet: matte-black wireless headphone on white seamless, top-three-quarter angle, softbox Vintage camera — being adjusted, close-up of lens, on a wooden table.Snippet: vintage rangefinder camera being held, shallow DOF, warm tabletop texture Antique pocket watch — opening, hands moving, close macro of gears.Snippet: antique pocket watch open, macro view of gears, warm tungsten glow \\
\bottomrule
\end{tabular}
\label{tab:physics_example}
\end{table*}

\begin{table*}[t]
\caption{\textbf{Robotics example.} A randomly sampled query with one positive and one negative document from MM-BRIGHT.}
\centering
\small
\begin{tabular}{p{1\textwidth}}
\toprule
\textit{\textbf{Query}} \\
\midrule
How can we use the accelerometer for altitude estimation?\\ 
I am currently implementing an autonomous quadcopter which I recently got flying and which was stable, but is unable to correct itself in the presence of significant external disturbances. I assume this is because of insufficiently tuned PID gains which have to be further tweaked inflight.\\ 
Current progress:\\ 
 - I ruled out a barometer since the scope of my research is only indoor flight and the barometer has a deviation of +-5 meters according to my colleague.\\ 
 - I am currently using an ultrasonic sensor (HC-SR04) for the altitude estimation which has a resolution of 0.3cm.  However I found that the ultrasonic sensor's refresh rate of 20Hz is too slow to get a fast enough response for altitude correction.\\ 
 - I tried to use the accelerations on the Z axis from the accelerometer to get height data by integrating the acceleration to get velocity to be used for the rate PID in a cascaded pid controller scheme. The current implementation for the altitude PID controller is a single loop pid controller using a P controller with the position input from the ultrasonic sensor.\\ 
 - I had taken into account the negative acceleration measurements due to gravity but no matter how much I compute the offset, there is still the existence of a negative acceleration (eg. -0.0034). I computed the gravitational offset by setting the quadcopter to be still on a flat surface then collecting 20,000 samples from the accelerometer z axis to be averaged to get the "offset" which is stored as a constant variable. This variable is then subtracted from the accelerometer z-axis output to remove the offset and get it to "zero" if it is not accelerating. As said in the question, there is still the existence of a negative acceleration (eg. -0.0034). My quad then proceeds to just constantly climb in altitude. With only the ultrasonic sensor P controller, my quad oscillates by 50 cm.\\ 
  How can this consistent negative acceleration reading be effectively dealt with?\\ 
Possible Solution: I am planning to do a cascading PID contoller for the altitude hold with the innerloop (PID controller) using the accelerometer and the outer loop (P controller) using the sonar sensor. My adviser said that even a single loop P controller is enough to make the quadcopter hold its altitude even with a slow sensor. Is this enough? I noticed that with only the P gain, the quadcopter would overshoot its altitude.\\ 
image: enter image description here\\ 
 - Leaky Integrator: I found this article explaining how he dealt with the negative accelerations using a leaky integrator however I have a bit of trouble understanding why would it work since I think the negative error would just turn to a positive error not solving the problem. I'm not quite sure.\\ 
 - Single loop PD controller with the ultrasonic sensor only:\\ 
Is this feasible using feedback from a slow sensor?\\ 
Sources:\\ 
 - LSM303DLHC Datasheet:\\ 
 - Leaky integrator:\\ 
 - ArduPilot PID Loop: \\
\midrule
\textit{\textbf{Query images}} \\
\midrule
\includegraphics[width=0.6\linewidth]{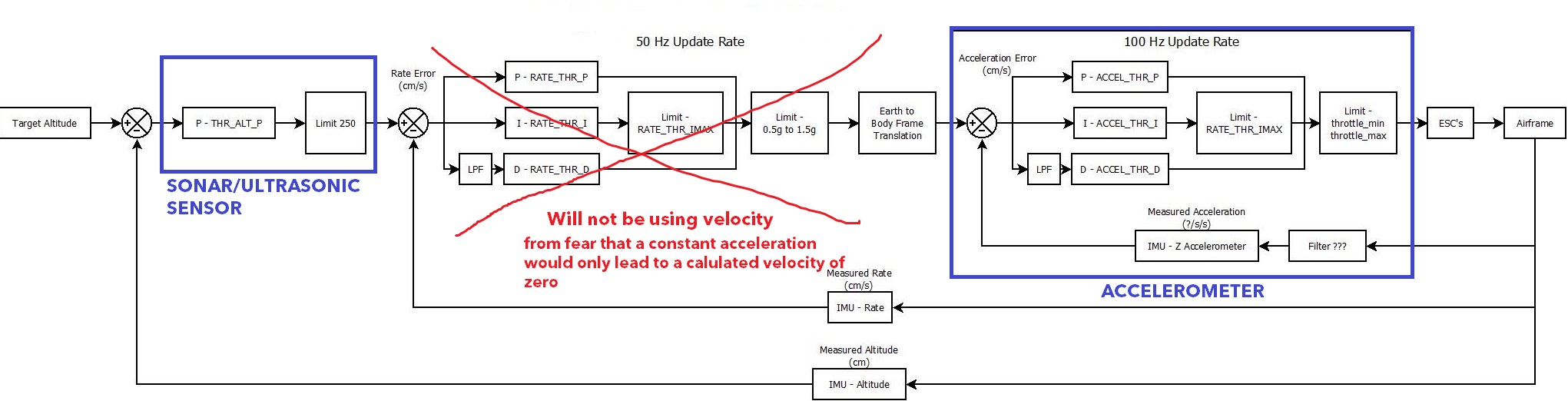}\\
\midrule
\textit{\textbf{Example positive document}} \\
\midrule
In robotics, localization refers to determining an object’s position and attitude in 2D or 3D space. This is a very important capability, especially in aerospace applications [1]. Using sensors, robots can maintain an estimate of their state, including velocity, attitude, and position [2]. Accelerometers within IMUs, which also include gyroscopes, are commonly used to calculate position by double integrating acceleration data [3]. However, IMU sensor readings inherently contain errors such as random noise and sensor bias, which degrade navigation performance [4]. Additionally, numerical integration introduces extra noise due to time discretization, exacerbating drift [5]. For instance, Euler integration results in local truncation errors, similar to those found in Taylor series approximations [6]. Thus, the double integration of accelerometer data to generate velocity and position leads to compounded drift errors [7], posing a significant challenge in autonomous UAV navigation. Recent experiments have shown the potential of fusing various sensors with IMUs to match the accuracy of IMU/GNSS systems [8]. This includes novel hardware combinations and algorithmic approaches to enhance sensor fusion accuracy in UAVs. Another approach in GPS-denied environments is the use of multiple UAVs, each with a range sensor and an IMU [9]. \\
\midrule
\textit{\textbf{Example negative document}} \\
\midrule
The use of the PID algorithm does not guarantee optimal control of the system or its control stability (see § Limitations, below). Situations may occur where there are excessive delays: the measurement of the process value is delayed, or the control action does not apply quickly enough. In these cases, lead–lag compensation is required to be effective. The response of the controller can be described in terms of its responsiveness to an error, the degree to which the system overshoots a setpoint, and the degree of any system oscillation. But the PID controller is broadly applicable since it relies only on the response of the measured process variable, not on knowledge or a model of the underlying process. \\
\bottomrule
\end{tabular}
\label{tab:robotics_example}
\end{table*}

\begin{table*}[t]
\caption{\textbf{Economics example.} A randomly sampled query with one positive and one negative document from MM-BRIGHT.}
\centering
\small
\begin{tabular}{p{1\textwidth}}
\toprule
\textit{\textbf{Query}} \\
\midrule
Why is railway electrification in North America far less common than in Europe?\\ 
Why is railway electrification in North America far less common than in Europe?  I suspect that the answer has something to do with the economics of electrification.  In Europe, railway electrification varies a lot between countries, ranging from 3\% in Ireland to 95\% in Luxembourg (not shown: 100\% in Switzerland):\\ 
image: Electrification\\ 
Source: European Commission: Mobility and Transport.\\ 
In the USA, less than 1000 km is electrified, or less than 0.5\% of the network length.\\ 
One might expect the answer to be related to population density, but electrification rate is 76\% in Sweden, 55\% in Finland, and nearly 50\% in Russia, three countries with lower or much lower population densities than the USA.  The US network is primarily freight whereas the European network is dual-use, but many freight-heavy railway lines in Europe (and Asian Russia) are electrified as well.  The Trans-Siberian railway electrification completed in the 21st century.  Why is it, then, that railway electrification is apparently economical in Europe, but not in the USA? \\
\midrule
\textit{\textbf{Query images}} \\
\midrule
\includegraphics[width=0.6\linewidth]{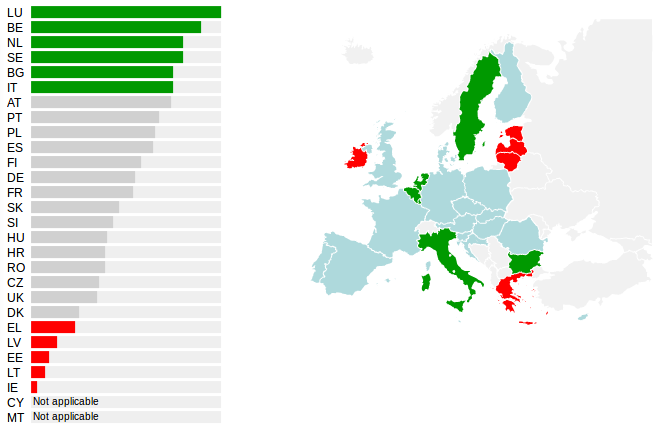}\\
\midrule
\textit{\textbf{Example positive document}} \\
\midrule
\textbf{economics/5373bf26\_0746.txt} \\
IElectric-Train-Germany-333.jpgf electric locomotives have so many advantages compared to diesel-powered locomotives, why aren't they more widespread in the United States? During much of the 20th century, U.S. railroads were the world leaders in innovation and in the use of cutting-edge technology. They now lag behind many other advanced nations, which have been investing for many years in electric-powered railroads. In the early to mid-20th century, steam engines were replaced by more efficient electric locomotives and diesel-electric (usually referred to as just diesel) locomotives. During that transition, U.S. railroad companies chose to switch to diesel over electric locomotives because of diesel’s much lower up-front costs, even though electric systems cost significantly less to operate and to maintain than diesel systems. Railroad operators in many other industrialized countries chose to switch to electric locomotives, partly because the railroads were owned by the governments of those countries, which could better afford the necessary transmission infrastructure. U.S. railroads have always been a regulated private sector industry, making it much harder for U.S. railroad companies to finance electrification upgrades than to build diesel-fueled systems. As a result, electrified rail is currently used on less than 1 percent of U.S. railroad tracks while electricity supplies more than one-third of the energy that powers trains globally.\\ 
A few passenger rail lines have been converted to electric power in the United States (Amtrak’s Northeast corridor and Harrisburg, PA, line), but the rest of passenger rail and all of freight rail is diesel-powered. The California commuter rail line (CalTrain) is currently being upgraded to very high speed rail (VHSR) service and will use electric power. The system is scheduled to be operational by 2022 and has an initial estimated cost of \$5 billion. Other electric VHSR systems (which would be electric-powered) are also being considered around the country, but do not yet have funding. \\
\midrule
\textit{\textbf{Example negative document}} \\
\midrule
\textbf{economics/3c97a732\_0780.txt} \\
A big problem for long-distance international freight services – despite the European Single Market allowing freedom of movement of goods, capital, labor and people and the Schengen area drastically reducing internal border controls – is the variety of differing standards for electrification, loading gauge, signaling, driver certificates and even gauge. Finland (Russian gauge), Portugal and Spain (Iberian gauge) use their own broad gauges, as do the Baltic States and several non-EU members (mostly Russian gauge). Rail Baltica is an EU-funded project to provide a standard gauge rail link in and through the Baltic countries, potentially connecting to a Helsinki-Tallinn tunnel. While attempts to unify the divergent standards date back to at least the 1880s with the Conférence internationale pour l'unité technique des chemins de fer (lit. 'international conference for the technical unity of railroads') in Bern, Switzerland, setting minimum standards for loading gauges (the so-called Berne gauge) and the so-called "Berne space" (the space reserved for railroad workers in buffer and chain couplers), most standards still differ widely between and even within countries as many rules only apply to newly-built infrastructure, as much of Europe's rail infrastructure was built in the 19th century, and upgrading it would be costly and disruptive. \\
\bottomrule
\end{tabular}
\label{tab:economics_example}
\end{table*}

\begin{table*}[t]
\caption{\textbf{Biology example.} A randomly sampled query with one positive and one negative document from MM-BRIGHT.}
\centering
\small
\begin{tabular}{p{1\textwidth}}
\toprule
\textit{\textbf{Query}} \\
\midrule
Spruce growing high up in a maple trunk: Can a partially rotted trunk completely sustain another tree?\\ 
I've come across a spruce tree that is growing 15ft up in the crotch of a sugar maple tree:\\ 
image: enter image description here\\ 
image: enter image description here\\ 
According to Google StreetView, the spruce has been there since before 2007 (over 10 years ago!).\\ 
I've seen other trees growing out of rotted stumps or in trunks near ground level before, but I've always just assumed that the roots of those trees reached down to soil -- in one way or another -- to get the nutrients they need for survival.\\ 
But with the subject tree, it's so high up that I would assume that that the roots do not have access to any soil. It's my guess that the spruce is getting all of its nutrients from decomposed wood in some rotten portion of the trunk.\\ 
I didn't know biology worked like this. I wouldn't have guessed that a live, partially decomposed trunk could contain 100\% of the nutrients that a tree needs to survive for 10+ years. I know soil is made, in part, from decomposed wood, but it's my understanding that there are a lot more ingredients too.\\ 
Can the decomposed wood in a partially rotted trunk completely sustain a tree? \\
\midrule
\textit{\textbf{Query images}} \\
\midrule

\noindent
\begin{minipage}{0.35\linewidth}
  \centering
  \includegraphics[width=\linewidth]{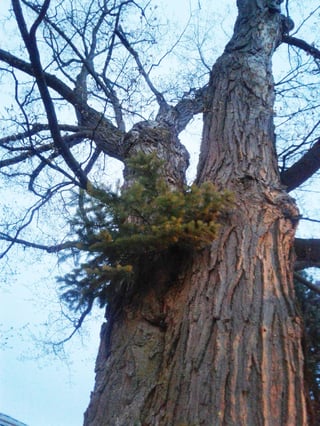}
\end{minipage}\hfill
\begin{minipage}{0.49\linewidth}
  \centering
  \includegraphics[width=\linewidth]{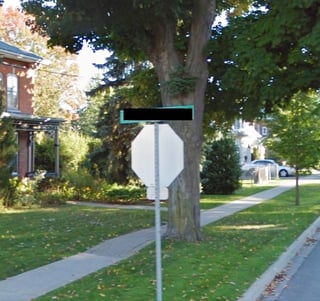}
\end{minipage}\\
\midrule
\textit{\textbf{Example positive document}} \\
\midrule 
The nurse log phenomenon is particularly visible in the northwest region of America, where Sitka spruces, hemlocks and Douglas-firs grow in abundance. As seedlings germinated on the nurse logs begin to sprout, their roots thicken as they creep down the log to the soil below. Simultaneously, fungi feast on the decaying tree, leading to the gradual rot of the nurse log that provides even more nutrient-rich sustenance for the young plants. It often takes several decades for a nurse log to decay completely, at which time the seedlings’ roots have become strong and thick enough to support themselves. And there you have it — another cycle begins! \\
\midrule
\textit{\textbf{Example negative document}} \\
\midrule
Wood-decay fungi can be classified according to the type of decay that they cause. The best-known types are brown rot, soft rot, and white rot.[4][5] Each produce different enzymes, can degrade different plant materials, and can colonise different environmental niches.[6] Brown rot and soft rot both digest a tree's cellulose and hemicellulose but not its lignin; white rot digests lignin as well. The residual products of decomposition from fungal action have variable pH, solubility and redox potentials. Over time this residue becomes incorporated in the soil and sediment so can have a noticeable effect on the environment of that area.[6]\\
\bottomrule
\end{tabular}
\label{tab:biology_example}
\end{table*}

\begin{table*}[t]
\caption{\textbf{Academia example.} A randomly sampled query with one positive and one negative document from MM-BRIGHT.}
\centering
\begin{tabular}{p{1\textwidth}}
\toprule
\textit{\textbf{Query}} \\
\midrule
An academic who died over 70 years ago still has Google Scholar profile with verified email. How can this be?\\ 
I'm surprised to see that an academic who died over 50 years ago has a Google Scholar profile with a verified email (\\ 
image: enter image description here\\ 
How is that possible? I see the email domain, melipona.org, is some online shop. \\
\midrule
\textit{\textbf{Query images}} \\
\midrule
\includegraphics[width=0.5\linewidth]{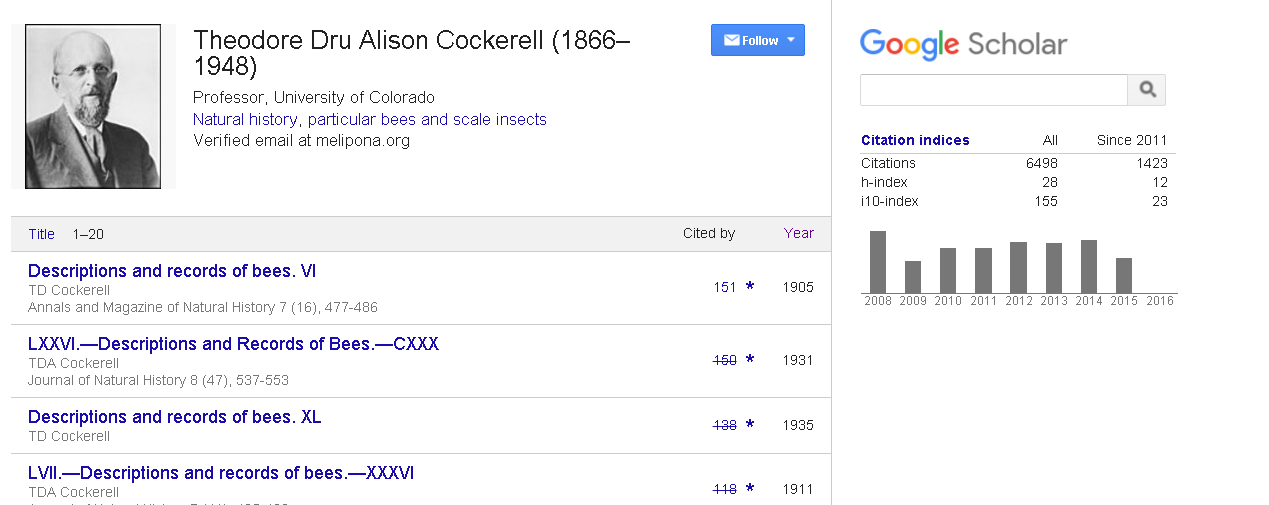}\\
\midrule
\textit{\textbf{Example positive document}} \\
\midrule
Like many viewers and commentators, we were left puzzled by Newton's profile. How did this happen?\\ 
Does someone at MIT, perhaps an official representative manage Google Scholar profiles for former professors?\\ 
BleepingComputer contacted MIT and Google multiple times well before publishing but did not hear back.\\ 
We reckon however that creating an author profile on Google Scholar and "verifying" the email address for it may be more straightforward than it seems.\\ 
In recent times, "verified" profiles on social media platforms such as X (Twitter) and Meta's Facebook and Instagram have generated much buzz, particularly after platforms have steered towards pay-for-blue-tick models and with scammers abusing the opportunity to mislead people.\\ 
"Verified" social media profiles have traditionally been associated with the likes of elite, famous, or notable public figures, and as such, these are generally checked for authenticity by a team behind the scenes. The same goes for accounts that pay for a blue tick—ultimately, there is a team of humans (paired with technology) running some basic checks to ensure that the person on social media is who they claim to be.\\ 
It's therefore understandable how the presence of the mere word "verified" on public profiles could be misinterpreted by some as a sign of the profile owner's identity having been checked.\\ 
A Google Scholar profile, on the other hand, makes no claims of Google verifying the identity of the profile owner. Instead, profiles state that their email address has been verified and hosted at the said institution. \\
\midrule
\textit{\textbf{Example negative document}} \\
\midrule
We recommend that you also enter your Western University email address and authorize Google Scholar to display your affiliation to Western University.  Later, check your email to complete the verification process. This will authorize Google Scholar to display your affiliation with the Western University as “verified.”On the next page, you'll see groups of articles written by people with names similar to yours. Add all articles that you have written. Keep in mind your articles may be in several different groups, and some groups may occasionally include articles by several different authors. If you publish under several different names, you may need to do several searches to add all your articles. Once you're done with adding articles, it will ask you what to do when the article data changes in Google Scholar. You can either have the updates applied to your profile automatically, or you can choose to review them beforehand. In either case, you can always go to your profile and make changes by hand.\\
\bottomrule
\end{tabular}
\label{tab:academia_example}
\end{table*}

\begin{table*}[t]
\caption{\textbf{Pm example.} A randomly sampled query with one positive and one negative document from MM-BRIGHT.}
\centering
\small
\begin{tabular}{p{1\textwidth}}
\toprule
\textit{\textbf{Query}} \\
\midrule
Re-Organize Task IDs in MS Project\\ 
In MS Project Professional (Online Desktop Client), I made several tasks that are all occurring simultaneously, with nested tasks for each one. To prioritize and organize my task list in order of importance, I dragged the tasks around. Now, my task ID list is all jumbled up and it won't save all of sub-task rearrangements that I do.\\ 
Is there any way that I can drag all my tasks in the order that I want and then set the task IDs to auto update in numerical order?\\ 
Thanks\\ 
image: Project with Jumbled IDs \\
\midrule
\textit{\textbf{Query images}} \\
\midrule
\includegraphics[width=0.3\linewidth]{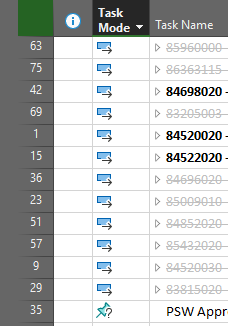}\\
\midrule
\textit{\textbf{Example positive document}} \\
\midrule
The Unique ID field contains the number that Microsoft Office Project automatically designates whenever a new task, resource, or assignment is created in the current project. This number indicates the sequence in which the task, resource, or assignment was created, regardless of placement in the schedule.\\ 
There are several categories of Unique ID fields.\\ 
Data Type    Integer\\ 
Unique ID (task field)\\ 
Entry Type    Calculated\\ 
How Calculated    As you create new tasks, Project adds a unique number to each task in the project. This number is unique in that it is never rearranged or reused if the task is moved or deleted.\\ 
Best Uses    Add the Unique ID field to a task sheet when you want to display or filter the unique ID for tasks.\\ 
Example    You want to review tasks by the order in which they were created. In the Task Sheet view, you sort the tasks by the Unique ID field.\\ 
Remarks    If tasks are moved, inserted, or deleted, the task IDs change to reflect the new task sequence. However, the Unique ID always remains constant, regardless of any editing.\\ 
Unique ID (resource field)\\ 
Entry Type    Calculated\\ 
How Calculated    As you add resources, Project assigns a unique number to each resource within the project. This number is unique in that it is never rearranged or reused if the resource is moved or deleted.\\ 
Best Uses    Add the Unique ID field to a resource sheet when you want to display or filter the unique ID for the resource.\\ 
Example    You want to review the project's resources by the order in which they were added to the project. In the Resource Sheet view, you sort the resources by the Unique ID field.\\ 
Remarks    If resources are moved, inserted, or deleted, the resource IDs change to reflect the new sequence. However, the Unique ID always remains constant, regardless of any editing.\\ 
\midrule
\textit{\textbf{Example negative document}} \\
\midrule
In the Add link to <Work Item> dialog, you can open the Choose Linked Work Items dialog to select one or more work items to link to. If you plan to find and list work items by using a saved query, first define the query. In the Add link to <Work Item> dialog, select Browse (Visual Studio): In the Choose Linked Work Items dialog, select the method to get the work items to link to: You configure the fields on this dialog in the same way as the Get Work Items dialog. For more information, see Add existing work items to your worksheet. In the Choose Linked Work Items dialog, select the method to get the work items to link to: You configure the fields on this dialog in the same way as the Get Work Items dialog. For more information, see Add existing work items to your worksheet. \\
\bottomrule
\end{tabular}
\label{tab:pm_example}
\end{table*}

\begin{table*}[t]
\caption{\textbf{Gaming example.} A randomly sampled query with one positive and one negative document from MM-BRIGHT.}
\centering
\small
\begin{tabular}{p{1\textwidth}}
\toprule
\textit{\textbf{Query}} \\
\midrule
How do I destroy Nekker nests?\\ 
In "The Nekker Contract" quest, I need to destroy all of the Nekker nests. I found my first Nekker nest, but when I try to destroy it, Geralt just says, "I've got to blow up this nest," and doesn't do anything.\\ 
So I assume that I need something to blow it up—what would that be?\\ 
Screenshot of a Nekker nest with the left-click "destroy nest" option:\\ 
image: a Nekker nest with the left-click \\
\midrule
\textit{\textbf{Query images}} \\
\midrule
\includegraphics[width=0.5\linewidth]{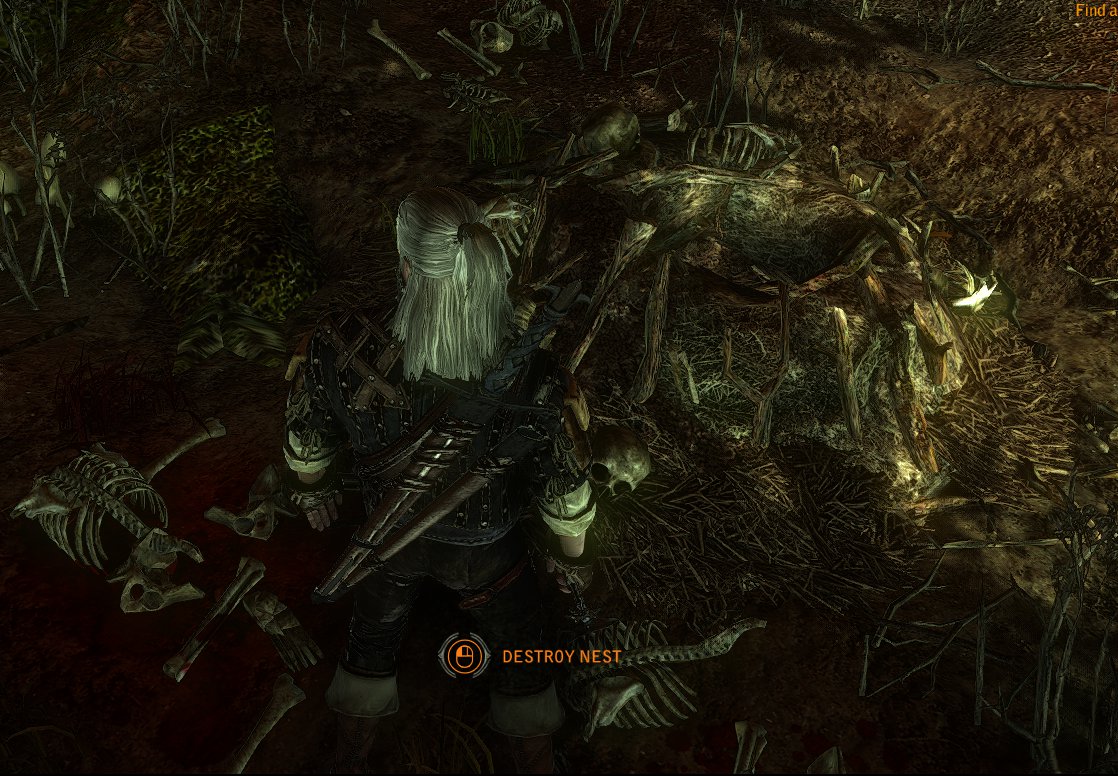}\\
\midrule
\textit{\textbf{Example positive document}} \\
\midrule
Note that only the Grapeshot bomb will work - the other bombs created cannot be used to destroy the Nekker nests.\\ 
There are four nekker tunnels that you'll need to find and destroy in total. They look like brown mounds of dirt with skulls lying on them.\\ 
Three of the nests are positioned close together near the large trees just outside the Flotsam city walls (all three are in sight of the city wall's torches) -- two are right next to one another, separated by a giant tree. The third can be found by going southwest (check your map) from the two grouped nests.\\ 
One final nest is located near the waterfalls along the side of the elven bath ruins. In fact, if you head north from the bottom of the trail leading to the bath ruins, you will probably run headlong into the nest.\\ 
As you are near nests, you'll come under attack by nekker monsters. Their purpose is to guard the nests, but they're actually helpful in that their presence lets you know when you're getting close to a target. Kill the nekkers, then destroy their nests. You don't have to guess about where to aim your bomb when dealing with the nests. Get near enough to a nest and adjust the camera so that you're looking down at it. You'll cause the needed prompt to come up on-screen and then you can simply press the indicated button to produce the explosion using a bomb from your inventory. Repeat the process for each nest.\\ 
Once you destroy the final nest, go see Louis Merse in his house in Flotsam about your reward. When you have the oren, the quest is complete. \\
\midrule
\textit{\textbf{Example negative document}} \\
\midrule
“People,” Geralt turned his head, “like to invent monsters and monstrosities. Then they seem less monstrous themselves. When they get blind-drunk, cheat, steal, beat their wives, starve an old woman, when they kill a trapped fox with an axe or riddle the last existing unicorn with arrows, they like to think that the Bane entering cottages at daybreak is more monstrous than they are. They feel better then. They find it easier to live.” –Andrzej Sapkowski, The Last Wish The Witcher role playing game is set in a world of dark, adult fantasy where happy endings are rare and actions have consequences, often swift and brutal. In the war-torn lands of the Continent, murder, assault, and theft are a daily threat and only the strong survive. Thieves run rampant and mercenaries are as numerous as the doctors who heal them or the priests who inter them. Only the bravest souls venture into the wilderness, where witchers armed with razor-sharp silver swords hunt monsters armed not only with tooth and claw but magic as well. In the cities, the poor scrabble to survive in filthy tenements and the rich perch high above on the unstable towers of their power, some advised by clever mages constantly looking for the best opportunities. Humans reign supreme, save for a few small communities. Once-proud elves and dwarves are kept in hovels and executed daily, often for crimes they have not even committed. Hatred and fear fuel this great blaze, and most will continue feeding the fire ‘til their last dying breath. There are no heroes, only people. Whether you’re a hard-bitten mercenary who lives from job to job never asking questions, or an idealistic bard traveling the land to spread some cheer and revelry in these dark days, you must fight if you want to survive in a world determined to break you.
 \\
\bottomrule
\end{tabular}
\label{tab:gaming_example}
\end{table*}

\begin{table*}[t]
\caption{\textbf{Travel example.} A randomly sampled query with one positive and one negative document from MM-BRIGHT.}
\centering
\small
\begin{tabular}{p{1\textwidth}}
\toprule
\textit{\textbf{Query}} \\
\midrule
What is the use of that Internal rail?\\ 
On Greek suburban railway (proastiakos) I noticed that on stations have an internal rail as long as the train platform:\\ 
image: Internal rail inage on proastiakos service greece\\ 
I was wondering what it the use for this internal rail? Also the suburban railway is electric and uses overhead lines and not a third rail layout.\\ 
The rail does not extend beyond the platform (I think it is even a bit shorter). \\
\midrule
\textit{\textbf{Query images}} \\
\midrule
\includegraphics[width=0.6\linewidth]{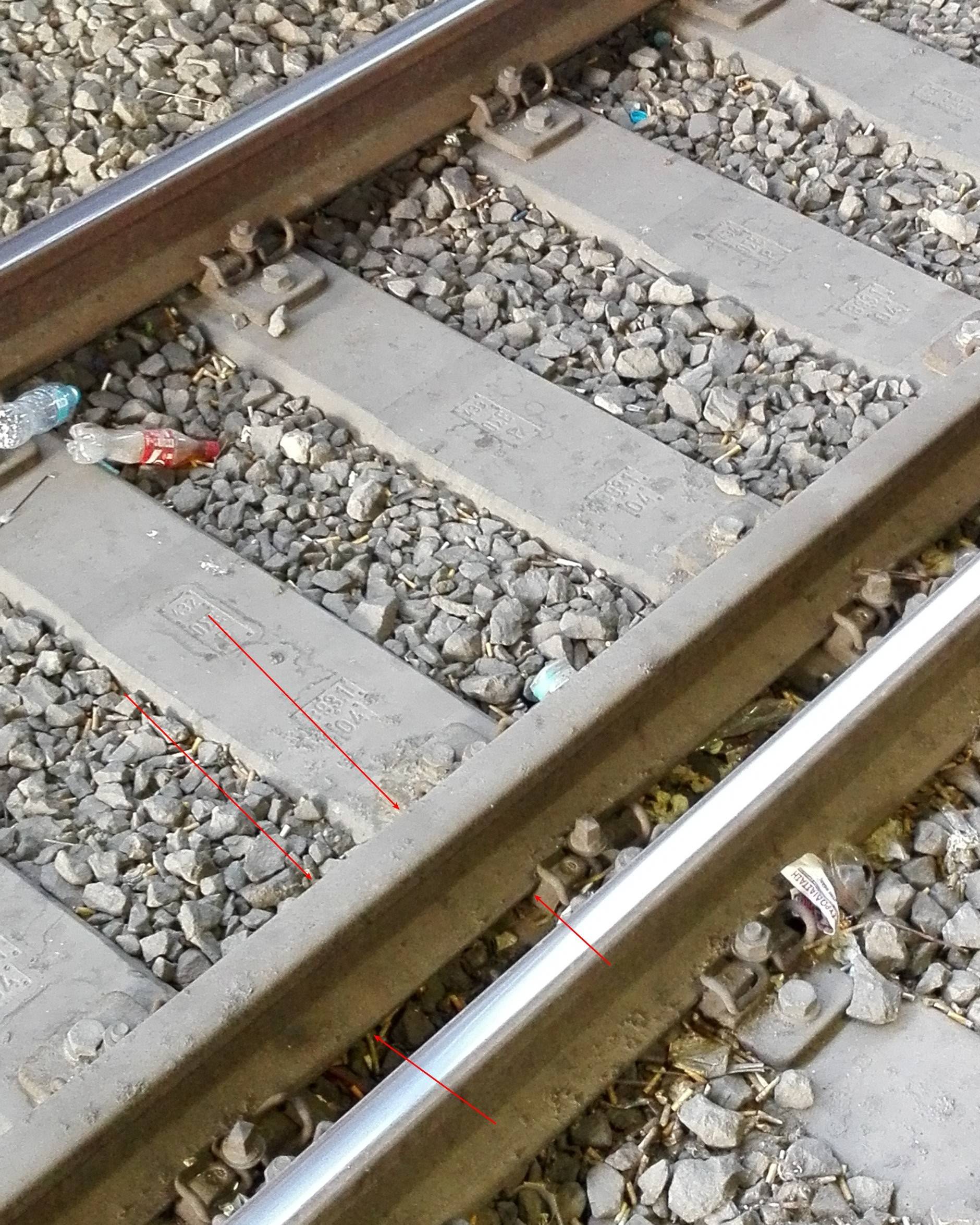}\\
\midrule
\textit{\textbf{Example positive document}} \\
\midrule
Sharp curves\\ 
On sharp curves, guard rails may be placed inside the inner rail, where they engage the back of the flange of the wheel on that side.[3]\\ 
Switches\\ 
Guard rails in use on a switch\\ 
Guard rails may be incorporated in switches, where they serve to prevent derailments caused by a train's wheels passing through the wrong side of the frog (the point where the straight and diverging rails cross).[4] Guard rails in this case are typically bolted to the traffic rails on each end, with a clamp placed towards the center to prevent movement.[4 \\
\midrule
\textit{\textbf{Example negative document}} \\
\midrule
 Internal Abseil System: Rail System for Internal Safety and Productivity
Internal Abseil System – DAVITPRORAIL IT system delivers a robust, all-in-one solution for improving safety and productivity in elevated work settings. Specifically designed for today’s construction and maintena -nce needs, it ensures a streamlined and efficient user experience, making it an ideal choice for high-performance work environments.\\
\bottomrule
\end{tabular}
\label{tab:travel_example}
\end{table*}

\end{document}